\documentclass{article}

\usepackage{preprint}

\usepackage{amsmath, amsthm, amssymb, amsfonts}

\usepackage[numbers,square]{natbib}
\usepackage[utf8]{inputenc}
\usepackage[T1]{fontenc}
\usepackage{xcolor}
\usepackage[colorlinks=true,
            linkcolor=purple,
            urlcolor=blue,
            citecolor=cyan,
            anchorcolor=black]{hyperref}
\usepackage{booktabs}
\usepackage{nicefrac}
\usepackage{microtype}
\usepackage{float}
\usepackage{graphicx}
\usepackage{multirow}
\usepackage{array}
\usepackage{longtable}
\usepackage{enumitem}

\usepackage{newfloat}
\DeclareFloatingEnvironment[name={Supplementary Figure}]{suppfigure}
\DeclareFloatingEnvironment[name={Supplementary Table}]{supptable}
\usepackage{sidecap}
\sidecaptionvpos{figure}{c}

\usepackage{titlesec}
\titleformat{\section}[block]{\Large\bfseries\raggedright}{}{0em}{}
\titleformat{\subsection}[block]{\large\bfseries\raggedright}{}{0em}{}
\titleformat{\subsubsection}[block]{\normalsize\bfseries\raggedright}{}{0em}{}
\titlespacing\section{0pt}{14pt plus 4pt minus 3pt}{4pt plus 2pt minus 1pt}
\titlespacing\subsection{0pt}{12pt plus 3pt minus 3pt}{2pt plus 1pt minus 1pt}
\titlespacing\subsubsection{0pt}{10pt plus 3pt minus 3pt}{1pt plus 1pt minus 1pt}

\usepackage{tikz}

\newcommand{\cedar}{\textsc{Cedar}}

\input{supp_xref}

\graphicspath{{figures/main_figures/}{figures/supplementary_figures/}}

\SetBgContents{}
\makeatletter
\renewcommand{\@maketitle}{%
  \vbox{%
    \hsize\textwidth
    \linewidth\hsize
    \centering
    {\LARGE\sc \@title\par}
    \vskip 0.15in
    \def\And{%
      \end{tabular}\hfil\linebreak[0]\hfil%
      \begin{tabular}[t]{c}\bf\rule{\z@}{24\p@}\ignorespaces%
    }
    \def\AND{%
      \end{tabular}\hfil\linebreak[4]\hfil%
      \begin{tabular}[t]{c}\bf\rule{\z@}{24\p@}\ignorespaces%
    }
    \begin{tabular}[t]{c}\bf\rule{\z@}{24\p@}\@author\end{tabular}%
  \vskip 0.2in
  }
}
\makeatother

\title{Agentic LLM reasoning in pediatric diabetes outcome prediction from multimodal medical records}

\usepackage{authblk}

\author[1,2,3,4]{Daniel Palacios}
\author[2,3,4,5]{Guantong Qi}
\author[6]{Anna Kadi}
\author[2]{Natali Borrego}
\author[2,3,7]{Sukru Aras}
\author[8,11]{Michael L. Ferm}
\author[8]{Elizabeth Kubota-Mishra}
\author[8,9,10]{Maria J. Redondo}
\author[1,2,3,4]{Zhandong Liu\textsuperscript{$*$}}
\author[8,9]{Mustafa Tosur\textsuperscript{$*$}}
\author[2,3,4]{Hu Chen\textsuperscript{$*$}}
\affil[1]{Quantitative and Computational Biosciences, Baylor College of Medicine, Houston, Texas, USA}
\affil[2]{Department of Pediatrics, Baylor College of Medicine, Houston, Texas, USA}
\affil[3]{Jan and Dan Duncan Neurological Research Institute, Texas Children's Hospital, Houston, Texas, USA}
\affil[4]{Data Science Center, Texas Children's Hospital, Houston, Texas, USA}
\affil[5]{Genetics and Genomics Graduate Program, Baylor College of Medicine, Houston, Texas, USA}
\affil[6]{School of Medicine, Baylor College of Medicine, Houston, Texas, USA}
\affil[7]{Department of Epidemiology and Population Health, Baylor College of Medicine, Houston, Texas, USA}
\affil[8]{Department of Pediatrics, Division of Pediatric Diabetes and Endocrinology, Baylor College of Medicine, Texas Children's Hospital, Houston, Texas, USA}
\affil[9]{Children's Nutrition Research Center, USDA/ARS, Department of Pediatrics, Baylor College of Medicine, Houston, Texas, USA}
\affil[10]{Immunobiology Center, Baylor College of Medicine, Texas Children's Hospital, Houston, Texas, USA}
\affil[11]{Division of Pediatric Endocrinology and Diabetes, Department of Pediatrics, NYU Grossman School of Medicine, New York, NY, USA}
\affil[ ]{\small\textsuperscript{$*$}Co-corresponding authors (senior authors);
H.C. is the lead corresponding author. Correspondence: Zhandong Liu
(\texttt{zhandong.liu@bcm.edu}), Mustafa Tosur
(\texttt{mustafa.tosur@bcm.edu}), Hu Chen (\texttt{hu.chen@bcm.edu}).}

\begin{document}

\maketitle

\begin{abstract}
Large language model (LLM) agents are proposed for clinical prediction but are
untested in pediatrics, where cohorts are small and outcomes are uncommon. Using records from 2{,}489 youth with type~2 diabetes (T2D) and 4{,}018
with type~1 diabetes (T1D), spanning structured data, note-derived social
determinants and, in T1D, continuous glucose monitoring data, we predicted seven
T2D outcomes, four also in T1D, at 24~months from 18~months of follow-up.
\cedar{} (Clinical Evidence-Driven Agent for Risk-prediction), an evidence-citing,
self-auditing LLM pipeline, outperformed trained machine learning in both cohorts
(pooled area under the receiver operating characteristic curve 0.815 versus
0.656--0.700 in T2D and 0.756 versus 0.603--0.631 in T1D), most on the sparsely
screened outcomes. Only 4 of 70{,}246 values it cited did not match the chart.
Reading only its cited evidence and reasoning, two pediatric endocrinologists
reached \cedar{}'s prediction in about nine of ten cases, agreeing with each other
on 90.5\%.
\end{abstract}

\vspace{0.35cm}

\section*{Introduction}
\label{sec:introduction}

Diabetes is among the most common chronic diseases of childhood in the
United States. Type~1 diabetes (T1D) remains the predominant pediatric form
(prevalence $\sim$215 per 100{,}000 under 20~years, incidence rising
$\sim$2\%/year~\citep{lawrence2021trends, wagenknecht2023trends,
divers2020trends}), while type~2 diabetes (T2D), once rare in youth, has
emerged as a parallel epidemic driven by childhood obesity (prevalence up
95\% from 2001--2017, incidence $\sim$18 per
100{,}000~\citep{lawrence2021trends, wagenknecht2023trends}). Youth of
Black, Hispanic, and American Indian descent bear a disproportionate burden
of both incidence and poor glycemic outcomes~\citep{copeland2011characteristics,
hood2014psychosocial}. U.S.\ diabetes-related costs exceeded \$400~billion in
2022 across all ages~\citep{parker2024economic}, and pediatric-onset disease
contributes disproportionately to lifetime cost through its longer duration of
exposure.

Pediatric diabetes also differs biologically from adult-onset disease. Youth with
T2D show earlier micro- and macrovascular complications~\citep{dart2014earlier,
tryggestad2015complications} and greater insulin resistance with hyperresponsive
beta cells relative to adults~\citep{rise2018metabolic}, and the TODAY trial found durable glycemic
control from metformin alone in only about half of adolescents with
T2D~\citep{zeitler2012today}. T1D poses distinct challenges as well, including
glycemic variability, recurrent hypoglycemia, and early cardiovascular risk
emergence~\citep{kim2019cvd, bjornstad2016dkd}. This heterogeneity motivates
precision approaches that stratify risk and guide individualized decisions.

Pediatric diabetes care is delivered by a multidisciplinary team. The pediatric
endocrinologist interprets a child's metabolic course alongside treatment history
and social circumstances, and integrates input from diabetes educators,
dietitians and social workers, each of whom contributes a different kind of
information.

Machine learning (ML), by contrast, collapses this reasoning into a single
function approximation. Such models are widely applied to clinical
prediction from electronic health records (EHR)~\citep{rajkomar2018scalable, zhu2021deep}, including in pediatric diabetes~\citep{subramanian2024dka,
javidi2024interpretable}, but treat prediction as pattern matching rather
than reasoning. They also leave largely untapped two modalities that plausibly
add value. Continuous glucose monitoring (CGM) captures temporal dynamics that static glycated hemoglobin (HbA1c) cannot~\citep{cichosz2024hypoglycemia, lutsker2026cgmfm}, and social
determinants of health (SDOH) are documented in narrative notes but rarely
available as structured data~\citep{sdoh_jmir2023, patra2021sdoh} and are now extractable at
scale with LLMs~\citep{sdoh_npj2025, guevara2024sdoh}. Each has demonstrated
prognostic value on its own: CGM-derived features predict short-term
hypoglycemia and glycemic variability in type~1
diabetes~\citep{cichosz2024hypoglycemia}, and psychosocial burden is associated with worsening
glycemic control in youth with diabetes~\citep{hood2014psychosocial}. What is
unresolved is how much either adds \emph{beyond} longitudinal structured EHR in
this setting, and in what representation. This gap between how clinicians
reason and how models predict motivates asking whether an LLM can reason
over the same multimodal evidence a clinician weighs.

Large language models (LLMs) with strong clinical knowledge~\citep{singhal2023medpalm, nori2023gpt4med}
enable explicit, chain-of-thought reasoning over patient data~\citep{wei2022cot},
and a natural next step is to decompose that reasoning across communicating
specialized agents~\citep{park2023generative, du2023debate}, by analogy to a care team.

Recent systems apply this idea to clinical prediction, each covering a slice of
our problem. CARE-AD~\citep{care_ad2025} predicts Alzheimer's onset
from clinical notes, ColaCare~\citep{colacare2025} predicts in-hospital
mortality and readmission through multi-agent collaboration between LLMs and
trained EHR models, MDAgents~\citep{mdagents2024} performs adaptive routing for
question answering and diagnosis, and MoMA~\citep{moma2025} coordinates
modality-specialist agents that convert non-text modalities to text for
multimodal clinical prediction. These establish the
promise of agentic clinical prediction, yet enthusiasm has outpaced evidence.
The gains attributed to multi-agent debate are contested: majority voting, rather
than debate itself, accounts for most of them~\citep{debateorvote2025}, decision
protocols help unevenly across task types~\citep{votingorconsensus2024}, and
debate strategies do not reliably outperform simpler prompting at substantially
higher cost~\citep{smit2024mad}. None of these systems has been evaluated in pediatric settings, where cohorts
are small and the outcomes of interest are uncommon, so a model has few training
examples to learn from. Pediatric diabetes is a demanding place to test them:
a child's record is longitudinal but sparse, the complications being predicted
are uncommon, and the evidence a clinician would weigh is spread across
social-worker notes, structured EHR and, in T1D, continuous glucose
monitoring data. Four questions therefore remain open. First, can LLM reasoning
predict a child's diabetes outcomes more accurately than a conventional
trained machine-learning model? Second, can the reasoning behind such a
prediction be \emph{trusted}, that is, are the cited chart values real rather
than hallucinated? Third, do those cited values actually drive the prediction, or merely accompany it? Fourth, can a clinician follow the reasoning well enough to reach the same conclusion?

\begin{figure}[H]
\centering
\includegraphics[width=\textwidth,height=0.9\textheight,keepaspectratio]{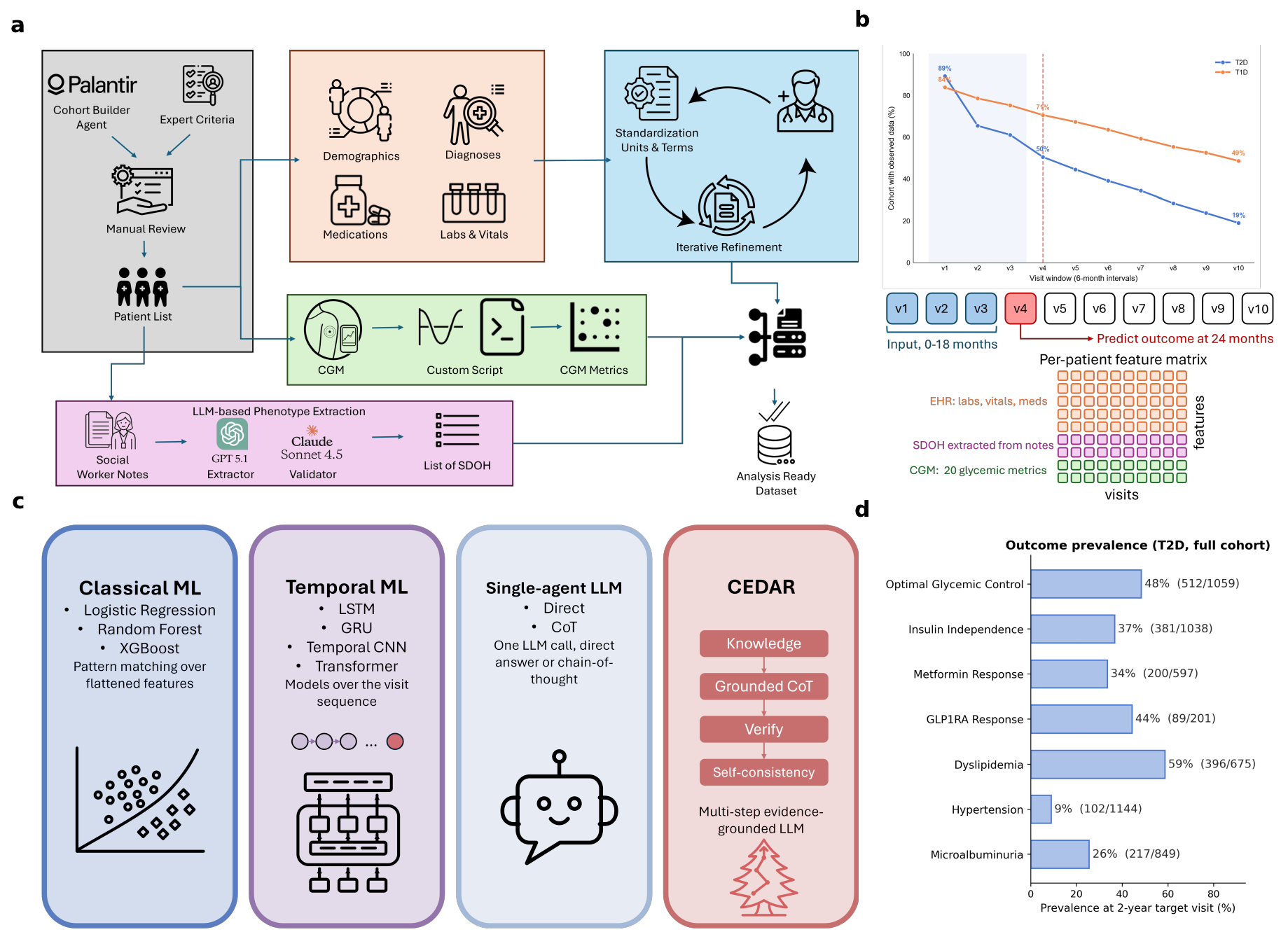}
\caption{\textbf{Study design, data construction, and methods.}
(a)~Data construction. A cohort-builder agent with expert-defined criteria plus
manual endocrinologist review yields the patient list, structured EHR is
standardized through iterative expert refinement, continuous glucose monitoring
(CGM) is processed into 20 per-visit glycemic metrics in T1D, and social
determinants of health (SDOH) are extracted from social-worker notes by a
two-stage LLM pipeline. All modalities are merged into the analysis-ready
dataset. (b)~Data structure and prediction task. Patients are followed over ten
6-month visit windows, and data availability declines over follow-up (top),
which motivates the 2-year primary horizon. The first three windows, months
0--18 and shaded, are the input and the outcome is predicted at visit~4. Each patient
is a feature~$\times$~visit matrix spanning EHR, SDOH and, in T1D, CGM
(schematic). (c)~The model families evaluated, from classical ML over flattened
features and temporal ML over visit sequences to the multi-step \cedar{}
pipeline. (d)~Prevalence of each
outcome at the 2-year target visit across the full assessable T2D population,
annotated with event and assessable counts.}
\label{fig:design}
\end{figure}

We compare LLM reasoning with classical and temporal ML for pediatric diabetes outcome prediction. Using longitudinal Texas Children's Hospital data, we predict seven type~2 and four type~1 diabetes outcomes at 24~months from the first 18~months of follow-up (Figure~\ref{fig:design}). We
introduce \cedar{} (Clinical Evidence-Driven Agent for Risk-prediction), a
single-model pipeline that reasons over structured EHR data, SDOH extracted from
social-worker notes and, for T1D, CGM, and cites the record values behind each
prediction so that a clinician can check them (Methods, Section~6).

We address these questions with four analyses. We compare \cedar{} with
classical and temporal ML across eleven cohort--outcome pairs in two cohorts, and
audit the faithfulness of the evidence it cites with established
rationale-evaluation protocols~\citep{deyoung2020eraser, jacovi2020faithfulness}.
Two pediatric endocrinologists review its reasoning blinded to the outcome, and an
information ablation measures how much of the predictive signal lies in the
outcome's own history. We also test whether supplying a trained model's score to
\cedar{} as a tool improves prediction.

\section*{Results}
\label{sec:results}

\subsection*{Discrimination of \cedar{} and machine learning in both pediatric
diabetes cohorts}
\label{sec:res_landscape}

We built a benchmark in which the first three 6-month visit windows, covering months 0--18, predict each outcome at 24~months, across seven T2D and four T1D outcomes (Figure~\ref{fig:design}; Methods, Sections~1 to~3). Prevalence ranges from 58.7\% for dyslipidemia to 8.9\% for hypertension in T2D, and from 24.6\% to 4.7\% for the same two outcomes in T1D (Supplementary Table~\ref{stab:outcome_prevalence}). Cohort composition and feature availability are in Supplementary Tables~\ref{stab:cohort_composition} and~\ref{stab:feature_availability}. Two companion studies describe these cohorts in more detail, in T2D (D.~Palacios et al., manuscript in preparation) and in T1D (M.~Ferm et al., manuscript in preparation).

\begin{figure}[H]
\centering
\includegraphics[width=\textwidth]{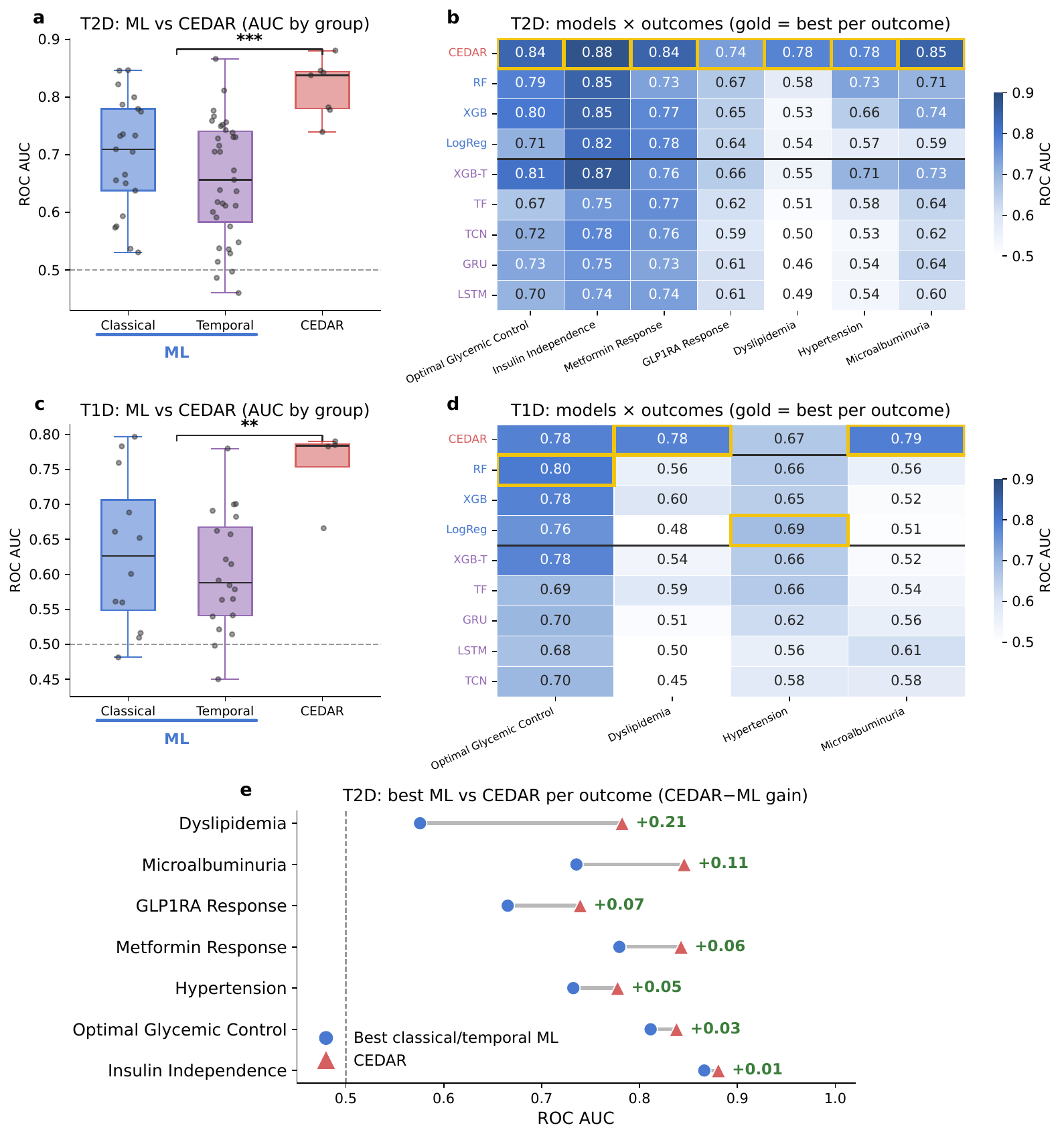}
\caption{\textbf{The performance landscape: \cedar{} beats machine learning in
both cohorts.} The trained statistical models form one family split into two
subgroups, classical ML over the flattened feature matrix and temporal ML over the
visit sequence, compared with the full \cedar{} pipeline.
\textbf{(a,~b)~T2D; (c,~d)~T1D.} (a,~c)~ROC-AUC by group, one point per model and
outcome. \cedar{} sits well above ML in both cohorts (Mann--Whitney~$U$ with one ROC-AUC per model and outcome, $^{***}p<0.001$ in T2D and $^{**}p<0.01$ in T1D, exact
statistics in Supplementary Table~\ref{stab:pooled_stats}; dashed line = chance).
(b,~d)~Model~$\times$~outcome ROC-AUC matrix, \cedar{} above the ML rows, with the
best model per outcome outlined in gold. \cedar{} is best on every T2D outcome and
on two of four in T1D. (e)~Best ML baseline (circle) versus \cedar{} (triangle) per
T2D outcome, ordered by the \cedar{}-over-ML gain, which is widest where signal is
scarce and narrowest on the best-powered outcomes. Per-outcome confidence intervals
are in Supplementary Table~\ref{stab:cedar_vs_best_ml}.}
\label{fig:landscape}
\end{figure}

The benchmark compares two machine-learning subgroups, \emph{classical} models
over the flattened feature matrix and \emph{temporal} models over the visit
sequence, with the full \cedar{} pipeline (Figure~\ref{fig:landscape}; Methods,
Sections~4 and~6). Across the seven T2D outcomes \cedar{} outperforms ML
(Figure~\ref{fig:landscape}a; Mann--Whitney~$U$ treating each model's score on each outcome as one observation, $p<0.001$). The pooled mean area under the receiver operating characteristic curve (ROC-AUC) was 0.815$\pm$0.049 against 0.700$\pm$0.100 for classical ML and 0.656$\pm$0.103 for
temporal ML. Exact test statistics for every pooled comparison are reported in
Supplementary Table~\ref{stab:pooled_stats}. We also benchmarked naive
single-call LLM implementations on the same base model and patients, which reached
discrimination comparable to \cedar{} (Supplementary Note~4). \cedar{}'s lead over
the best ML model stayed positive on all four base models tested, including the
open-weight Llama~3.3~70B that an institution could run locally (Supplementary
Figure~\ref{sfig:backbone_vs_ml}).

The same ordering holds in T1D (Figure~\ref{fig:landscape}c,~d), where \cedar{}
reaches 0.756$\pm$0.060 against 0.631$\pm$0.110 for classical ML and
0.603$\pm$0.083 for temporal ML ($p=0.007$ against ML as a whole, $p=0.058$
against classical ML alone, where only four outcomes are evaluable). The advantage is nonetheless uneven
across outcomes. In T2D the per-outcome heatmap (Figure~\ref{fig:landscape}b) and
the best-ML-versus-\cedar{} comparison (Figure~\ref{fig:landscape}e) show margins
as wide as $+0.21$ on dyslipidemia and $+0.11$ on microalbuminuria but as narrow as
$+0.03$ on optimal glycemic control and $+0.01$ on insulin independence. In T1D
individual ML models match or lead \cedar{}: random forest leads on optimal
glycemic control at 0.797 against 0.785, and logistic regression reaches 0.688 on
hypertension against 0.666.

At the level of individual outcomes the evidence is weaker than the pooled
contrast suggests, since the 95\% bootstrap intervals of \cedar{} and the best ML
model separate only on dyslipidemia, in both cohorts (Supplementary
Table~\ref{stab:cedar_vs_best_ml}). And although \cedar{} needs no training, its
inference costs about \$0.10 per patient against less than \$$10^{-6}$ for a
trained model (Methods, Section~8; Supplementary Figure~\ref{sfig:economics}).

\subsection*{Faithfulness of the evidence \cedar{} cites, and clinician review}
\label{sec:res_faithfulness}

\begin{figure}[H]
\centering
\includegraphics[width=\textwidth]{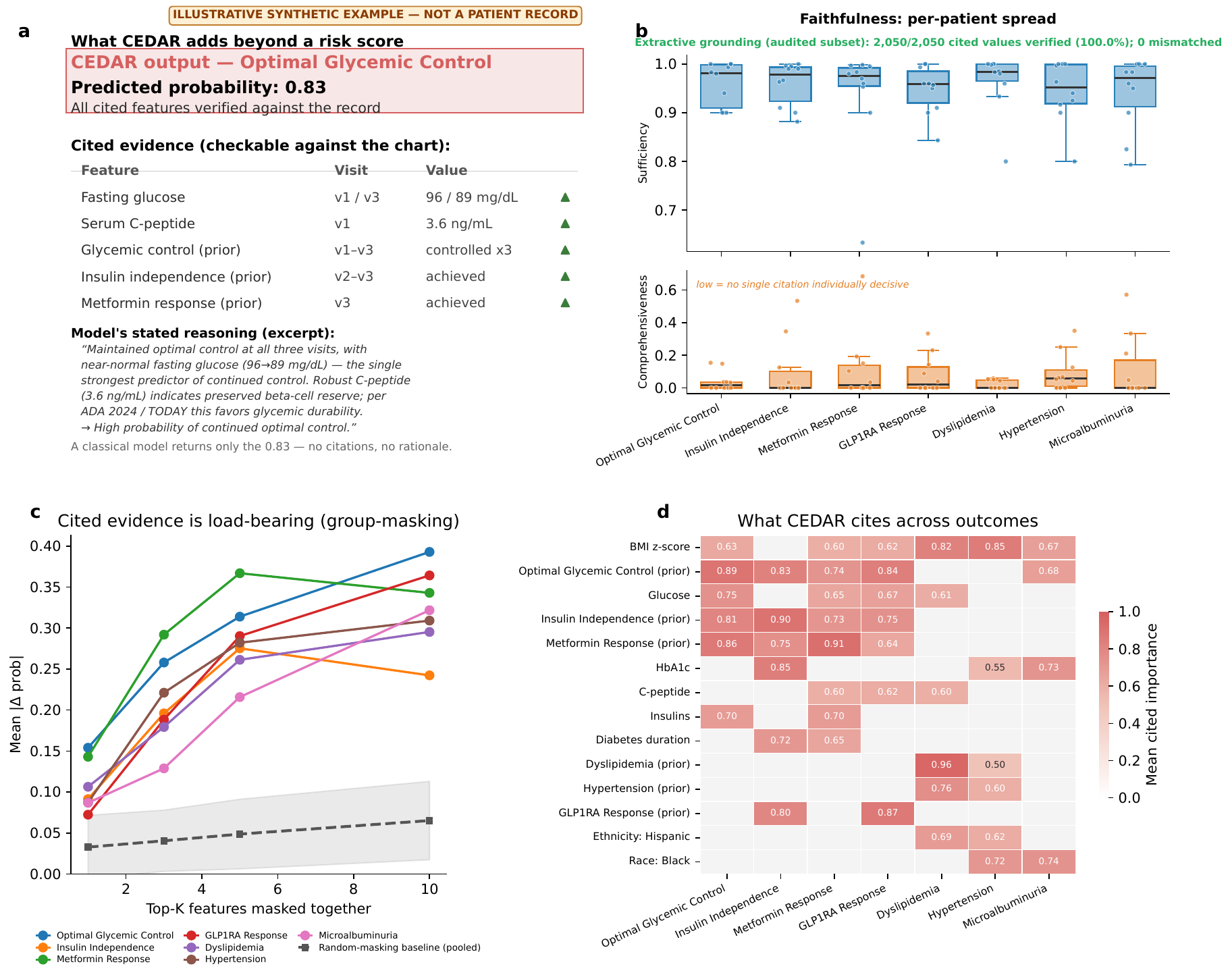}
\caption{\textbf{What \cedar{} adds beyond a risk score: faithful, auditable
reasoning (T2D).}
(a)~A \cedar{} output alongside its probability, showing the cited record
values it reasoned from with feature, visit and value, each checkable against the
chart, where a classical model emits only a risk score. Every value in this panel
is a synthetic composite constructed to represent a high-grounding case, not any
patient's record. (b)~Faithfulness per outcome across audited patients, as box
plots over individual patients. \emph{Sufficiency} is high because the cited
evidence alone reproduces the prediction, whereas \emph{comprehensiveness} is low
because no single citation is individually decisive. \emph{Extractive grounding}
is a near-ceiling fabrication check, so it is annotated as a count rather than
plotted; the count shown covers the 70-record counterfactual subsample, 2{,}050
citations, whereas the cohort-wide check over all 1{,}600 T2D records, 47{,}803
citations with 4 mismatches, is reported in the Results. (c)~Masking the cumulative top-$K$ cited features moves the prediction
steadily further for every outcome, far above a size-matched random-masking
baseline (dashed gray, pooled), so the prediction depends specifically on
the cited evidence rather than on removing any features at all. (d)~Mean cited
importance for the features \cedar{} cites across the most outcomes, with gray cells not among an outcome's top citations. \cedar{} relies most on clinically
coherent drivers, BMI z-score, the outcome's own recent history, and the glucose
and HbA1c axis.}
\label{fig:faithfulness}
\end{figure}

Where a classical model returns only a risk score, \cedar{} also returns the
record values it reasoned from, each traceable to the chart
(Figure~\ref{fig:faithfulness}a). The record shown there is synthetic. Any chain-of-thought prompt narrates a rationale; what
distinguishes \cedar{} is that the rationale arrives as a fixed table of cited
features, visits and values (Methods, Section~6), so each claim resolves to a
record value a clinician can look up.

Two audits at different scales, both built on established rationale-evaluation protocols~\citep{deyoung2020eraser, jacovi2020faithfulness}, test whether that trace holds up.

We first checked every scored patient-outcome record for fabricated evidence, 1{,}600 in T2D and 890 in T1D, comparing each cited feature, visit and value against the chart. Of 47{,}803 cited values in T2D
only 4 fail to match the record, three on metformin response and one on
microalbuminuria, and in T1D all 22{,}443 match. Fabricated evidence is therefore
essentially absent. This is a check on transcription rather than on clinical
judgment.

We then asked whether the cited values drive the prediction or merely decorate
it, on a deeper counterfactual audit of up to 10 patients per outcome, giving 70
records and 2{,}050 cited features in T2D and 40 records and 1{,}046 cited
features in T1D. The cited evidence alone reproduces the prediction, with mean
sufficiency 0.95 in both cohorts, while no single citation is decisive, with mean
comprehensiveness 0.08 in T2D and 0.05 in T1D
(Figure~\ref{fig:faithfulness}b). Masking the top cited values cumulatively
moves the prediction steadily further, from a mean change of 0.11 at the top
citation to 0.32 at the top ten, about five times the shift from masking the
same number of \emph{randomly chosen} available values, 0.065 at the top ten
(Figure~\ref{fig:faithfulness}c). Cited features therefore carry more of the
prediction than an equally sized arbitrary set, and the prediction rests on the
combined weight of many mutually supporting clues rather than on any one of
them, much as a clinician reasons.

Finally, the evidence \cedar{} highlights is clinically sensible
(Figure~\ref{fig:faithfulness}d). Across outcomes it relies most on recognized
drivers such as BMI z-score, the outcome's own recent history, and the glucose and
HbA1c axis.
To test whether a clinician can follow that reasoning, two board-certified
pediatric endocrinologists reviewed 63 T2D patient-outcome cases, sampled across
the four confusion cells and presented as \cedar{}'s citation table and
reasoning chain with the probability withheld. Each rated clinical usefulness
from 1 to 5 and named the prediction the rationale supports, blinded to the
recorded outcome and to whether \cedar{} was correct; both completed all 63.
Working independently, each agreed with \cedar{}'s prediction from the rationale
alone in 92.1\% and 88.9\% of cases (Cohen's $\kappa=0.84$ and $0.78$), and the two
agreed with each other on 90.5\% (57/63, $\kappa=0.81$;
Supplementary Figure~\ref{sfig:annotation}a). Mean usefulness was 4.68 and 3.95 of 5 (Supplementary Figure~\ref{sfig:annotation}c). The cases \cedar{} scored
incorrectly were largely the cases the specialists also scored incorrectly: on 29
of 31 errors at least one expert reached the same incorrect call, and on 24 both
did (Supplementary Figure~\ref{sfig:annotation}b). Because the specialists saw only the evidence \cedar{} cited, this shows that its
errors cannot be detected from its rationale, not that clinicians reading the full
record would make the same errors. Alongside its accuracy gain over ML
(Figure~\ref{fig:landscape}), what \cedar{} adds
is a rationale a clinician can read and check. Further configurations tested on top of \cedar{} and the full per-citation audit are in Supplementary Figures~\ref{sfig:omitted_configs} and~\ref{sfig:faithfulness}.

\subsection*{Error overlap between \cedar{} and machine learning, and machine-learning
scores as a tool}
\label{sec:res_errors}

\begin{figure}[H]
\centering
\includegraphics[width=\textwidth]{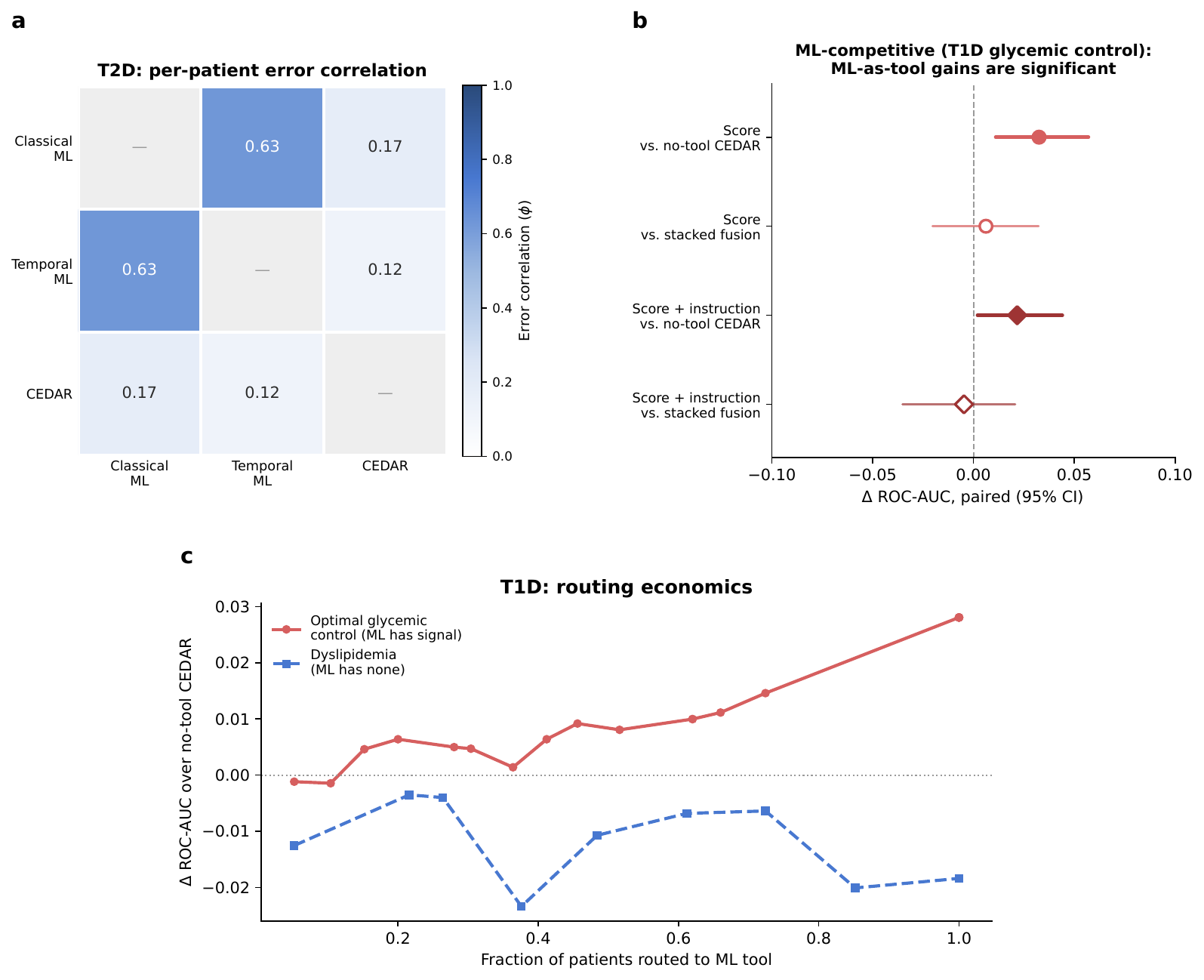}
\caption{\textbf{\cedar{} and machine-learning errors are decorrelated, and
targeted tool use exploits that.}
(a)~Per-patient error correlation among classical ML, temporal ML and \cedar{} in
T2D, as $\phi$ of the misclassification indicators at a 0.5 threshold. Errors
overlap strongly within the ML family yet are largely independent between ML and
\cedar{}, so the two families make mistakes on different patients. T1D shows the same
pattern, 0.61 within ML and 0.03 to 0.06 between ML and \cedar{}.
(b)~Supplying the best-per-outcome ML model's frozen score to \cedar{} as an
in-context tool on T1D optimal glycemic control, the outcome where a standalone ML
model was strongest, either alone (circles) or with an instruction to weigh it
against the record (diamonds). Paired-bootstrap $\Delta$ROC-AUC against no-tool
\cedar{} and the stacked late-fusion combiner, where a filled marker means the 95\% confidence interval excludes zero. The same analysis on a representative outcome on which ML performs poorly, T2D dyslipidemia,
finds nothing significant (Supplementary Figure~\ref{sfig:hybrid_b}a).
(c)~Routing economics in T1D: the gain over no-tool \cedar{} as the fraction of
patients routed to the ML tool rises. Optimal glycemic control, where the ML model
holds complementary signal, climbs with more routing, while dyslipidemia stays flat
to negative. The learned routing policy evaluated on every outcome is in
Supplementary Figure~\ref{sfig:routing_policy}.}
\label{fig:errors}
\end{figure}

We next asked whether the ML models and \cedar{} make the \emph{same} mistakes
(Figure~\ref{fig:errors}a shows T2D). Within the ML family errors overlap strongly, with an error correlation ($\phi$,
the correlation between two models' misclassification indicators) of 0.63 in T2D and 0.61 in T1D between the best classical and best temporal models. Between \cedar{} and ML they are largely independent, with $\phi$ of 0.12 to 0.17 in T2D and 0.03 to 0.06 in T1D, so the two families make mistakes on different patients. Error rates also diverge with prevalence (Supplementary Figure~\ref{sfig:error_analysis}a shows T2D). On the outcomes with low prevalence classical ML makes fewer errors, 0.15 against 0.58 for \cedar{} in T2D and 0.05 against 0.31 in T1D, because it predicts the majority class while \cedar{} attempts the rare positive calls. In T2D, \cedar{} errs less on the more balanced outcomes.

That complementarity invites combining them, but \emph{how} matters. Rather than
a black-box combiner, we gave \cedar{} the frozen ML score as an in-context
tool. T1D optimal glycemic control is the outcome where a standalone statistical
model is strongest relative to \cedar{}, with random forest numerically ahead, and
logistic regression is likewise ahead on T1D hypertension, within noise. On
optimal glycemic control the tool gained $+0.03$ ROC-AUC over no-tool \cedar{},
with a bootstrap interval excluding zero, matching a stacked late-fusion combiner
while keeping a checkable rationale (Figure~\ref{fig:errors}b). On a
representative outcome on which ML performs poorly, T2D dyslipidemia, no comparison reached
significance (Supplementary
Figure~\ref{sfig:hybrid_b}a). The hybrid approach therefore helps where the ML
tool is competitive and gives no benefit, at times a small loss, where it is not.

Declining to score the least confident 40\% of patients, ranked by \cedar{}'s own
probability, raised ROC-AUC from 0.767 to 0.825 in T2D (Supplementary
Figure~\ref{sfig:error_analysis}e; Supplementary Note~1). Offering the ML score to
more patients kept helping on T1D optimal glycemic control but not on
dyslipidemia (Figure~\ref{fig:errors}c), and a learned per-patient policy for when
to offer it was safe on every outcome (Supplementary
Figure~\ref{sfig:routing_policy}).

Against that one targeted win, no other attempt to improve \cedar{}'s discrimination moved it beyond noise. These were a pre-specified screen of verifier-gated
self-refinement and a diverse-reasoner ensemble (Supplementary
Figure~\ref{sfig:omitted_configs}), LLM-guided feature engineering
(Supplementary Table~\ref{stab:fe_poc}) and a prompt-level refinement supplying the empirical transition rates (Supplementary
Table~\ref{stab:refinement2}). Accuracy
here appears limited by the information in the data rather than by the strength of
the model. A case-level review of \cedar{}'s T2D errors locates that limit (Supplementary Note~2, Supplementary Figure~\ref{sfig:cedar_refinement} and Supplementary Table~\ref{stab:error_taxonomy}). \cedar{} mostly carries the last observed
outcome state forward and cannot anticipate a change, carrying it forward about as often when the state changed as when it held ($p=0.39$), so most errors
are unpredictable transitions rather than reasoning failures. Recalibration would
improve its probabilities but cannot change any ROC-AUC reported here
(Supplementary Table~\ref{stab:refinement}). A baseline that simply carries the
last observed outcome state forward is beaten by \cedar{} on six of seven T2D
outcomes, by $+0.035$ to $+0.083$ ROC-AUC (Supplementary Note~2).

Further analyses, all reported in the supplement, support these results. Fusing
\cedar{} with the best ML model improved calibration but not discrimination
(Supplementary Figures~\ref{sfig:omitted_configs} and~\ref{sfig:calibration}).
CGM improved prediction only on T1D optimal glycemic control, and SDOH on no
outcome in any model family (Supplementary Figures~\ref{sfig:regime}
and~\ref{sfig:modality}). Every family held its accuracy when trained on
earlier-diagnosed patients and tested on more recently diagnosed ones
(Supplementary Figure~\ref{sfig:temporal_val}). On the adult critical-care
database MIMIC-IV (Medical Information Mart for Intensive Care), \cedar{} led
every model on one-year mortality but trailed supervised ML on readmission and
in-hospital mortality (Supplementary Note~3, Supplementary
Table~\ref{stab:mimic}).

\subsection*{Contribution of outcome history and clinical context}
\label{sec:res_ablation}

\begin{figure}[H]
\centering
\includegraphics[width=\textwidth]{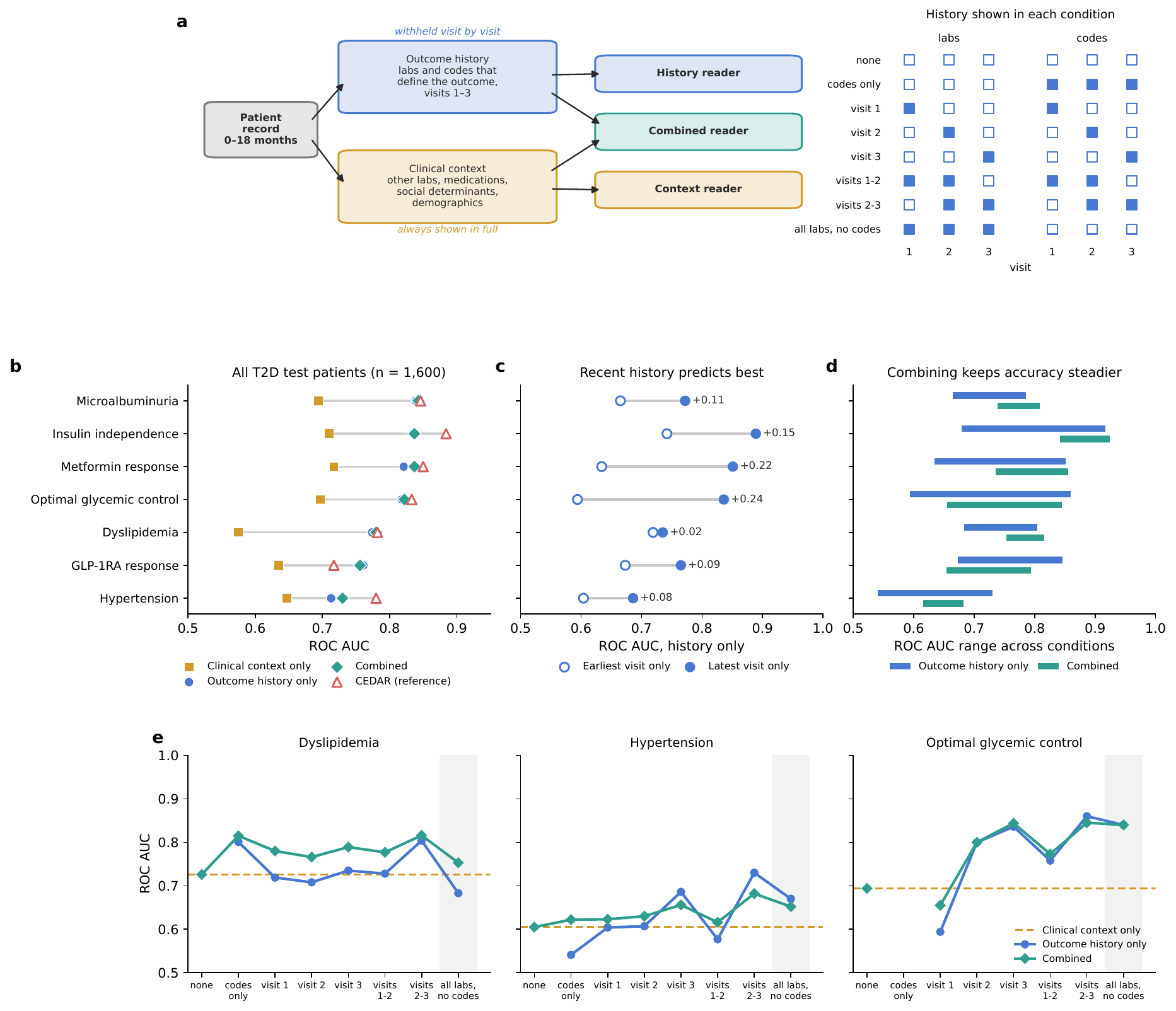}
\caption{\textbf{Recent outcome history drives prediction, and clinical context
recovers accuracy when that history is missing.} (a)~Design. Each record is split
into the outcome history, the laboratory values and diagnosis codes that define the
outcome at visits 1 to 3, and the clinical context, everything else. Three readers
on the same base model score the history alone, the context alone, or both; the
grid shows which visits of the history each condition reveals (filled) or withholds
(open). (b)~All 1{,}600 T2D test patient-outcomes with complete records: the
history alone approaches \cedar{} (hollow triangle), and the context alone is
lower. (c)~History restricted to the earliest or the latest visit, 100 patients per
outcome: the latest visit predicts better on every outcome. (d)~Range of ROC-AUC
across the history conditions for the history alone and for both together; reading
both narrows the range on every outcome. (e)~ROC-AUC by condition for three
outcomes, ordered from no history to the full laboratory history (shaded, not
nested with the others). When the history is thin, reading both recovers toward or
above the context-only level (dashed). All seven outcomes are in Supplementary
Figure~\ref{sfig:ablation_all}, with estimates and 95\% confidence intervals in Supplementary
Table~\ref{stab:ablation}.}
\label{fig:ablation}
\end{figure}

The error review suggests that prediction rests on the outcome's own recent course.
To measure how much of the signal that history carries, we separated each record
into two parts and scored them apart (Figure~\ref{fig:ablation}a; Methods,
Section~11). The \emph{outcome history} is the laboratory values and diagnosis codes
that define the outcome, at visits 1 to 3. The \emph{clinical context} is everything
else, including other laboratory values, medications, social determinants and
demographics. Three separate LLM prompts on the same base model, which we call readers, scored each patient from the
history alone, from the context alone, or from both. On all 1{,}600 T2D test
patient-outcomes, the history alone reached a mean ROC-AUC of 0.79 across the
seven outcomes, close to \cedar{}'s 0.81 on the same patients, whereas the context
alone reached 0.67 (Figure~\ref{fig:ablation}b). Adding the context to a complete
history changed discrimination by $-0.005$ to $+0.018$. Most of the predictive
signal therefore sits in the outcome's own recent course.

We then withheld that history visit by visit, in seven conditions applied to 100
patients per outcome (Figure~\ref{fig:ablation}a, right). Recency mattered more than
amount. A history showing only the latest visit beat one showing only the earliest
visit on all seven outcomes, by $+0.02$ to $+0.24$ ROC-AUC
(Figure~\ref{fig:ablation}c), and the two most recent visits matched or beat the
full laboratory history on five of seven. When the history was thin, the clinical
context recovered much of what was lost. In all 11 outcome-condition pairs where the
history alone fell below the context alone, reading both improved on the history,
by a mean of $+0.076$ ROC-AUC (Figure~\ref{fig:ablation}e and Supplementary
Figure~\ref{sfig:ablation_all}). Reading both also kept accuracy steadier: across
conditions its ROC-AUC spanned 0.06 to 0.19 depending on the outcome, against 0.12
to 0.27 for the history alone, and it was narrower on all seven outcomes
(Figure~\ref{fig:ablation}d). Per-condition estimates with confidence intervals are
in Supplementary Table~\ref{stab:ablation}, and the design's limits in
Supplementary Note~5.

\section*{Discussion}
\label{sec:discussion}

We set out to test whether an evidence-citing LLM pipeline predicts pediatric diabetes outcomes better than trained machine learning, and whether its reasoning can be trusted. Across eleven cohort--outcome pairs in
two pediatric diabetes cohorts, \cedar{} beat trained classical and temporal ML,
with pooled T2D ROC-AUC 0.82 against 0.66--0.70. The accuracy benefit is
outcome-dependent, largest on sparsely screened endpoints such as dyslipidemia and
microalbuminuria and near zero on the best-powered outcomes, where a classical
model stays competitive at a fraction of the cost. Beyond accuracy, what \cedar{}
adds is reasoning a clinician can use: two pediatric endocrinologists reading only its rationale reached the same
prediction in about nine cases in ten, and only 4 of
70{,}246 cited values did not match the record. Supplying the ML score as a tool added accuracy where the statistical model was strong, $+0.03$ ROC-AUC on that outcome.

The pattern of gains across outcomes suggests an explanation. They are largest on dyslipidemia and microalbuminuria, outcomes that are screened sparsely, so few patients carry the measurement a trained model would learn from, whereas \cedar{} brings clinical priors through its guideline knowledge and its base model. On rare outcomes the trained models default to the majority class and rank near chance, and the two families make mistakes on different patients, which is why an ML score supplied as a tool helps only where that model is already strong. The information ablation supports this account. Most of the signal lies in the
outcome's own recent history, consistent with trained models doing well where that
history is dense and poorly where it is sparse, and the clinical context matters
most when the history is thin. Reasoning over the whole record can therefore hold
accuracy for patients whose recent measurements are missing, a group that grows as
follow-up lengthens (Figure~\ref{fig:design}b).

Several limitations bound these conclusions. The study is single-center and
retrospective, and the \cedar{} evaluation, though bootstrapped, uses adaptive
per-outcome test samples rather than the full cohort. Tool use raised accuracy on only one outcome, so its benefit is outcome-specific. Cost
estimates are based on vendor list prices, so the tiers should be read as orders of
magnitude. The temporal and external checks summarized in the Results do not replace
multi-site pediatric validation. Additional limitations and future directions are
detailed in Supplementary Note~1.

Four directions follow. First, the MIMIC-IV
benchmark~\citep{johnson2023mimic4} still lacks
the head-to-head comparison against ColaCare~\citep{colacare2025},
MDAgents~\citep{mdagents2024} and MoMA~\citep{moma2025}. Second, a larger review in which clinicians predict from the full record, with
and without \cedar{}'s rationale, would test whether the rationale helps them
catch its errors.
Third,
the 24-month horizon is one snapshot, so predicting at successive visits and
horizons would show when each outcome first becomes predictable. Fourth, accuracy is limited by the information in the data, so new signal should help more than new machinery. Two candidates are the raw social-worker notes in place of binarized SDOH factors, and a continuous-glucose foundation model~\citep{lutsker2026cgmfm} applied to the raw sensor trace.

In this setting of small cohorts and uncommon outcomes \cedar{} outperforms the
trained statistical models we tested while producing reasoning that two pediatric
endocrinologists could follow and largely agreed with, and reasoning over the
whole record recovers accuracy when a patient's recent measurements are missing.
Evidence-citing clinical reasoning is therefore a
practical foundation for pediatric prediction, and the direction worth building on
is richer evidence for the model to reason over.
\section*{Methods}
\label{sec:methods}

All models are compared on a shared, held-out test set. For each
outcome, a single deterministic, prevalence-stratified train/test split
defines the test patients (random seed~42), and
every model family is scored on those identical patients with the same feature representation,
collinearity exclusions, and metrics. Trainable models are fit on the
training partition; the LLM configurations are inference-only. This matched
comparison is the basis for every head-to-head claim in this paper.

\subsection*{1. Study population}
\label{sec:data}

We use data from Texas Children's Hospital, one of the largest pediatric
health systems in the United States. The primary cohort comprises 2{,}489
patients with T2D. A parallel cohort of 4{,}018 patients with T1D serves as a
second, larger cohort in which the target outcomes are rarer and fewer are
clinically applicable (Supplementary Figure~\ref{sfig:prevalence}). All patients were diagnosed at least 2~years before data extraction, but not all
have recorded data at the 24-month visit, so each outcome is analyzed in the
patients assessable at that visit (Figure~\ref{fig:design}b,~d; Supplementary
Table~\ref{stab:outcome_prevalence}).

The established PheKB computable phenotype algorithms for type~1 and type~2
diabetes~\citep{phekb_t2dm, phekb_t1dm} cross-contaminated cohorts on our
pediatric Observational Medical Outcomes Partnership (OMOP) data, so we identified candidates with the Palantir Foundry
Cohort Builder agent system instead. The agent searched the diabetes Epic
flowsheets for patients carrying a recorded type~1 or type~2 diabetes diagnosis.
Where a patient had conflicting diagnosis entries or diagnosis dates across
encounters, we kept the most recent reliable record. A medical expert team at Texas Children's Hospital validated these
candidates by manual chart review, removing misclassified patients and correcting
diagnosis dates. Candidates added in a later rebuild of the cohort were confirmed with the same criteria, so the phenotype definition did not change between builds.
The medical expert team comprised three board-certified pediatric
endocrinologists and three trainees (a pediatric endocrinology fellow, a
pediatric resident, and a medical student) who worked under their supervision. We then accessed the full OMOP records, demographics,
medications, laboratory results, and International Classification of Diseases (ICD) codes, and curated a clinically
relevant variable set with domain experts (Supplementary
Table~\ref{stab:feature_summary}). We collected the corresponding social-worker
notes for SDOH extraction, and continuous glucose monitoring (CGM) data for about 2{,}000 T1D patients. CGM coverage in the T2D cohort was too sparse to be representative, so CGM is not used there.

This retrospective study was approved by the Baylor College of Medicine
Institutional Review Board (protocols H-55929 and H-53771), which granted a
waiver of informed consent for secondary analysis of existing clinical data.

\subsection*{2. Data structure and modalities}

Each patient has a matrix of features across 10 visit windows of 182~days (6-month intervals, 0--5~years post-diagnosis), organized into modality groups (Supplementary Table~\ref{stab:feature_summary}). These are labs and vitals (18), medications (3, namely insulin, biguanide and glucagon-like peptide-1 receptor agonist, GLP-1RA) and conditions (4, namely diabetic ketoacidosis, DKA, ketosis, retinopathy and neuropathy). The remaining groups are demographics (9, namely age at diagnosis, sex, Hispanic ethnicity, four race indicators, primary language and diabetes duration), SDOH (13 factors, enumerated below) and, for T1D only, CGM (20 per-visit glycemic metrics). The analysis-ready representation contains 47 features for T2D and
67 for T1D (cohort composition and feature availability in Supplementary
Tables~\ref{stab:cohort_composition} and~\ref{stab:feature_availability}).

SDOH factors were extracted from social-worker notes with a two-stage LLM
pipeline. An OpenAI GPT~5.1 extractor (Palantir Language Model Service
identifier \texttt{openai-gpt-5-1}, version~1.151.0, accessed 2026-07-03)
produced structured JSON, and an Anthropic Claude~Sonnet~4.5 validator
(\texttt{anthropic-claude-4-5-sonnet}, v1.151.0) confirmed each field. The 13
extracted factors are eight binary indicators (food insecurity, adverse childhood
experiences, housing instability, transportation barriers, physical or sexual
abuse, alcohol abuse, substance abuse, smoking), four binarized categorical
factors (parental education, parental employment, financial strain and family
support) and insurance, encoded as an ordinal category. All are
treated as static across visits. A further extracted factor, physical activity,
was excluded for missingness. Clinical experts reviewed the extracted factors
against the source social-worker notes, approximately ten patients per factor,
and the disagreements identified were used to revise the extractor prompts. This
review was iterative rather than an independent held-out test, so no agreement
rate was recorded.

CGM data are available only for T1D, which makes the two cohorts a natural modality contrast. Vendor exports differ in format and sampling, and none provides visit-aligned summaries, so we wrote our own extraction routine. It reads the raw sensor traces and reduces each patient's readings to 20 metrics per 182-day visit window, covering glucose statistics, time in, above and below range, hypoglycemia episode counts and monitoring coverage (Supplementary Table~\ref{stab:feature_summary}). The metrics and the glucose thresholds that define the time-in-range bands follow international consensus recommendations~\citep{battelino2019timeinrange} rather than choices made for this study, so they are the features a clinician reading a CGM report would recognize. Because sensor wear is intermittent, the metric set also carries reading counts, monitoring duration and the fraction of the window missing, so every model can see how much data each metric rests on.

\subsection*{3. Prediction targets, outcome definitions, and horizon}

We predict seven binary outcomes. Three are complications: hypertension (ICD diagnosis, or three consecutive outpatient blood-pressure readings above AAP-2017 age and sex thresholds) and dyslipidemia (non-HDL cholesterol $\geq$145~mg/dL, an ICD E78.x diagnosis, or lipid-lowering therapy). The third is microalbuminuria (ICD diagnosis or urine albumin-to-creatinine ratio (UACR) $\geq$30~mg/g confirmed on two or more measurements, capped at 300). Four are treatment goals: optimal glycemic control (HbA1c $<$7\%), insulin independence (off insulin, HbA1c $<$7\%, no DKA), metformin response (on metformin alone with HbA1c $<$7\%) and GLP-1RA response (on a GLP-1RA with HbA1c $<$7\%). Each label is
1, 0, or not-assessable (null), the last excluded from that outcome's
analysis. Because laboratory screening is not completed for every patient, the
assessable patients may differ from those who were not screened, which may limit
how far the results generalize.

To prevent collinearity, the laboratory value or medication that defines an outcome is removed from that outcome's feature set, identically for every model family (Supplementary Table~\ref{stab:collinear}). These are HbA1c for glycemic control, all blood-pressure features for hypertension, the lipid panel for dyslipidemia, UACR for microalbuminuria, insulin for insulin independence, and biguanide plus HbA1c for metformin and GLP-1RA response.

The primary analysis uses a 2-year horizon, so the first three visit windows
covering months 0--18 are the input and the outcome is evaluated at visit~4, at
approximately 24~months. A 5-year horizon is defined analogously (input visits
1--5, target visit~10) and is deferred to future work. Outcome prevalence at
the target visit is shown in Figure~\ref{fig:design}d.

\subsection*{4. Reference baselines: classical ML and temporal ML}
\label{sec:baselines}

Classical ML comprises three classifiers on the flattened feature matrix (all features by input visits). These are logistic regression (elastic net, L1 ratio~0.5, inverse regularization strength~1.0, SAGA solver), random forest (200 trees, max depth~10, min-leaf~5) and XGBoost (200 estimators, max depth~5, learning rate~0.1, subsample and column-sample~0.8). Temporal ML treats the visit history as a sequence. XGBoost-Temporal augments XGBoost with five engineered temporal features per base feature: last value, mean, standard deviation, first-to-last delta and ordinary least-squares slope. Four deep-sequence models complete the family. These are long short-term memory (LSTM) and gated recurrent unit (GRU) networks (2 layers, hidden size~64, dropout~0.3), a temporal convolutional neural network (CNN; 1-D convolution, 64 channels, adaptive pooling) and a Transformer (4-head attention, 2 layers, model dimension~64, learned positional embeddings). All four are trained for 100 epochs with Adam (learning rate~0.001, weight decay~0.0001) and class-weighted binary cross-entropy. Classical models use median imputation and standardization fit
on the training partition, and feature importances are collected on the
held-out folds.

\subsection*{5. LLM configurations}
\label{sec:llm_configs}

\cedar{} and every LLM configuration use Anthropic Claude~Sonnet~4.6 (Palantir Language Model Service identifier \texttt{anthropic-claude-4-6-sonnet}, version~1.151.0, accessed 2026-07-03), referred to below by its vendor short name \texttt{claude-sonnet-4-6}. Models are called through the Palantir Language Model Service, within the
institution's environment approved for protected health information, at temperature~0, except the self-consistency samples described below. The Results report the full \cedar{} pipeline (Section~6). For the reference benchmark in Supplementary Note~4 we also scored naive single-call configurations on the same patients. A direct configuration (``Vanilla'') gives one model the full multimodal record and asks for a probability, a binary prediction and a brief rationale. A chain-of-thought configuration (CoT) uses the same input with a structured step-by-step scaffold (metabolic, treatment, social, CGM for T1D, then severity), and Vanilla variants withhold the demographics and SDOH blocks or, in T1D, the CGM block.
A naive multi-agent configuration, in which modality-specialist agents' free-text
findings are fused by an LLM synthesizer, underperformed and is reported only in
Supplementary Figure~\ref{sfig:omitted_configs}. To test whether results depend on
the base model, \cedar{} and the Vanilla configuration were re-run on all 11
cohort--outcome pairs with three further models spanning vendors and sizes,
\texttt{claude-haiku-4-5}, \texttt{gemini-3-1-pro} and the open-weight
\texttt{llama-3-3-70b}, holding prompts, inputs and test sets fixed.

\subsection*{6. \cedar{}: evidence-grounded, self-consistent clinical reasoning}
\label{sec:cedar}

\cedar{} is a four-stage pipeline run per patient
(Figure~\ref{fig:design}c). It runs on a single underlying model, and each stage targets one failure mode of naive prompting: missing domain priors, ungrounded reasoning, hallucinated evidence or miscalibration.

\textbf{S0, knowledge injection.} Two to three one-sentence guideline snippets,
selected by outcome, are prepended to the prompt as a fixed per-outcome knowledge
base rather than as retrieval over a corpus. They were curated by the authors
from the American Diabetes Association Standards of Care~\citep{ada2024standards}, the 2022 International Society for Pediatric and Adolescent Diabetes Clinical Practice Consensus Guidelines~\citep{ispad2022} and two landmark youth-onset T2D trials~\citep{zeitler2012today, tamborlane2019ellipse}. Statements of accepted pathophysiology carrying no specific citation complete the set. The complete S0 text
for every outcome, reproduced exactly as it is inserted into the prompt and
attributed line by line, is given in Supplementary
Tables~\ref{stab:s0_snippets} and~\ref{stab:s0_snippets_t1d} for T2D and T1D
respectively. One snippet compresses the TODAY trial to
``$\sim$50\% of youth with T2D lose glycemic control on metformin alone within
3~years''. The published result is a 51.7\% metformin-monotherapy failure rate
over a mean follow-up of 3.9~years~\citep{zeitler2012today}, so the injected
wording rounds the rate and shortens the horizon.

\textbf{S1, evidence-grounded chain-of-thought.} The model
reasons step by step over each modality (temperature~0) and emits a strict
plain-text record. The probability comes first so it is never truncated, then a
table of cited evidence (\texttt{FEATURE | VISIT | VALUE | DIRECTION |
IMPORTANCE}) with values at explicit visit tokens such as
\texttt{v3(18mo)=8.50} so citations can be checked against the record, and
finally the free-text reasoning. \textbf{S2, verification.} Every citation is checked against the record by the
deterministic tolerance test of Methods, Section~7, and unmatched citations are
removed. In parallel a clinical evidence auditor pass reviews the same citations
and may issue a single revised probability. Both signals are reported separately in Supplementary Note~1. \textbf{S3, self-consistency calibration.} The
S1 prompt is resampled $K=3$ times at temperature~0.7, and the mean of the
parsed probabilities is the self-consistency (SC) estimate. Although the pipeline has
four stages, only S1 to S3 issue model calls, five per patient, since S0 is
prepended context.

The final score averages the verified probability and the self-consistency
mean, $p_{\text{final}} = \tfrac{1}{2}\!\left(p_{\text{verify}} +
p_{\text{SC}}\right)$, and is used directly for ROC-AUC and Brier score. Two ablations, reported in Supplementary Note~4, isolate each mechanism's contribution: CoT+Verify (S1 and S2 only) and CoT+SC (S1 and S3 only).

Relative to a single prompted call, which returns only a probability, \cedar{}
adds guideline priming, forced grounding in citable record values, verification
of those claims, and stability across resamples (Supplementary
Table~\ref{stab:cedar_vs_vanilla}). The individual techniques are not
themselves new; the contribution is their composition into a workflow
purpose-built for auditable clinical outcome prediction.

\subsection*{7. Faithfulness and interpretability}
\label{sec:faithfulness}

For \cedar{}, we audit whether cited evidence corresponds to real record
values and is collectively relevant to the prediction (change
threshold~0.10 throughout), with five complementary analyses. The audit targets
faithfulness, how accurately a rationale reflects the reasoning that produced
the prediction, as distinct from plausibility, how convincing that rationale is
to a reader~\citep[\S2]{jacovi2020faithfulness}. We therefore intervene on the model's inputs rather than inspect its prose, and leave reader-facing judgement to the expert review of Section~10. Extractive grounding, being cheap, runs on
every scored patient-outcome record; the four counterfactual analyses each re-run
S1 under masking and run on a random subsample of patients per outcome.

Extractive grounding compares each cited (feature, visit, value) triple with the
actual record. A citation is verified if it matches within tolerance
($|\text{cited}-\text{actual}|\le\max(0.5,\,0.05\,|\text{actual}|)$), and
unmatched or non-existent citations are counted as
hallucinations~\citep{asgari2025hallucination}. Sufficiency and comprehensiveness
are the two faithfulness metrics of the ERASER rationale-evaluation
framework~\citep[\S4.2]{deyoung2020eraser}, adapted here from extractive
rationales over text to a citation table over structured records. Sufficiency
masks all non-cited features and re-runs S1, reporting
$1-|p_{\text{baseline}}-p_{\text{cited-only}}|$, so higher is better.
Comprehensiveness masks each cited feature individually and reports the fraction
whose removal moves the probability by $\geq$0.10. Both depart from ERASER's
published form, which is detailed with the rationale for each departure in
Supplementary Note~1.

The group-masking curve ranks cited features by importance and masks the
cumulative top-$K$ ($K\in\{1,3,5,10\}$) together, tracing mean $|\Delta p|$
against $K$, following the area-over-the-perturbation-curve construction of
ERASER~\citep[\S4.2, Eq.~3]{deyoung2020eraser} with cited features in place of
token bins. It is paired with a size-matched random-masking control. For the same patients and the same $K$, $K$ features are drawn uniformly at random from those available for the patient, under the same outcome-specific exclusions, and the draw is repeated 20 times per patient and $K$ and averaged. The evidence
audit plots each cited feature's model-reported importance against its actual
counterfactual impact when masked, colored by verification status. A model's stated reasoning need not correspond to what drives its output~\citep[\S3.2]{turpin2023unfaithful}, and intervening on the reasoning is the established way to measure that gap~\citep[\S2]{lanham2023measuring}.

\subsection*{8. Cost and clinical economics}
\label{sec:cost}

Cost here is a token- and compute-based estimate, not a vendor bill, and it is
reported to separate the tiers by order of magnitude rather than to predict any
site's budget. For LLM configurations, input and output tokens are tracked per call, priced at \$3 and \$15 per million tokens, the list price of the Claude Sonnet tier used here, and divided by the number of patients scored. Latency and call counts are logged separately. At these prices \cedar{}'s five calls per patient (Section~6), with their long prompts and rationales, cost \$0.099 per patient. That figure is indicative only, since actual spend depends on the provider, contract, model tier and any caching or batching, none of which are properties of the method. Classical and temporal models issue no tokens, so we report a deliberately conservative inference-equivalent cost instead: wall-clock training time priced at \$$5\times10^{-5}$ per second ($\approx$\$0.18/CPU-hour) and amortized over the test cohort. This gives \$$1.8\times10^{-7}$ per patient for classical ML and \$$4.2\times10^{-6}$ for temporal ML, which overstates their true marginal cost and so understates rather than inflates the \cedar{}-versus-ML gap.

We compute the cost-performance Pareto frontier across all families and, per outcome, the ROC-AUC gain over the best classical or temporal model divided by cost per patient (Supplementary Figure~\ref{sfig:economics}). A gain counts as statistically distinguishable from ML when the two models' 95\% confidence intervals do not overlap.

\subsection*{9. Extensions: tool use and routing, selective prediction, external
benchmark, and temporal validation}
\label{sec:extensions}

\textbf{ML score as an in-context tool, and per-patient routing.} The frozen statistical model's predicted probability for the patient is supplied to \cedar{} inside the S1 prompt as an explicitly labeled tool reading, which the model can cite, discount or contradict in its citation table. No model is retrained and no scores are combined arithmetically. Four policies were compared per outcome on the shared held-out set. These were never route (\cedar{} without the tool), always route, a pre-specified rule offering the tool only where standalone ML is competitive, and a learned per-patient policy gated jointly on \cedar{}'s self-consistency spread and the ML model's own confidence. For each we report
ROC-AUC with bootstrap confidence intervals, the fraction of patients routed and
the cost per patient, since routing adds tokens. The learned
policy's gating threshold was then swept so that the routed fraction covered the
full 0--100\% range, reporting at each point the ROC-AUC difference from
never-route on a positive-control outcome, T1D optimal glycemic control, and on a
null-contrast outcome, T1D dyslipidemia.

\textbf{Selective prediction.} Patients are ranked by a confidence signal and the
least confident withheld rather than scored. Four signals were compared: the
spread of the $K{=}3$ self-consistency samples, the fraction of cited evidence
the S2 verifier flags, the predicted probability's distance from the 0.5 decision
boundary, and a composite of the three. Coverage was swept from 100\% down to 10\% in 10-point steps. At each level, ROC-AUC, accuracy and Brier score were recomputed on the retained subset only, both pooled over records and averaged
over outcomes, and summarized as the area under the risk--coverage
curve~\citep{geifman2017selective}, for which lower is better. The identical
protocol was applied to \cedar{} and to the Vanilla single call, so that the
comparison isolates the signal rather than the system.

\textbf{External benchmark on MIMIC-IV.} \cedar{} was run unchanged on
MIMIC-IV~v3.1~\citep{johnson2023mimic4}, a single-center adult critical-care
database, under its PhysioNet data use agreement. Three tasks were sized to
mirror the T2D cohort number for number, 2{,}489 patients with 2{,}239 in
training and 250 held out, prevalence-stratified with the same seed: 30-day
readmission at discharge, death within 365~days of discharge, and in-hospital
death from the first 48~ICU hours. S1 to S3 were identical to the pediatric
pipeline and only the S0 knowledge base was replaced, with ten one-sentence
snippets drawn from the severity-scoring, sepsis and readmission literature,
frozen and hashed before any test prediction was scored. Comparators were the
published training-free instruments clinicians already
use~\citep{vanwalraven2010lace, donze2013hospital, charlson1987comorbidity,
vincent1996sofa, johnson2013oasis, legall1993sapsii}, the Vanilla single call, and the eight trained families used throughout this paper (Supplementary Note~3).

\textbf{Temporal validation.} We split each cohort on
diabetes duration rather than at random, training on longer-duration
(earlier-diagnosed) patients and validating on more recently diagnosed ones, with
the cut placed so the two partitions do not overlap in duration (partition sizes
and duration ranges in Supplementary Note~1). Up to 200 validation patients per
outcome were scored by the LLM configurations to bound inference cost. The statistical families were
refitted on the training partition and the LLM configurations re-scored the
validation partition with prompts unchanged, and each family's ROC-AUC
distribution was compared with that from the primary split (Supplementary
Figure~\ref{sfig:temporal_val}). The T1D split follows the same rule.

\subsection*{10. Evaluation and statistics}
\label{sec:eval}

The primary metric is ROC-AUC, which is threshold-free and comparable across outcomes of very different prevalence, and Brier score is reported as an additional summary. ROC-AUC summarizes ranking across all thresholds, so it can stay high when decisions at a usable threshold are poor on a rare outcome. The benchmark was therefore repeated under Matthews correlation, balanced accuracy and positive-class $F_1$, each at a threshold selected on the training folds, which give the same ordering (Supplementary Note~1). Uncertainty is quantified by 2{,}000-iteration
patient-level bootstrap 95\% confidence intervals. The shared held-out test
set is sized adaptively to the eligible population (50--250 patients, the
same test patients used for every model family) to keep per-outcome estimates
stable under imbalance, and tasks with fewer than 40 assessable patients or a
single class are skipped. The evaluation covers all seven outcomes for T2D
and the four outcomes that are clinically applicable in T1D (glycemic control,
hypertension, microalbuminuria, dyslipidemia); insulin independence, metformin
response and GLP-1RA response are T2D-specific. Classical and temporal ML baselines are run on the same eleven cohort--outcome pairs.

The family-level comparisons are exploratory. Each is a Mann--Whitney~$U$
test in which one observation is one model's ROC-AUC on one outcome, computed on
overlapping patients and correlated outcomes, so we report them as descriptive
effect sizes and use the patient-paired bootstrap for per-outcome comparisons
(Supplementary Note~1). The base-model comparison of Supplementary Note~4 uses a paired Wilcoxon signed-rank test on the 11 per-outcome ROC-AUC pairs within each base model, Holm--Bonferroni corrected across the four base models. Throughout, the main text rounds any $p$-value below 0.001 to
$p<0.001$, and the exact statistics for all pooled and multi-model tests are
reported in Supplementary Table~\ref{stab:pooled_stats}. A modality ablation (EHR
only, EHR+SDOH, and EHR+SDOH+CGM for
T1D) is run for every model family on a fixed matched held-out set of 100 patients per outcome
(Supplementary Table~\ref{stab:sample_sizes}).

For the clinician review of \cedar{}'s reasoning chains
(Results; Supplementary Figure~\ref{sfig:annotation}), two board-certified
pediatric endocrinologists each received the same deck of 63 T2D patient-outcome cases, drawn
to span the four confusion cells across all seven outcomes. Each case showed
\cedar{}'s citation table and reasoning chain with the predicted probability
removed, and the raters saw neither the recorded outcome nor whether \cedar{}
was correct. For every case a rater gave a 1--5 clinical-usefulness score and
named the prediction the rationale supports, so agreement is scored on a forced
binary call rather than on free text. Both raters are authors of this study. Agreement is reported as raw percentage
and Cohen's $\kappa$, with 2{,}000-iteration bootstrap 95\% confidence
intervals over cases. Because the 63 cases are a stratified draw across the
confusion cells rather than a random sample, neither rater's accuracy nor
\cedar{}'s on this subset estimates population accuracy, and we report only
agreement and usefulness.

\subsection*{11. Information-ablation experiment}
\label{sec:ablation}

This experiment asks how much of each prediction rests on the outcome's own
history. Each record is split into two channels. The outcome history holds the
laboratory values and diagnosis codes that define the outcome, at visits 1 to 3.
For the three treatment goals defined by glycemic control this is HbA1c, and for
insulin independence HbA1c with DKA codes. For the complications it is blood
pressure with the hypertension code, UACR, urine microalbumin and urine creatinine
with the microalbuminuria code, and the lipid panel with the dyslipidemia code. The clinical context holds every other
feature, namely demographics, SDOH, other laboratory values, other diagnosis codes
and medications. Medications that define an outcome are withheld from both
channels. Unlike the benchmark, which removes the defining laboratory value to
prevent collinearity (Section~3), the history reader sees it on purpose, since its
contribution is what the experiment measures. The label is always taken at
visit~4, after every visit shown.

Three readers run on \texttt{claude-sonnet-4-6} at temperature~0 (Section~5). The
history reader sees only the outcome history and the context reader only the
clinical context. The combined reader sees both, together with the other two
readers' scores and reasoning, and is instructed to weight the history by how much
of it is available. Each returns a probability from 0 to 100 with cited evidence,
and every citation is checked against the supplied features. The context reader is
called once per patient and its output is reused in every condition, so only the
history changes between conditions.

Two designs were run. The first scores the 1{,}600 T2D test patient-outcomes of the
benchmark with every available history value. The second withholds the history in seven conditions. Two show all three visits
of one source, laboratory values without codes or codes without laboratory values.
The other five restrict the history to visit~1, 2 or 3 alone, visits 1 and 2, or
visits 2 and 3, showing both laboratory values and codes. A withheld value is marked as unknown rather than normal, and the codes-only
condition does not apply to the three HbA1c-defined outcomes, which have no defining
code. This design uses a seeded, label-stratified sample of 50 positive and 50
negative patients per outcome, drawn once from patients with history evidence at
each of visits 1 to 3, with T1D and T2D pooled within outcome and the same patients
in every condition. Discrimination is ROC-AUC per outcome, never pooled across
outcomes, with 1{,}000-resample paired bootstrap 95\% confidence intervals that use
the same resampled patients for every reader.

\subsection*{Reporting and use of large language models}

The study is reported in accordance with the TRIPOD+AI statement; a completed
checklist is provided in the Supplementary Information.

\textbf{Use of large language models.} LLMs are the object of study in this work
and were additionally used for cohort construction and for extraction of social determinants of health from clinical notes. All model identifiers, versions and access dates are reported in Methods, Sections~1, 2 and~5. Beyond that analytical
use, no AI-assisted tool contributed to study design or to primary manuscript
drafting; generative AI tools were used only for proofreading and typographical
and grammatical correction of author-written text. In accordance with COPE's
position and npj Digital Medicine's AI editorial policy, no AI tool is listed as
an author; the authors reviewed and verified all text and take full
responsibility for the integrity, accuracy and originality of all content.

\section*{Data availability}
The underlying clinical and administrative data cannot be shared because they
are protected patient health information subject to institutional and HIPAA
data-use restrictions, and are not available for public deposition. Requests for
access to de-identified derived data may be directed to the corresponding author
and will be considered subject to institutional review, an executed data use
agreement, and applicable HIPAA restrictions. De-identified aggregate metrics
that support the findings of this study are provided within the article and its
Supplementary Information, and the analysis code is available as described below.

\section*{Code availability}
The analysis and figure-generation code (\texttt{manuscript\_npj\_cedar/}) and
the upstream cohort-construction and Foundry data-transform pipeline
(\texttt{palantir/t1d/}, \texttt{palantir/t2d/}, and related directories) are
available at \url{https://github.com/DanPalBCM/clinical-diabetes-t1d-t2d}.
Publication figures were produced with Matplotlib; their visual design (color palette, typography, and soft-bar styling) was created with and inspired by the
PubliPlots library~\citep{botas2025publiplots}.

\section*{Acknowledgments}
This research was supported by a fellowship from the Gulf Coast Consortia on
the NLM Training Program in Biomedical Informatics and Data Science
(T15~LM007093) and by the National Science Foundation Graduate Research Fellowship Program (NSF GRFP Fellow ID 2024370642). Further support came from the Cancer Prevention and Research Institute of Texas (CPRIT, RP240131), the Chan Zuckerberg Initiative (2023-332162) and the National Institutes of Health (NIH, U54NS093793 and OT2OD040565). The Eunice Kennedy Shriver National Institute of Child Health and Human Development of the NIH (P50HD103555), the Chao Endowment, the Huffington Foundation and the Jan and Dan Duncan Neurological Research Institute at Texas Children's Hospital also supported this work.
This study was also supported by the National Institutes of Health through the
National Institute of Diabetes and Digestive and Kidney Diseases (dkNET Award
3U24DK097771-12S1, Subaward 707871; M.T., M.J.R., Z.L., H.C., D.P. and S.A.) and
K23DK129821 (M.T.).

\section*{Author contributions}
Following the CRediT taxonomy:
\textbf{D.P. (Palacios):} conceptualization, methodology, software, formal
analysis, investigation, data curation, visualization, writing -- original
draft.
\textbf{G.Q. (Qi):} methodology, software, formal analysis, investigation,
writing -- review and editing. G.Q. designed and ran the information-ablation
experiment (Methods, Section~11).
\textbf{M.T. (Tosur):} conceptualization, validation, investigation, data
curation, resources, supervision, project administration, funding acquisition,
writing -- review and editing. M.T. led the medical expert team and its
annotation, chart-review and spot-checking work, defined and adjudicated the
clinical questions and outcome definitions, assessed the clinical relevance of
the model's reasoning, and directed the quality control of cohort construction
and of the curated variable set. M.T. and E.K.M. were the two specialists who
independently rated \cedar{}'s reasoning chains for clinical usefulness and
reconstructed its predictions (Methods, Section~10).
\textbf{A.K. (Kadi), N.B. (Borrego), M.J.R. (Redondo), and M.F. (Ferm):}
clinical expertise, validation, investigation, guidance on clinical questions
and outcome definitions, validation of pipeline components (e.g., cohort chart
review and SDOH annotation), writing -- review and editing.
\textbf{E.K.M. (Kubota-Mishra):} clinical expertise, validation,
investigation, writing -- review and editing.
\textbf{S.A. (Aras):} statistical and epidemiological methodology, data
curation, validation, supervision, writing -- review and editing. S.A.
contributed foundational work on the underlying dataset and served as the senior
statistical reviewer, independently checking the analyses reported here.
\textbf{Z.L. (Liu) and H.C. (Chen):} conceptualization, methodology,
investigation, supervision, project administration, funding acquisition, writing
-- review and editing. Z.L. and H.C. directed the design and development of the
agent architecture and the computational experiments.

\section*{Competing interests}
The authors declare no competing interests.

\bibliography{references}

\clearpage

\end{document}

\begin{center}
{\large\bfseries Supplementary Information}\\[6pt]
{\normalsize\itshape for}\\[2pt]
{\normalsize\bfseries Agentic LLM reasoning in pediatric diabetes outcome
prediction from multimodal medical records}
\end{center}
\bigskip

\noindent The completed TRIPOD+AI reporting checklist is Supplementary
Table~\ref{stab:tripod_ai}.
\medskip

\section*{Supplementary Notes}

\subsection*{Supplementary Note 1: Extended Discussion}
\label{sec:ext_discussion}

This note expands the main-text Discussion, giving the mechanism behind each
finding and the analyses that support it.

\paragraph{Why extra reasoning stages do not help.} A capable LLM brings clinical
priors that substitute for the training signal a data-starved statistical model
lacks. One competent reasoning pass therefore captures most of the available
signal, and further stages have little left to add: \cedar{} and its CoT+Verify
and CoT+SC ablations differ neither from each other nor from a single call. Two
further designs failed the same way. An early naive multi-agent decomposition,
which split the patient across modality specialists, scored below the single-LLM
baseline and was set aside, and a pre-specified screen of verifier-gated
self-refinement and a diverse-reasoner ensemble also failed to beat a single call
(Supplementary Figure~\ref{sfig:omitted_configs}). Both are consistent with
evidence that self-correction needs reliable external feedback to
help~\citep{huang2024selfcorrect, kamoi2024selfcorrect} and that multi-agent
debate does not reliably beat simpler prompting~\citep{smit2024mad}. This is a
boundary of the approach rather than of our particular pipeline.

The boundary does move with the base model. Across four models the null held on
the three commercial ones but not on the open-weight \texttt{llama-3-3-70b}, where
\cedar{} gained $+0.099$ ROC-AUC over a single call (Supplementary
Figure~\ref{sfig:llm_benchmark}f). The extra stages
thus substitute for single-pass reliability rather than adding to it, which predicts that their value grows as models shrink, the relevant regime for
institutions that must run inference locally on protected health data. Two versions of this comparison appear in Supplementary Note~4 because they average differently. The pooled subgroup means cover three configurations per side in T2D and four on the single-LLM side in T1D ($+0.006$). The paired base-model comparison contrasts the full pipeline against a single Vanilla call on identical
patients ($+0.009$; Supplementary Figure~\ref{sfig:llm_benchmark}f, Supplementary
Table~\ref{stab:pooled_stats}b).

\paragraph{Why we did not try a multi-model consensus.} Ensembles improve on their
members only when those members make decorrelated
errors~\citep{krogh1994ensembles, dietterich2000ensemble, kuncheva2003diversity}.
Our error analysis shows a single LLM and \cedar{} misclassify almost the same
patients ($\phi=0.87$; Supplementary Figure~\ref{sfig:llm_errors}a, Supplementary
Figure~\ref{sfig:error_analysis}), and different frontier models reasoning over
the same inputs would likely be correlated the same way, so a consensus across
vendors should mostly reproduce their shared mistakes. Our four-model comparison
is consistent with this, since the three commercial models reproduced the same
pattern, though we did not measure per-patient error correlation across vendors.
A consensus would also trade the single-model reproducibility of this pipeline for
dependence on several proprietary models and their versions. Disagreement signals
are more plausibly useful for uncertainty estimation than for accuracy, and a
genuinely heterogeneous disagreement signal remains open.

The one place our data show decorrelated errors is \emph{across} families, between
the LLMs and classical ML ($\phi\leq0.17$). We tested that fusion directly: it
improved calibration but not discrimination, and helped only on the outcomes where
standalone ML was already competitive (Supplementary
Figures~\ref{sfig:omitted_configs} and~\ref{sfig:calibration}).

\paragraph{What \cedar{} adds is trust, not accuracy.} The auditability claim
rests on a grounding rate $\geq$0.9999 with essentially no fabrication, cited
evidence that is collectively relevant to the prediction, and cited features that
are clinically coherent. The 4 mismatches in 70{,}246 reported in the main text
are counted on S1's citation table, which is also the final one: the deterministic
tolerance check removes those 4 and S2's auditor flags no further citations in
either cohort. S2 therefore contributes its revised probability, isolated in the
CoT+Verify ablation, rather than error detection, the same external-feedback limit noted above.

We also asked whether \cedar{}'s extra stages give a better \emph{uncertainty}
signal. Abstaining on the least-confident patients improves accuracy on those
retained for both a single LLM and \cedar{}, but only when the ranking uses the
output probability itself; \cedar{}'s own confidence signals rank hard cases worse
rather than better. Its verifier flag almost never fires, precisely because
grounding is near-perfect, so it carries little information and defers patients
the model already scores well (Supplementary Figure~\ref{sfig:error_analysis}).
Useful abstention is thus a generic benefit of a usable probability rather than
a \cedar{}-specific one.

One clarification about the faithfulness audit. Counterfactual intervention
measures faithfulness, how accurately a rationale reflects the computation that produced the prediction, whereas clinician rating measures plausibility and
clinical usefulness. The two can come apart~\citep[\S2]{jacovi2020faithfulness},
so the two-specialist review reported in the Results
(Supplementary Figure~\ref{sfig:annotation}) measures a different construct, not
a stronger version of the audit. The review bears this out: the two
endocrinologists agreed with \cedar{}'s prediction on about nine cases in ten, yet on 29 of its 31
errors at least one of them reached the same incorrect call. A rationale can be
legible and well rated while the prediction it supports is wrong, which is why
the grounding and counterfactual audits, not the ratings, carry the
faithfulness claim. Because the two specialists largely reproduced \cedar{}'s
errors, a rationale a clinician endorses is not yet a safeguard against a wrong
prediction. A larger review, and abstention when the record lacks the measurement
that defines the outcome, would test whether that gap can be closed.

\paragraph{Why the faithfulness metrics depart from ERASER's published form.}
Sufficiency and comprehensiveness are ERASER's metric
names~\citep[\S4.2]{deyoung2020eraser}, and we keep the names because the
constructions are the same in spirit, but two deviations are worth stating.
ERASER defines sufficiency as the probability drop when only the rationale is
retained, so that a lower value is better; we report $1-|\Delta p|$ instead, so
that higher is better for both metrics and neither has to be read backwards
against the other. ERASER defines comprehensiveness by deleting the whole
rationale at once, which suits a short span of text. \cedar{} cites tens of
features per patient, so deleting the rationale entire would remove most of the
input and measure little, and we therefore mask each cited feature individually
and keep ERASER's cumulative-deletion form separately as the group-masking curve.
The random-masking control exists because the group-masking curve on its own
cannot show that the movement is specific to the citations rather than to
removing that many features at all, so the control draws from exactly the
candidate pool the model could have cited, under the same outcome-specific
exclusions.

\paragraph{Methods for the refinement and ensembling screen and for blind late
fusion.} The main text reports these two experiments only by their result, so
their methods are given here.
\textbf{Verifier-gated refinement and diverse-reasoner ensembling
(``\cedar{}~v2'').} A pre-specified screen asked whether two further
\emph{internal} mechanisms improve on the pipeline of Methods, Section~6, with the null
hypothesis that neither beats a single call by a meaningful margin. Five systems
were scored on the \emph{same} fixed set of 60 held-out patients per outcome: a
Vanilla single call, \cedar{} as described, and three variants. Verifier-gated self-refinement re-runs the extractive-grounding check on S1's citation table. If any cited triple is unverified, the offending citations are returned to the model by name, with an instruction to remove or correct them, re-examine the record and revise the probability, for at most two rounds, after which S2 and S3 proceed unchanged. In practice this gate almost never opened: first-pass
extractive grounding is already at ceiling, so no refinement round was triggered
in the screen (Supplementary Note~1). The diverse-reasoner ensemble replaces S1's single
system prompt with three structurally distinct ones over an identical patient
prompt, directing the model to reason primarily from the longitudinal laboratory
and vital-sign trajectory, primarily from demographics, comorbid conditions and
social determinants, or primarily by applying the injected guideline knowledge. A meta-reasoner then receives all three reasoning chains with their probabilities and returns one integrated probability, instructed to weigh agreement and disagreement rather than to average. The unweighted mean of the three member
probabilities is retained as a secondary reference. The third variant combines
both mechanisms. The screen ran on the three highest-headroom T2D outcomes
(GLP-1RA response, hypertension, dyslipidemia), with both anchors re-scored on
the same 60 patients; promotion required a paired ROC-AUC difference whose
2{,}000-iteration bootstrap 95\% interval excluded zero against \emph{both}
anchors. Neither addition met that rule (Supplementary
Figure~\ref{sfig:omitted_configs}a).

\textbf{Blind late fusion.} Separately from the tool-use experiments, \cedar{}'s
probability was combined post hoc with the statistical models' probabilities,
blind in the sense that the combiner is never told which family is stronger on
the outcome. Four combiners were evaluated on the full held-out set against \cedar{}, the single call, and the best classical and best temporal models. These were the unweighted mean of \cedar{} and the best ML model, the unweighted mean of
\cedar{} with both the best classical and the best temporal model, a logistic
stacker over the two scores, and a routed combiner selecting one source per
outcome from training-fold performance. The router's per-cohort choices, and why
its pooled T1D ROC-AUC falls below \cedar{}'s, are given in the next paragraph.

\paragraph{Design details behind the extension experiments.}
Four choices in those experiments need their reasoning stated. First, the
verifier-gated refinement arm returned a null that should be read as
uninformative rather than negative. First-pass extractive grounding is already at ceiling, so no refinement round was ever triggered in the screen, and the null says nothing about refinement in lower-grounding settings. Second, routing adds tokens, and across outcomes the
routed fraction spanned 0 to 1 at a cost of \$0.091 to \$0.104 per patient, so
the routing decision is not cost-free even where it is accuracy-neutral. Third,
the routed late-fusion combiner chose \cedar{} on all seven T2D outcomes, making
its T2D scores numerically identical to \cedar{}'s, whereas in T1D it chose
classical ML on hypertension and on optimal glycemic control, which is why its
pooled T1D ROC-AUC (0.734) falls below \cedar{}'s (0.759) in that cohort. Fourth,
the temporal split was cut so the partitions do not overlap in diabetes duration. In T2D the minimum training duration exceeded the maximum validation duration for all seven outcomes, with median duration 4.1--7.7~years in training against 2.4--3.4~years in validation. Training partitions held 140--801 patients
against 61--343 in validation.

\paragraph{Why the family-level comparisons are exploratory.} Each pooled
comparison is a Mann--Whitney~$U$ test in which one observation is one model's
ROC-AUC on one outcome. Those observations are not fully independent, being
computed on overlapping patients and correlated outcomes, so nominal significance
is inflated and we report the comparisons as descriptive effect sizes. The
patient-paired bootstrap on the shared test set carries the more rigorous
per-outcome comparisons instead. Where many pairwise tests are reported we say so
rather than implying family-wise error control, and the multi-model comparison is
exploratory in the same sense.

\paragraph{Class-imbalance metrics give the same ordering as ROC-AUC.} ROC-AUC summarizes ranking across all thresholds, so it can stay high when decisions at a usable threshold are poor, and several of the outcomes here are imbalanced (Supplementary Table~\ref{stab:outcome_prevalence}). We therefore repeated the full benchmark of Supplementary Figure~\ref{sfig:llm_benchmark} under three threshold-based metrics: Matthews correlation coefficient (MCC), balanced accuracy, and $F_1$ on the positive class. We did not use weighted
$F_1$ or raw accuracy, both of which can be maximized by predicting the majority
class. All three are threshold-dependent, so each model's decision threshold was
selected on the training folds alone by maximizing out-of-fold MCC, and every
model was then scored on the same held-out patients as in the primary analysis.
The LLM configurations have no training partition and therefore borrow the
threshold selected by the strongest statistical model on that outcome, which
denies them any tuning advantage. Classical models were additionally refit with
balanced class weights so that all eight statistical models carry the weighting
the deep-sequence models already used; this changed their ROC-AUC by $+0.002$ and
their MCC by $-0.005$ on average, because a tuned threshold already absorbs most
of the effect.

Both conclusions of that benchmark hold. The LLM family leads the statistical family
on every metric and on 9 to 11 of the 11 cohort--outcome pairs
(Supplementary Table~\ref{stab:imbalance_ml_vs_llm}), and a single LLM call and
\cedar{} remain indistinguishable, with $p\geq0.69$ on all three new metrics
against $p=0.26$ for ROC-AUC (Supplementary
Table~\ref{stab:imbalance_cedar_vs_single}). The equivalence is therefore firmer
under these metrics than under ROC-AUC, not weaker. The margin between families
does narrow: comparing the \emph{best} model in each family per outcome, rather
than the family mean, the LLM lead stays positive but is no longer statistically
distinguishable on the three new metrics, and in T1D, where only four outcomes
are evaluable, no family contrast reaches significance. We therefore read these
metrics as confirming the ordering while bounding how strongly the size of the
gap should be stated, and we report ROC-AUC in the main text because it is the
metric the comparison was designed around and is threshold-free.

\paragraph{The numbers behind selective prediction and routing (Supplementary Figure~\ref{sfig:error_analysis}e and Figure~4c).}
Ranking patients by \cedar{}'s bare probability gives area under the
risk-coverage curve 0.148 in T2D and 0.185 in T1D, against 0.205 and 0.233 for
self-consistency spread. The spread gains at 80\% coverage in T2D, 0.803, then
plateaus and reaches only 0.809 by 60\%, and in T1D it loses accuracy outright
(0.760 then 0.744).
Verification failure rate is worse than not abstaining at all (0.758 then 0.729 in
T2D, 0.738 then 0.717 in T1D). A single LLM's bare probability separates confident
from uncertain cases equally well (0.171 in T2D, 0.182 in T1D). Averaging over
outcomes rather than pooling over records preserves the ordering in attenuated
form: the bare probability reaches $+0.079$ ROC-AUC at 40\% coverage, spread peaks
at $+0.036$ near 60\% and then decays, and verification failure rate stays near
$+0.01$ throughout (Supplementary Figure~\ref{sfig:error_analysis}e).

The verifier fails for an instructive reason. It flags nothing for essentially
every patient, so its ranking is close to arbitrary, and ROC-AUC on the deferred
20\% is in fact \emph{higher} than on the retained 80\% (0.831 against 0.758 in
T2D, 0.784 against 0.738 in T1D): the signal defers easy cases rather than hard
ones. For the routing sweep, the null-contrast outcome T1D dyslipidemia moves from
$-0.004$ to $-0.023$ across the same threshold range. Per-outcome intervals in
that panel are wide given our sample sizes (Supplementary
Table~\ref{stab:sample_sizes}), so it shows the learned policy does not hurt
anywhere rather than carrying the statistical evidence for the tool-use effect,
which rests on the tighter paired estimate.

\paragraph{Where agentic design does help: targeted tool use.} Re-reasoning over the same inputs and re-encoding them both failed, because neither escapes the LLM's shared error set. The one exception crosses family boundaries. T1D optimal
glycemic control was the only outcome where a standalone statistical model
out-predicted the LLM by a margin whose bootstrap interval excluded zero;
logistic regression was also numerically ahead on T1D hypertension, but within
noise. When we exposed that model's score to \cedar{} as an in-context tool,
applied only where standalone ML is independently competitive, \cedar{}
incorporated it and beat a single LLM on that outcome ($+0.05$ ROC-AUC, 95\% CI
$[0.023, 0.081]$; $+0.03$ over \cedar{} without the tool). The tool gave no
benefit, and occasionally some harm, where it carried no complementary signal
(Supplementary Figure~\ref{sfig:hybrid_b}).

Two features make this more than a lucky subgroup. \cedar{} matched a black-box
logistic stacker on the same outcome, but unlike the stacker it exposed why it
combined the two sources. And its behavior was appropriately conditional: it moved
toward the tool's estimate when the tool was good and largely ignored it when
weak, over-trusting it only on the noisiest, lowest-prevalence outcomes
(Supplementary Figure~\ref{sfig:hybrid_b}). That behavior is both the mechanism of the gain and a caution for deployment.

Because the initial gain came from a single pre-specified outcome, we replaced the
pre-specification with a learned per-patient policy, gated on \cedar{}'s
self-consistency spread and the tool's own confidence, and evaluated it on every
outcome in both cohorts. It held: the policy tracked the always-route and
hand-picked-rule gain where complementary signal existed and stayed inert, at no
added cost, where it did not (Supplementary Figure~\ref{sfig:routing_policy}). Sweeping the routed fraction separates
the two regimes. On T1D optimal glycemic control the gain rises with the
fraction routed, reaching $+0.028$ when all patients are routed, whereas the same
sweep on T1D dyslipidemia runs flat to negative (Figure~4c).

This identifies the productive role for agentic architecture. Its value is not
reasoning harder over the same data, but serving as an auditable substrate that
integrates genuinely complementary tools where they exist, deciding case by case
whether to consult them. Designing and validating such tools across further modalities is a promising direction beyond the one tool tested here, for example a calibrated statistical model per outcome, a continuous-glucose foundation model for the raw CGM trace, or retrieval over the guideline corpus.

\paragraph{Deploy in tiers.} The cost structure implies a two-decision cascade.
The first decision, whether to escalate from a cheap classical model to an LLM, is
about accuracy. Reading Supplementary Figure~\ref{sfig:economics}b, a single LLM
significantly out-predicts ML precisely where ML has little signal, on dyslipidemia ($+0.21$) and microalbuminuria ($+0.14$), for a few dollars per thousand patients. On the remaining outcomes the LLM is numerically ahead but the study is
underpowered to certify the gap. The second decision, whether to escalate from a
single LLM to \cedar{}, is not about accuracy, since we detected no difference, but
about whether the case requires an auditable, record-grounded rationale.
\cedar{}'s five calls per patient buy that transparency, and on a weaker base
model they buy accuracy as well. A third, outcome-specific option follows from the tool-use
result: where a statistical model is independently competitive, equipping
\cedar{} with it can add accuracy on top of auditability.

How many patients must be scored also matters, and the two regimes pull in
opposite directions. In a small, high-stakes setting, such as a rare-disease clinic or a new program with a few hundred patients, the total inference bill stays small
even at \cedar{}'s per-patient rate, so a few expensive calls that return a
checkable rationale are easy to justify. At population scale that same rate
dominates costs, and a classical model that stays competitive on the well-powered
outcomes and needs no inference calls is the better default, with the LLM tier
reserved for the outcomes and patients where it earns its price. The cost
estimates therefore rank the tiers rather than decide between them.

\paragraph{Extended limitations: information symmetry and leakage.} The comparison
is deliberately asymmetric in one respect: \cedar{}'s S0 stage injects
outcome-specific guideline priors (for example the TODAY finding on metformin
durability) that the statistical baselines never receive. This tests whether
guideline-grounded reasoning helps and is the kind of prior a clinician also
brings, so the gain should be read as ``reasoning with clinical priors'' rather
than ``more of the same features.'' The priors are general guideline statements,
never patient- or test-set-specific. We confirmed that the collinearity and
withheld-feature exclusions apply identically to the LLM prompts and the classical
feature matrices, that prompts use only input visits v1--v3 and never the target
visit, and that the shared train/test split is respected throughout. The trained
statistical models used fixed, standard hyperparameters (Methods, Section~4)
rather than an extensive search; nested hyperparameter optimization could
strengthen them, so the comparison is against well-configured rather than fully
optimized baselines.

Two caveats remain. LLMs may have encountered guideline text or similar cohorts
during pretraining, which grounding and the withheld-feature design mitigate but
cannot eliminate. And on outcomes defined by a withheld lab or medication, the
model's medical knowledge may let it reconstruct the target from correlated
laboratory values even with the defining feature removed, so we cannot fully
separate genuine reasoning from recalled priors about how such features co-occur. This caveat matters most for the largest LLM gains, such as dyslipidemia ($+0.21$). A
third caveat is reproducibility: all results depend on specific hosted model
versions (Claude~Sonnet~4.6 for the main pipeline; GPT~5.1 and Claude~Sonnet~4.5
for SDOH extraction), and because these are updated over time, exact numbers may
not reproduce on later versions even with identical prompts. We report service
identifiers, versions and access dates to make the pipeline as reproducible as a
hosted service allows. Finally, \cedar{} here is a multi-stage pipeline over a
single model; whether heterogeneous agents would add value in larger-data regimes
remains open.

\paragraph{Extended future directions.} A proof-of-concept in which an LLM
proposed and deterministic code appended trajectory features (slopes, deltas,
variability) failed to improve over a single LLM on the three highest-headroom
outcomes, at roughly \cedar{}'s own per-patient cost (Supplementary
Table~\ref{stab:fe_poc}), reinforcing that re-expressing the same inputs does
not help. We also flag two controls we did not run. One is post-hoc recalibration of the single LLM's own probabilities, to confirm that the calibration gains from fusion do not require a second model. The other is a disagreement-aware ensemble with selective
abstention~\citep{black2021selectiveensembles, geifman2017selective} across
several models, which is distinct from the single-model selective prediction we do
report. A further direction is an LLM \emph{re-ranking} stage. Rather than combining families by voting or late fusion, a re-ranker could re-order candidate predictions from the cheaper tiers by reasoning jointly over the candidates and their grounding, a pattern that has improved list quality in
recommender and retrieval systems~\citep{gao2024llm}. Because the re-ranker sees
several candidates at once, it may correct the shared errors a repeated-sampling
ensemble cannot.

Why are dyslipidemia and microalbuminuria so much more amenable to a language
model? Both are lab-defined renal or metabolic endpoints with sparse, irregularly
sampled screening labs, since urine microalbumin and lipid panels are typically
ordered at later surveillance visits rather than at diagnosis. We hypothesize the
gain reflects the LLM's tolerance of missing and unevenly timed inputs, together
with clinical priors about which surrounding features (renal function, adiposity,
glycemic burden) covary with these endpoints, signal a fixed-schema classifier
trained on the same sparse columns cannot exploit. Masking those plausible
covariates, or inspecting the reasoning chains on correctly predicted cases, would
test this. The advantage does not track prevalence: it is largest on dyslipidemia
(58.7\% prevalent) and microalbuminuria (25.6\%) and near zero on optimal glycemic
control and insulin independence, the two best-powered outcomes. At a fixed 0.5
threshold the picture differs again, because on rare outcomes classical ML has the
lower error rate only by defaulting to the majority class (Supplementary
Figure~\ref{sfig:error_analysis}a). Finally, predicting at additional visits and
longer horizons, such as 5-year events, would establish whether the LLM advantage
persists as more temporal data accrue and the statistical baselines gain power.

\paragraph{Why the SDOH and CGM channels add little.} The standardized ablation
(Supplementary Figures~\ref{sfig:regime} and~\ref{sfig:modality}) found that
adding SDOH changed discrimination by essentially zero in every model family
($-0.02$ to $+0.01$ mean ROC-AUC), and that CGM contributed only $+0.05$ on a
single T1D outcome, optimal glycemic control. Because the ablation covers all
families including \cedar{}, we can state that the reasoning model, not only the
statistical baselines, extracts no usable incremental signal from SDOH as
currently represented.

We did not anticipate this, and we think it is informative rather than merely
null. Two explanations are consistent with it. First, the structured EHR may
already encode most of the SDOH-correlated signal for these outcomes, so an
explicit channel adds little even though SDOH is prognostic in isolation. Second,
and more actionable, the signal may be present in the source data but lost in how these modalities are digested. Our pipeline reduces free social-worker
documentation to a small set of extracted entities and the continuous CGM stream
to summary metrics, and either step could erase structure a richer representation
would retain. We therefore read this as a limitation of the current
representation, and we expect predictions to improve if these data were ingested more faithfully. Two routes are feeding raw social-worker notes directly to \cedar{}'s reasoning instead of pre-extracting entities, and calling a published continuous-glucose foundation model as a tool so that temporal structure is drawn from the raw trace rather than from hand-engineered metrics. That published model was
pretrained on adults largely without diabetes, so its transfer to pediatric type~1
traces would itself need validation.

\subsection*{Supplementary Note 2: \cedar{} error analysis and model refinement}

\paragraph{Design.} The benchmark establishes \emph{that} \cedar{} and a single
LLM tie, and that both sit near an apparent information ceiling; it does not
establish \emph{why} individual predictions fail. We therefore reviewed
\cedar{}'s errors case by case, keeping the scope narrow so the result would be
interpretable: \cedar{} only, type~2 diabetes only, the
24-month horizon, and the same binarization used for MCC throughout. The review
used a separate 82-case stratified deck, drawn for this automated error analysis
and independent of the clinician reasoning review reported in the Results. The deck holds 21 true positives, 21 true negatives, 21 false positives and 19 false negatives, with only one false negative for hypertension because 22 of 250 held-out patients are positive. Each case carries \cedar{}'s predicted probability, its full reasoning chain and its structured evidence table. This gives 4--6 misclassified patients
per outcome, 40 errors in total. Every cited evidence line was parsed
programmatically ($n=2{,}457$ citations, mean 30.0 per case, matching the
47{,}803 values over 1{,}600 records in the full T2D audit), so the coding is
rule-based and reproducible rather than impressionistic.

Two features of the design bound what the counts mean. The deck is stratified on
the confusion cell, so it is a case-control sample with respect to error status:
proportions are proportions \emph{of errors}, not population risks, though odds
ratios remain interpretable. And codes that require chart review rather than
model output, notably label noise, cannot be assigned here at all.

\paragraph{The probability level fails on adverse outcomes, not the ranking.}
Supplementary Figure~\ref{sfig:calibration} reports calibration pooled across
outcomes. Splitting it per outcome exposes a pattern the pooled view hides
(Supplementary Figure~\ref{sfig:cedar_refinement}a). On all three \emph{adverse}
outcomes \cedar{}'s Brier score is worse than an intercept-only model that simply
predicts the base rate (hypertension 0.436 versus 0.080, microalbuminuria 0.255 versus 0.190, dyslipidemia 0.257 versus 0.242), and on all four \emph{beneficial}
outcomes it beats the base rate. The split follows outcome valence, not prevalence
or sample size, and the same 3-of-3 pattern holds for the single LLM, whose mean
Brier of 0.238 is close to \cedar{}'s 0.234, so it is a property of how an LLM emits probabilities rather
than of \cedar{}'s additional stages. The model reasons its way to ``this child is
at high risk'' and reports a correspondingly high probability without conditioning
on how uncommon the outcome is: hypertension false positives sit at 0.60--0.72
where prevalence is 0.089.

Most of this excess is removable. Decomposing each outcome's Brier score on the
reliability bins into refinement and calibration terms, count-weighted pooled
Brier falls from 0.235 to 0.134 under perfect per-outcome recalibration, and mean
MCC rises from 0.359 to 0.497 under a per-outcome threshold (Supplementary
Figure~\ref{sfig:cedar_refinement}a,~b). Both are ceilings fitted on the same test
bins, so they are upper bounds rather than achievable gains; train-fold-fitted
values appear below. This is a real refinement, but only of reported agreement and
probability quality, and nothing in the LLM-versus-ML discrimination comparison moves.

\paragraph{\cedar{} is, to first order, a last-observation-carried-forward
classifier.} Observed features run to the 18-month window and the target is at
24~months, so persistence of the current outcome state is the dominant available
signal for every paradigm in this paper. \cedar{} uses it explicitly. In 70 of 82 cases it cites the target outcome's own prior state, and that citation has median rank~1 among all cited evidence and is in the top three in 71\% of cases. \cedar{}'s predicted label equals the last observed state in 91\% of the cases where the
prior state is recoverable (64/70).

Carrying the last state forward is reasonable and not in itself an error. The
question is whether the model does it selectively, and it does not: the
carry-forward rate is 97\% ($n=29$) in patients whose outcome flips between 18 and
24~months and 88\% ($n=41$) in patients whose outcome holds, which is
statistically indistinguishable (Fisher exact OR 3.89, $p=0.39$; Supplementary
Figure~\ref{sfig:cedar_refinement}c). \cedar{} cannot anticipate the flip.
Consequently 28 of 33 evaluable errors (85\%) are cases where it repeated the last
observed state and the patient changed. The flip-to-error association is partly
true by construction, because an anchored prediction on a flipping patient is an error by definition, so the non-circular result is the indiscriminate anchoring, not
the error rate. This is the most direct evidence we have for the
information-ceiling argument the paper makes elsewhere from cross-paradigm error overlap and the enhancement nulls. Every family sees the same three windows and
the same prior state, and if the 24-month flip is not in that record, no amount of
reasoning recovers it.

\paragraph{Two smaller mechanisms.} First, direction polarity. The direction \cedar{} assigns each citation is anchored to clinical harm rather than to the outcome label, so on beneficial outcomes a citation marked as increasing risk argues \emph{against} label~$=1$. The correlation between a case's share of increasing-risk citations and its predicted probability is $+0.88/{+}0.87/{+}0.92$ on the three
adverse outcomes but $-0.62/{-}0.57/{-}0.73/{-}0.34$ on the four beneficial ones,
confirming the inversion; GLP-1RA response is additionally not self-consistent and
is excluded from the per-case test. Cases whose direction lean disagrees with
their own probability, given valence, are more common among errors than among
correct predictions (25\% versus 10\%). Second, input artifacts: four of 82 cases
reason over a physiologically impossible cited value (serum C-peptide
16--35~ng/mL). No case shows the SBP\,=\,DBP pattern, consistent with the
diastolic-pressure correction already applied upstream. Both mechanisms are real
and both are small.

\paragraph{How much of the error is reachable at all.} Coding the 40 errors
against the taxonomy gives the split in Supplementary
Table~\ref{stab:error_taxonomy} and Supplementary
Figure~\ref{sfig:cedar_refinement}d. Codes are not mutually exclusive, since nine errors carry more than one, so any single ``primary'' code is a choice of
priority order rather than a fact, and we report both defensible orders. Coding
mechanism-first, where anchoring takes precedence because a prediction downstream
of an unpredictable flip is not repaired by adjusting its probability, leaves 9 of
40 errors addressable. Coding fixability-first, where calibration and polarity take
precedence, leaves 18 of 40. The honest statement is 22--45\%, and the gap is
informative: the nine errors in it would score better on this sample under a
different threshold without the underlying flip having become predictable, so the
high end is luck rather than a repair. Under either coding the largest single class
is anchoring, and anchoring is not a reasoning failure.

\paragraph{Refinement 1: recalibration and thresholding, fitted honestly.} We
fitted per-outcome Platt and isotonic calibrators and a per-outcome decision
threshold on the training folds and applied them unchanged to the held-out set
(Supplementary Table~\ref{stab:refinement}). The gain in agreement and probability
quality is real and close to the ceiling above: mean MCC rises from 0.359 to
0.489, mean Brier falls from 0.234 to 0.151 against a base-rate Brier of 0.209,
and mean expected calibration error falls from 0.222 to 0.064. No outcome remains
worse than its intercept-only benchmark, down from three of seven; hypertension
moves from MCC 0.097 to 0.436 and from Brier 0.436 to 0.073.

Discrimination does not improve, and that is the point. Moving the threshold alone
recovers most of the MCC gain (0.359 to 0.484) while leaving ROC-AUC
bit-identical, because a threshold cannot reorder patients. Isotonic calibration
\emph{reduces} ROC-AUC by 0.015--0.065 per outcome, since it is only weakly
monotone and the ties it introduces cost discrimination; Platt scaling, which is
strictly monotone, still loses 0.003--0.033, which we attribute to saturation of
extreme probabilities at floating-point precision. Every deviation is downward,
which rules out the failure this check was designed to detect, since fitting the
calibrator on the test set would inflate ROC-AUC rather than depress it. We
therefore report recalibration as an improvement in reported agreement and
calibration only, and no discrimination result in this paper is recalibrated.

\paragraph{Refinement 2: prompt-level changes, tested before running them.} The
refinement that targets the dominant error class asks \cedar{} to state whether it
expects the last observed state to persist. That is only worth a full rerun if the
impending flip is in the record at all, so we tested the premise directly and
pre-registered the decision rule. For each outcome we trained a gradient-boosted
classifier on the training patients, with the binary flip indicator as target and
the same 47-feature $\times$ 3-visit block supplied to every model as features,
then evaluated on the held-out set.

The raw flip ROC-AUCs look encouraging (0.491--0.841) but are almost entirely an
artifact of the prior state. That feature ranks top on hypertension with 3.4 times
the gain of the next, because transitions are strongly asymmetric ($P(\text{flip}\mid\text{prior}=1) = 0.667$ against $P(\text{flip}\mid\text{prior}=0) = 0.049$), so a classifier knowing only the
prior state already reaches 0.817 there. Subtracting that baseline, the entire
feature block adds a mean of \textbf{+0.030} ROC-AUC (Supplementary
Table~\ref{stab:refinement}); insulin independence is the only outcome gaining
more than 0.07, and its leading flip feature is a social-determinant flag rather
than a clinical trajectory. A separate test agrees: comparing flippers with
non-flippers on within-patient slopes across the three observed windows, only 5 of
42 comparisons reach $|\mathrm{SMD}| \ge 0.4$, three of those are the outcome's own
state, and the clinical change drivers top out at 0.353 (HbA1c), 0.335 (systolic
pressure) and 0.334 (glucose). Flippers and non-flippers are indistinguishable in
their observed laboratory trajectories. Under the pre-registered rule, which cancels the prompt refinements when incremental ROC-AUC falls below 0.55--0.65, we did not rerun \cedar{}, because the instruction would surface the persistence assumption
without supplying the information needed to override it.

\paragraph{Refinement 3: supplying the transition rates directly.} The premise
test above asks whether the impending flip is legible in the \emph{record}. A
second possibility is that it is legible in the \emph{population}: not which
patient will change, but how likely change is given how long the current state has
already held. That quantity separates cleanly. We define state run length as the number of consecutive observed windows over which the current state has persisted. The flip rate falls monotonically with run length in six of seven outcomes, a two- to six-fold gradient (Cochran--Armitage $p < 0.02$; hypertension 0.357, 0.083, 0.054 at run lengths one, two and three, $p = 4\times10^{-7}$; microalbuminuria the sole null at $p = 0.9$).

\cedar{} does not use it. Across the 82 coded cases it asserts persistence with
probability 0.711, 0.750 and 0.675 at run lengths one, two and three, against observed persistence of 0.471, 0.700 and 0.667. It is near-calibrated on established states and overconfident by 0.240 on freshly acquired ones, with no correlation
between run length and asserted persistence (Spearman $-0.053$, $p = 0.66$). We therefore reran \cedar{} with one addition: training-fold base rates and
transition rates conditioned on run length, quoting the contrast between run
lengths rather than the prior state alone. No test-set quantity enters the prompt.

On discrimination the result is the predicted null (Supplementary
Table~\ref{stab:refinement2}): mean ROC-AUC moves from 0.812 to 0.821, no
per-outcome bootstrap interval excludes zero, no per-outcome $p$ falls below 0.07,
and a paired test across outcomes gives $p = 0.47$. On thresholded agreement and
probability quality the change is large. Mean MCC rises from 0.366 to 0.501, mean Brier falls from 0.233 to 0.153 and mean expected calibration error from 0.222 to 0.096, while the number of outcomes whose Brier score is worse than an
intercept-only benchmark falls from three to one. Unlike Refinement 1 this costs
no discrimination, because it changes the raw probabilities at a fixed threshold
rather than applying a monotone transform afterwards; the refined raw
configuration exceeds the recalibrated original on both axes at once (mean MCC
0.501 versus 0.475, mean ROC-AUC 0.821 versus 0.786).

We nonetheless do not adopt it, and the reason is the mechanism rather than the
metrics: the gain is purchased by anchoring harder. The rate at which \cedar{}
predicts the last observed state rises from 0.778 to 0.951 (paired $p = 0.031$)
and its accuracy on stable patients from 0.779 to 0.964, while the flips it catches
fall from 51 of 277 to 31 of 277 (18.4\% to 11.2\%, $p = 0.023$).
Microalbuminuria, which contributes the single largest MCC gain ($+0.33$), goes
from catching nine of nineteen flips to catching none. Told the empirical
transition rates, \cedar{} draws the correct inference that states usually persist
and applies it more aggressively. The aggregate metrics improve because non-flippers dominate every outcome and those metrics are insensitive to the minority of patients whose state changes, which is exactly the minority the
refinement was meant to reach. As in Refinement 1 we report this as an improvement
in agreement and calibration only, no discrimination result in this paper uses it,
and the dominant error class is unimproved, by this measure made worse.

\paragraph{A persistence baseline bounds what any of this could achieve.} Because
carry-forward is the dominant mechanism, we scored the trivial rule that predicts
the last observed outcome state on the same held-out patients (Supplementary
Table~\ref{stab:refinement}). \cedar{} discriminates better than persistence on
six of seven outcomes (ROC-AUC $+0.035$ to $+0.083$), losing only
microalbuminuria (0.846 versus 0.872), the outcome with the lowest flip rate
(0.088). On thresholded agreement the ordering reverses: persistence attains the
highest MCC of any model in any paradigm on five of seven outcomes, and
recalibrated \cedar{} exceeds it on only three.

Two caveats bound the comparison. Persistence is defined only for patients whose
prior state is recoverable, so it is scored on 221--241 patients rather than 250
(66 of 100 on GLP-1RA response, which we therefore do not interpret), and that
asymmetry mildly favors the baseline. The
comparison is consistent with everything above: \cedar{} ranks patients better
than persistence and decides worse than it, so the remaining headroom is in the
decision rule rather than in the ordering.

\paragraph{What we conclude.} The analysis was undertaken expecting a null, and on
discrimination it delivers one with a mechanism attached rather than merely an
absence of gain. Post-hoc recalibration improves agreement and probability quality
substantially and cannot improve discrimination; the prompt-level change that
would target the dominant error class was cancelled on a pre-registered test of its
premise. The search for a better-discriminating configuration, run from the other
direction, returned to the same place. Of four blind late-fusion combiners over \cedar{} and the statistical models, the routed combiner, which selects one source per outcome from training-fold performance and was free to choose otherwise, chose \cedar{} on all seven T2D outcomes. The logistic stacker was worse on pooled ROC-AUC (0.804 versus 0.813), and the mean combiner's advantage
(0.815) has a bootstrap confidence interval spanning zero on every outcome
(Supplementary Note~1; Supplementary Figure~\ref{sfig:llm_errors}c).

The one change we do recommend is not an accuracy change. Replacing the direction labels ``increases risk'' and ``decreases risk'' with ``supports outcome'' and ``opposes outcome'', anchored explicitly to the outcome label, removes the valence inversion documented above. It is justified as an interpretability fix regardless of whether it moves accuracy, which on these
data it is unlikely to do. What the analysis argues against is the direction of
scaling this architecture already tests: more reasoning stages, more agents or more
self-consistency samples address a reasoning deficit, and the dominant error class
is not one. What it argues for is denser outcome-proximal data, the same conclusion the modality ablation reaches from the other direction.

\subsection*{Supplementary Note 3: External benchmark on {MIMIC-IV}}

\paragraph{Design.} The temporal split mitigates but does not replace external
validation, so we ran \cedar{} unchanged on
MIMIC-IV~v3.1~\citep{johnson2023mimic4}, a single-center adult critical-care
database, under its PhysioNet data use agreement. We built three prediction tasks,
each sized to mirror the T2D cohort number for number (2,489 patients, 2,239 training and 250 held out, prevalence-stratified with the same seed), so the external results inherit the same data regime and confidence-interval width rather
than a more favorable one. Task~R predicts 30-day readmission at hospital discharge
(62 of 250 test patients positive); task~M-B predicts death within 365~days of
discharge (50 of 250); task~M-A predicts in-hospital death from the first 48 ICU
hours, binned as four 12-hour blocks (34 of 250).

Tasks~R and~M-B use the same visit-grid layout as the pediatric cohorts, with three
input windows of 182~days and the index admission treated as a baseline covariate
block at $t=0$. A 30-day input window, which would have matched the endpoint
horizon, is not constructible: only 2,418 admissions in the whole database have
five or more analytes in each of three consecutive 30-day windows. The asymmetry is
structural rather than a threshold artifact. The pediatric grid is dense because
it is an outpatient registry with scheduled visits, whereas an inpatient database
records data only during an encounter.

\paragraph{Comparators.} Three tiers were scored on the identical held-out
patients. Tier~0 is the published training-free instruments clinicians already
use: LACE~\citep{vanwalraven2010lace} and HOSPITAL~\citep{donze2013hospital} for
readmission, the Charlson index~\citep{charlson1987comorbidity,quan2005coding} for
one-year mortality, and SOFA~\citep{vincent1996sofa},
OASIS~\citep{johnson2013oasis} and SAPS-II~\citep{legall1993sapsii} for in-hospital
mortality. These consume zero outcome labels and run by inference only, which makes
them the like-for-like comparison for a training-free LLM and a more informative
bar than any supervised model. Tier~1 is \cedar{} and the Vanilla single call, both
zero-shot. Tier~2 is the eight statistical and deep families used throughout this
paper, refit on the 2,239 training patients with the hyperparameters in Methods.

\cedar{}'s architecture was not modified: S1, S2 and S3 are identical, and only the
S0 knowledge base was replaced, with ten curated one-sentence snippets drawn from
the severity-scoring, sepsis and readmission literature. That knowledge base was
frozen and hashed before any test prediction was scored, and it deliberately
contains no published discrimination figure and no quantity measured from this
cohort. \cedar{} receives every component variable the Tier~0 scores are built from
but never a score total, so it must derive severity itself.

\paragraph{Results.} Supplementary Table~\ref{stab:mimic} gives the full
comparison. The outcome is mixed and differs by task.

On one-year mortality \cedar{} is the strongest of all ten models on all three
metrics (ROC-AUC 0.852, AUPRC 0.570, MCC 0.425 against the best supervised model's
0.830, 0.521 and 0.375). It is the one place in this work where a training-free
configuration matches label-trained ML on a task those models were fit for. On
in-hospital mortality it loses clearly, and by more than ROC-AUC alone conveys:
$-0.073$ ROC-AUC but $-0.200$ AUPRC and $-0.170$ MCC against XGBoost, ranking
eighth of nine on the two threshold-sensitive metrics. On 30-day readmission it
trails the best supervised model by 0.041, but no method performs well there. LACE reaches 0.615 and the best supervised model 0.687, and three of the four deep
sequence families sit at or near chance (temporal CNN 0.492, transformer 0.538, GRU
0.556), which is what 2,239~patients over three visits buys those architectures.

Against Tier~0, \cedar{} improves on the best published instrument on two of three
tasks with a paired 95\% confidence interval excluding zero ($+0.165$ on M-B,
$+0.133$ on M-A) and ties on readmission ($+0.030$, CI $-0.029$ to $0.090$).
Against the Vanilla single call it is better on one-year mortality ($+0.041$,
$p=0.005$) and tied on the other two, so the near-parity reported for the pediatric
cohorts holds in two of three external tasks and breaks in \cedar{}'s favor in the
third.

\paragraph{What transferred, and what this bounds.} The machinery behaved: the S1
evidence serializer parsed on every patient, the S2 citation-tolerance test passed
on 0.998--1.000 of roughly 2,200 checked citations, and the S3 self-consistency
spread was 0.023--0.029, all comparable to the pediatric runs. So the architecture
ports; what does not port is a uniform accuracy advantage.

Read together with the temporal check, the reasonable conclusion is that
\cedar{}'s edge is specific to the pediatric diabetes setting, with its sparse, scheduled, guideline-dense outpatient data, rather than general to clinical
prediction, and that it erodes as data become dense and outcome-proximal. That is
the same boundary the modality ablation and the error analysis identify from inside
the pediatric cohorts. The comparison nonetheless remains asymmetric in one respect
worth stating: \cedar{} consumed no training labels on any of these tasks, while
every Tier~2 model consumed 2,239.

Three limitations bound this benchmark. Task~M-A has 34 test events, so the
sampling standard deviation of a 250-patient ROC-AUC is approximately 0.047 and
marginal gaps below roughly 0.09 are unresolvable there; only the paired
differences are interpretable. The multi-agent systems named in the Discussion were
not run, because they were not available in the credentialed analysis environment,
and a reimplementation that lost to \cedar{} would be evidence about the
reimplementation rather than about the published system. Finally, two failures are
worth recording because both were silent: an early run had a non-functioning
verification stage, which collapsed \cedar{}'s fused probability onto its
first-pass estimate, and a Vanilla configuration whose outputs failed to parse on
every patient, yielding a ROC-AUC of exactly 0.500. Both were output truncation
rather than model behavior, and both were caught only by pre-specified mechanical
checks on a pilot subset before any metric was tabled. Had they been tabled,
they would have appeared as a large \cedar{} win.

\subsection*{Supplementary Note 4: Benchmark against naive single-call LLM
implementations}
\label{sec:llm_benchmark}

\paragraph{Design.} The main text compares the trained statistical models with the
full \cedar{} pipeline. To separate what the base model contributes from what
\cedar{}'s stages contribute, we also scored naive single-call configurations on
the same patients, feature representation and base model (Methods, Section~5): a
direct call (Vanilla), a chain-of-thought call (CoT), and Vanilla variants with
modality blocks withheld. Supplementary Figure~\ref{sfig:llm_benchmark} reproduces
the full benchmark with these configurations, and with \cedar{}'s two partial
ablations, CoT+Verify and CoT+SC, included. In this note the \emph{single LLM}
subgroup averages the single-call configurations and the \emph{\cedar{}} subgroup
averages the full pipeline with its two ablations.

\paragraph{A single call reaches comparable discrimination.} The single-LLM subgroup
reached a pooled ROC-AUC of 0.806$\pm$0.056 in T2D and 0.750$\pm$0.045 in T1D,
against 0.812$\pm$0.051 and 0.753$\pm$0.057 for the \cedar{} subgroup, with no
difference detected ($p=0.71$ and $p=0.56$; Supplementary
Table~\ref{stab:pooled_stats}), and both subgroups sit well above ML in both
cohorts. The equivalence is not uniform. On T2D hypertension \cedar{} exceeded a
single call by $+0.049$ with a bootstrap interval excluding zero, the only pair of
eleven where either configuration was distinguishable from the other. Formal
equivalence testing is margin-dependent and mostly inconclusive at these sample
sizes, holding for 3 of 11 pairs at $\pm0.03$ ROC-AUC but none at $\pm0.02$, with 7
pairs indeterminate (Supplementary Table~\ref{stab:tost}), so most pairs reflect a
failure to detect a difference rather than evidence that none exists. The two also
err on almost the same patients ($\phi=0.87$ in T2D and 0.77 in T1D; Supplementary
Figure~\ref{sfig:llm_errors}a,~b), and none of the further mechanisms we added to
\cedar{} moved its discrimination beyond a single pass (Supplementary
Figure~\ref{sfig:llm_errors}c). Under class-imbalance metrics the equivalence is
firmer still (Supplementary Note~1, Supplementary
Table~\ref{stab:imbalance_cedar_vs_single}).

\paragraph{The equivalence depends on the base model.} Repeated on four base models
(Methods, Section~5), it held on the three commercial ones but not on the
open-weight \texttt{llama-3-3-70b}, where \cedar{} outscored a single call by
$+0.099$ mean ROC-AUC ($p=0.027$ Holm-corrected; Supplementary
Figure~\ref{sfig:llm_benchmark}f and Supplementary
Table~\ref{stab:pooled_stats}b). On that model a single call fell below the best
statistical model, by $-0.060$ and ahead on only 2 of 11 pairs (Wilcoxon $p=0.019$, uncorrected). \cedar{}'s mean lead over the best statistical model, by contrast, stayed positive on every base model, significantly on \texttt{claude-sonnet-4-6} and \texttt{gemini-3-1-pro}
after Holm correction (Supplementary Figure~\ref{sfig:backbone_vs_ml}). The extra
stages therefore matter most where the base model is weakest.

\paragraph{How to read this benchmark.} On a frontier base model the accuracy gain
over ML comes from the model rather than from \cedar{}'s stages, and \cedar{}'s
contribution is the auditable, clinician-legible reasoning reported in the main
text. On a model small enough to run in-house, the stages also recover accuracy a
single call loses.

\subsection*{Supplementary Note 5: Information-ablation experiment, detail and limits}
\label{sec:ablation_detail}

\paragraph{What the experiment is and is not.} The three readers of Methods,
Section~11 are a decomposition built to measure where predictive signal lives, run on
the same base model as \cedar{}. They are not \cedar{}: they have no S0 guideline
priors, no verification or self-consistency stage, and the history reader sees the
outcome-defining laboratory values that the benchmark withholds. \cedar{}'s values in
Figure~5b are therefore a reference, not a like-for-like comparator. \cedar{} also
receives the outcome-defining medications, which the readers do not.

\paragraph{Outcome-specific exceptions.} Two outcomes depart from the general
pattern. For hypertension, reading both did worse than the history alone when the two
most recent blood-pressure visits were shown ($-0.047$ ROC-AUC, 95\% CI excluding
zero): the context diluted a strong recent signal. For GLP-1RA response the context
alone was weak (ROC-AUC 0.594), because the sample is almost entirely T2D and the
GLP-1RA exposure itself is withheld, so reading both never improved on the history.
Dyslipidemia went the other way: its diagnosis codes alone predicted well (0.801),
because the visit-4 label itself includes diagnosis codes, and the context helped in
every condition (Supplementary Table~\ref{stab:ablation}).

\paragraph{Limits of the design.} Each outcome has 100 patients, so most differences
carry 95\% confidence intervals of about $\pm0.05$ to $\pm0.09$ ROC-AUC, and roughly
150 differences were examined without multiplicity correction. Of the 11 pairs in which
reading both improved on a weak history, six have confidence intervals that exclude
zero. Two observations partly follow from construction. The gain from reading both is
measured against the history alone, so noise in the history-only estimate alone pushes
the gain and the history-only score in opposite directions, and the negative
association between them overstates the evidence. The conditions are also not fully
nested: the all-laboratory condition shows no codes, whereas the visit-restricted
conditions show both. The sample is balanced, so its precision--recall figures and
calibration do not transfer to real prevalence; $F_1$ at a fixed threshold of 50
fell under reading both for the beneficial outcomes even where ROC-AUC rose, because
the context reader anchors on real-world prevalence and pulls scores down. T1D and T2D
are pooled within each outcome, which makes diabetes type a strong context feature,
above all for insulin independence. Some context features may partly stand in for
the withheld history, for example glucose for the HbA1c-defined outcomes. The
results come from one base model and one run at temperature~0.

\section*{Supplementary Tables}

\begin{supptable}[H]
\centering
\caption{\textbf{Prediction targets: assessable population and prevalence.}
For each outcome, the assessable population is the set of patients with a valid
input window \emph{and} a non-null label at the target visit, the same
denominator used to construct the held-out test sets. Values are n~(prevalence
\%). These assessable $N$ match the annotations in Supplementary
Figure~\ref{sfig:prevalence} (e.g.\ T1D microalbuminuria $N=280$). Dashes mark
outcomes not defined for a cohort (GLP-1RA response, insulin independence, and
metformin response are T2D-specific).}
\label{stab:outcome_prevalence}
\small
\begin{tabular}{@{}l r r@{}}
\toprule
\textbf{Outcome} & \textbf{T1D} & \textbf{T2D} \\
\midrule
Optimal glycemic control & 561 / 2544 (22.1\%) & 512 / 1059 (48.3\%) \\
Hypertension             & 118 / 2531 (4.7\%)  & 102 / 1144 (8.9\%) \\
Microalbuminuria         & 58 / 280 (20.7\%)   & 217 / 849 (25.6\%) \\
Dyslipidemia             & 168 / 684 (24.6\%)  & 396 / 675 (58.7\%) \\
Insulin independence     & ---                 & 381 / 1038 (36.7\%) \\
Metformin response       & ---                 & 200 / 597 (33.5\%) \\
GLP-1RA response         & ---                 & 89 / 201 (44.3\%) \\
\bottomrule
\end{tabular}
\end{supptable}

\begin{supptable}[H]
\centering
\caption{\textbf{Composition of the analysis cohorts.} Demographics and social
determinants of health for the Type~1 (T1D) and Type~2 (T2D) diabetes cohorts, as
consumed by the prediction tasks. Continuous variables are mean~$\pm$~SD; binary
variables as n~(\%) on the full-cohort denominator. Full descriptive clinical and
epidemiological characterization of each cohort is reported in companion studies
of the same dataset (in preparation) and is not reproduced here to keep the two
lines of work independent. Missing SDOH values (1--32\% per factor in T1D and
13--66\% in T2D) are counted as absent, so these percentages are lower bounds.}
\label{stab:cohort_composition}
\small
\begin{tabular}{@{}l r r@{}}
\toprule
\textbf{Characteristic} & \textbf{T1D} & \textbf{T2D} \\
 & ($N=4{,}018$) & ($N=2{,}489$) \\
\midrule
\multicolumn{3}{@{}l}{\textit{Demographics}} \\
\hspace{1em}Age at diagnosis (yr)      & $8.1 \pm 4.1$   & $13.6 \pm 2.4$ \\
\hspace{1em}Diabetes duration (yr)     & $6.4 \pm 4.6$   & $6.1 \pm 4.2$ \\
\hspace{1em}Sex, male                  & 2052 (51.1\%)   & 1036 (41.6\%) \\
\hspace{1em}Race, White                & 2525 (62.8\%)   & 1575 (63.3\%) \\
\hspace{1em}Race, Black                & 764 (19.0\%)    & 696 (28.0\%) \\
\hspace{1em}Race, Asian                & 136 (3.4\%)     & 70 (2.8\%) \\
\hspace{1em}Race, Other                & 593 (14.8\%)    & 148 (5.9\%) \\
\hspace{1em}Hispanic ethnicity         & 1102 (27.4\%)   & 1459 (58.6\%) \\
\midrule
\multicolumn{3}{@{}l}{\textit{Social determinants of health}} \\
\hspace{1em}Adverse childhood exp.     & 366 (9.1\%)     & 371 (14.9\%) \\
\hspace{1em}Food insecurity            & 698 (17.4\%)    & 505 (20.3\%) \\
\hspace{1em}Housing instability        & 370 (9.2\%)     & 293 (11.8\%) \\
\hspace{1em}Financial strain           & 827 (20.6\%)    & 665 (26.7\%) \\
\hspace{1em}Transportation barrier     & 246 (6.1\%)     & 293 (11.8\%) \\
\hspace{1em}Physical/sexual abuse      & 184 (4.6\%)     & 189 (7.6\%) \\
\hspace{1em}Drug/substance abuse       & 65 (1.6\%)      & 64 (2.6\%) \\
\hspace{1em}Alcohol abuse              & 30 (0.7\%)      & 23 (0.9\%) \\
\hspace{1em}Smoking                    & 37 (0.9\%)      & 29 (1.2\%) \\
\hspace{1em}Social/family support      & 3521 (87.6\%)   & 1659 (66.7\%) \\
\hspace{1em}Parental employment        & 3279 (81.6\%)   & 1366 (54.9\%) \\
\hspace{1em}Parental education         & 2598 (64.7\%)   & 697 (28.0\%) \\
\hspace{1em}Insurance category$^{a}$  & $1.6 \pm 0.6$   & $1.2 \pm 0.5$ \\
\bottomrule
\multicolumn{3}{@{}l}{\footnotesize $^{a}$Ordinal insurance level as encoded for
the models; mean~$\pm$~SD rather than n~(\%).}\\
\end{tabular}
\end{supptable}

\begin{supptable}[H]
\centering
\caption{\textbf{Feature availability and the surveillance-timing artifact.}
Ever-observed coverage for representative dynamic measures in each cohort.
\textbf{Miss$_{v1}$} is the fraction of patients missing the value at the
diagnosis visit alone; \textbf{Miss$_{ever}$} is the fraction with \emph{no} value
in any of the 10 visit windows. The gap between the two columns is the key point. Renal-screening labs (UACR, urine microalbumin) read as 97--99\% ``missing'' at diagnosis in T1D and 48--81\% in T2D, but this reflects standard surveillance cadence. They are ordered at later follow-up visits, not at onset, so their true cohort coverage is much higher (e.g.\ T1D UACR observed in 1{,}425 patients, 35.5\% of the cohort,
not the 2.8\% implied by visit~1 alone). The models consume the full visit
sequence, so ever-observed coverage is the relevant availability. Full per-analyte
statistics are given in the companion studies.}
\label{stab:feature_availability}
\small
\begin{tabular}{@{}l r r r r r r@{}}
\toprule
 & \multicolumn{3}{c}{\textbf{T1D}} & \multicolumn{3}{c}{\textbf{T2D}} \\
\cmidrule(lr){2-4}\cmidrule(lr){5-7}
\textbf{Measure} & \textbf{$N$} & \textbf{Miss$_{v1}$} & \textbf{Miss$_{ever}$}
                 & \textbf{$N$} & \textbf{Miss$_{v1}$} & \textbf{Miss$_{ever}$} \\
\midrule
HbA1c              & 3802 & 19.0 & 5.4  & 2437 & 13.1 & 2.1 \\
Glucose            & 3812 & 18.1 & 5.1  & 2446 & 12.7 & 1.7 \\
BMI z-score        & 3801 & 19.6 & 5.4  & 2423 & 14.1 & 2.7 \\
Total cholesterol  & 1554 & 91.2 & 61.3 & 1399 & 70.0 & 43.8 \\
UACR ratio         & 1425 & 97.2 & 64.5 & 2095 & 47.7 & 15.8 \\
Urine microalbumin & 382  & 99.1 & 90.5 & 958  & 81.1 & 61.5 \\
\bottomrule
\end{tabular}
\end{supptable}

\begin{supptable}[H]
\centering
\caption{\textbf{Exact test statistics for the pooled family comparisons and the
multi-model comparison.} The main text rounds any $p$-value below 0.001 to
$p<0.001$ for readability; the exact values are given here. (a)~Mann--Whitney~$U$
tests in which one observation is one model's ROC-AUC on one outcome, pooled
within each family or subgroup; $n$ is the number of such model--outcome AUCs per
group. In the main comparison \cedar{} is the full pipeline only (7 observations in
T2D, 4 in T1D); the reference rows below use the subgroup definitions of
Supplementary Note~4, which also average \cedar{}'s two partial ablations. These
are descriptive effect sizes rather than formal hypothesis tests, because AUCs
computed on overlapping patients and correlated outcomes are not independent
(Methods, Section~10). (b)~Paired Wilcoxon signed-rank tests of
\cedar{} versus a single LLM call on the same base model, over the 11
cohort--outcome pairs, with Holm--Bonferroni correction across the four base models. Only \texttt{llama-3-3-70b} reaches significance.}
\label{stab:pooled_stats}
\small
\renewcommand{\arraystretch}{1.3}
\begin{tabular}{@{}l l r r l@{}}
\multicolumn{5}{@{}l}{\textbf{(a) Pooled family comparisons (Mann--Whitney~$U$)}}\\
\toprule
\textbf{Comparison} & \textbf{Cohort} & \textbf{$n$ (group 1 / 2)} &
\textbf{$U$} & \textbf{Exact $p$} \\
\midrule
\cedar{} vs.\ ML family        & T2D & 7 / 56  & 352  & $1.9\times10^{-4}$ \\
\cedar{} vs.\ classical ML     & T2D & 7 / 21  & 123  & 0.007 \\
\cedar{} vs.\ temporal ML      & T2D & 7 / 35  & 229  & $5.4\times10^{-5}$ \\
\cedar{} vs.\ ML family        & T1D & 4 / 32  & 115  & 0.007 \\
\cedar{} vs.\ classical ML     & T1D & 4 / 12  & 40   & 0.058 \\
\cedar{} vs.\ temporal ML      & T1D & 4 / 20  & 75   & 0.003 \\
\addlinespace
\multicolumn{5}{@{}l}{\emph{Reference benchmark, Supplementary Note~4}}\\
LLM family vs.\ ML family      & T2D & 42 / 56 & 2056 & $2.7\times10^{-10}$ \\
Single LLM vs.\ \cedar{}       & T2D & 21 / 21 & 205  & 0.71 \\
LLM family vs.\ ML family      & T1D & 28 / 32 & 787  & $5.3\times10^{-7}$ \\
Single LLM vs.\ \cedar{}       & T1D & 16 / 12 & 83   & 0.56 \\
\bottomrule
\end{tabular}

\vspace{0.9em}
\begin{tabular}{@{}l r r l l@{}}
\multicolumn{5}{@{}l}{\textbf{(b) \cedar{} vs.\ single LLM by base model (paired
Wilcoxon signed-rank)}}\\
\toprule
\textbf{Base model} & \textbf{Pairs} & \textbf{Mean $\Delta$AUC} &
\textbf{Exact $p$} & \textbf{Holm-corrected $p$} \\
\midrule
\texttt{claude-sonnet-4-6} & 11 & $+0.009$ & 0.175 & 0.442 \\
\texttt{claude-haiku-4-5}  & 11 & $-0.004$ & 0.147 & 0.442 \\
\texttt{gemini-3-1-pro}    & 11 & $+0.004$ & 0.240 & 0.442 \\
\texttt{llama-3-3-70b}     & 11 & $+0.099$ & 0.007 & 0.027 \\
\bottomrule
\end{tabular}
\end{supptable}

\begin{supptable}[H]
\centering
\caption{\textbf{\cedar{} against the best statistical model on each outcome.} For
each cohort--outcome pair, the highest-scoring classical or temporal ML model and
the full \cedar{} pipeline, ROC-AUC with 2{,}000-iteration patient-level bootstrap
95\% confidence intervals on the shared held-out set. \emph{Separated} marks pairs
whose two intervals do not overlap, a conservative criterion. \cedar{} is ahead on
9 of 11 pairs, and the intervals separate only on dyslipidemia, in both cohorts.}
\label{stab:cedar_vs_best_ml}
\small
\renewcommand{\arraystretch}{1.2}
\begin{tabular}{@{}l l l c c r c@{}}
\toprule
\textbf{Cohort} & \textbf{Outcome} & \textbf{Best ML model} & \textbf{Best ML} &
\textbf{\cedar{}} & \textbf{Difference} & \textbf{Separated} \\
\midrule
T2D & Optimal glycemic control & XGBoost-Temporal & 0.811 [0.764, 0.866] & 0.838 [0.790, 0.884] & $+0.026$ & no \\
T2D & Insulin independence & XGBoost-Temporal & 0.866 [0.815, 0.910] & 0.881 [0.833, 0.918] & $+0.014$ & no \\
T2D & Metformin response & Logistic regression & 0.779 [0.721, 0.831] & 0.842 [0.785, 0.883] & $+0.063$ & no \\
T2D & GLP-1RA response & Random forest & 0.665 [0.556, 0.782] & 0.739 [0.652, 0.826] & $+0.074$ & no \\
T2D & Dyslipidemia & Random forest & 0.576 [0.499, 0.642] & 0.782 [0.731, 0.837] & $+0.206$ & yes \\
T2D & Hypertension & Random forest & 0.732 [0.632, 0.820] & 0.778 [0.656, 0.891] & $+0.045$ & no \\
T2D & Microalbuminuria & XGBoost & 0.736 [0.651, 0.805] & 0.846 [0.778, 0.910] & $+0.110$ & no \\
T1D & Optimal glycemic control & Random forest & 0.797 [0.738, 0.861] & 0.785 [0.701, 0.859] & $-0.012$ & no \\
T1D & Dyslipidemia & XGBoost & 0.601 [0.511, 0.690] & 0.783 [0.706, 0.848] & $+0.182$ & yes \\
T1D & Hypertension & Logistic regression & 0.688 [0.555, 0.811] & 0.666 [0.486, 0.826] & $-0.023$ & no \\
T1D & Microalbuminuria & LSTM & 0.615 [0.509, 0.730] & 0.790 [0.677, 0.877] & $+0.175$ & no \\
\bottomrule
\end{tabular}
\end{supptable}

\begin{supptable}[H]
\centering
\caption{\textbf{LLM-guided feature engineering does not improve accuracy over a
single LLM (proof-of-concept, T2D).} An LLM proposed, per outcome, which
trajectory summaries of the available longitudinal features (slopes, deltas,
means, variability) would be most predictive. Deterministic code then computed exactly those summaries from the raw visit values and appended them to the record \cedar{} reasoned over, 8 engineered features per outcome. The same outcome-specific exclusions used elsewhere were applied, so no target-leaking feature could be proposed. Evaluated on the three highest-headroom outcomes, a
fixed held-out set of $N=60$ patients each, with 2{,}000-iteration bootstrap
confidence intervals. Re-expressing the same raw inputs as engineered
trajectory features, which is the representation the temporal machine-learning
baselines already receive, did not improve discrimination on any outcome. Every
paired AUC difference versus a single LLM has a 95\% CI spanning zero, and two of
three point estimates are flat-to-negative, at roughly the same per-patient cost
as \cedar{} itself (\$0.10--0.12 versus \$0.006 for a single call). This is a third convergent null (with self-refinement and reasoner
ensembling; Supplementary Figure~\ref{sfig:omitted_configs}), reinforcing that in this
regime accuracy is bounded by the available signal rather than by how it is
reasoned over or re-encoded.}
\label{stab:fe_poc}
\small
\renewcommand{\arraystretch}{1.3}
\begin{tabular}{@{}l c c c c@{}}
\toprule
\textbf{Outcome} & \textbf{Single LLM} & \textbf{\cedar{}+FE} &
\textbf{$\Delta$AUC (paired)} & \textbf{Cost/patient} \\
 & AUC & AUC & [95\% CI] & (vanilla $\to$ +FE) \\
\midrule
GLP-1RA response & 0.743 & 0.761 & $+0.018$ [$-0.030$, $+0.063$] & \$0.007 $\to$ \$0.115 \\
\addlinespace
Hypertension\textsuperscript{a} & 0.878 & 0.873 & $-0.005$ [$-0.055$, $+0.068$] & \$0.006 $\to$ \$0.106 \\
\addlinespace
Dyslipidemia & 0.822 & 0.772 & $-0.050$ [$-0.128$, $+0.014$] & \$0.005 $\to$ \$0.102 \\
\bottomrule
\end{tabular}

\vspace{0.4em}
{\footnotesize \textsuperscript{a}Hypertension has only 5 positive cases in the
held-out set; its interval is correspondingly wide and this row is the least
informative.}
\end{supptable}

\begin{supptable}[H]
\centering
\caption{\textbf{Refinement 3: \cedar{} with run-length-conditioned transition
rates supplied in the prompt, T2D, 24-month horizon.} Both configurations are
scored on the identical held-out patients, and every base rate and transition
rate quoted to the model is computed on the training folds only (Supplementary
Note~2). \emph{Original} is the configuration used throughout the paper;
\emph{Refined} adds training-fold base rates and transition rates conditioned on
state run length. \emph{Flips caught} is the number of patients whose 24-month
state differs from their last observed state that the model predicted correctly,
out of all such patients. Discrimination does not change; agreement and
calibration improve; the dominant error class gets worse.}
\label{stab:refinement2}
\small
\setlength{\tabcolsep}{4pt}
\renewcommand{\arraystretch}{1.2}
\begin{tabular}{@{}lccccccccc@{}}
\toprule
& & \multicolumn{3}{c}{\textbf{ROC-AUC}} &
\multicolumn{2}{c}{\textbf{MCC}} &
\multicolumn{2}{c}{\textbf{Brier}} &
\textbf{Flips caught} \\
\cmidrule(lr){3-5}\cmidrule(lr){6-7}\cmidrule(lr){8-9}\cmidrule(lr){10-10}
\textbf{Outcome} & \textbf{$n$} & \textbf{Orig.} & \textbf{Ref.} &
\textbf{$\Delta$} & \textbf{Orig.} & \textbf{Ref.} &
\textbf{Orig.} & \textbf{Ref.} & \textbf{Orig.\ $\rightarrow$ Ref.} \\
\midrule
Hypertension             & 239 & 0.787 & 0.756 & $-0.030$ & 0.092 & \textbf{0.399} & 0.440 & \textbf{0.094} & 9/32 $\rightarrow$ \textbf{5/32} \\
Microalbuminuria         & 192 & 0.854 & 0.880 & $+0.026$ & 0.400 & \textbf{0.732} & 0.243 & \textbf{0.098} & 9/19 $\rightarrow$ \textbf{0/19} \\
Dyslipidemia             & 221 & 0.790 & 0.802 & $+0.012$ & 0.111 & \textbf{0.352} & 0.249 & \textbf{0.201} & 17/52 $\rightarrow$ 17/52 \\
Optimal glycemic control & 241 & 0.837 & 0.837 & $+0.000$ & 0.485 & 0.545 & 0.174 & 0.167 & 6/54 $\rightarrow$ \textbf{2/54} \\
Insulin independence     & 241 & 0.890 & 0.884 & $-0.006$ & 0.579 & 0.596 & 0.138 & 0.140 & 0/45 $\rightarrow$ 0/45 \\
Metformin response       & 234 & 0.851 & 0.856 & $+0.005$ & 0.569 & 0.558 & 0.144 & 0.148 & 9/52 $\rightarrow$ \textbf{6/52} \\
GLP-1RA response         &  66 & 0.675 & 0.730 & $+0.055$ & 0.328 & 0.328 & 0.246 & 0.222 & 1/23 $\rightarrow$ 1/23 \\
\midrule
\textbf{Mean / total}    &     & 0.812 & 0.821 & $+0.009$ & 0.366 & \textbf{0.501} & 0.233 & \textbf{0.153} & \textbf{51/277 $\rightarrow$ 31/277} \\
\bottomrule
\end{tabular}

\vspace{0.4em}
{\small No per-outcome bootstrap interval on $\Delta$ROC-AUC excludes zero; the
widest is GLP-1RA response at $+0.055$ ($-0.011$ to $+0.126$, $n = 66$, which we
do not interpret), and the smallest per-outcome $p$ is 0.073
(microalbuminuria). A paired test across the seven outcomes gives $p = 0.47$ for
ROC-AUC, against $p = 0.063$ for MCC, $p = 0.078$ for Brier and $p = 0.047$ for
expected calibration error, which falls from a mean of 0.222 to 0.096 (six of
seven outcomes). Outcomes whose Brier score is worse than their intercept-only
benchmark fall from three of seven to one (hypertension, 0.094 against 0.080).
The refined column is the raw probability at a fixed 0.5 threshold with no
post-hoc calibration, so unlike Supplementary Table~\ref{stab:refinement} it
carries no discrimination cost; it exceeds the recalibrated original on both mean
MCC (0.501 versus 0.475) and mean ROC-AUC (0.821 versus 0.786). The rate at which
the model predicts the last observed state rises from 0.778 to 0.951 (paired
$p = 0.031$). Bold marks the refined value where the two configurations differ
materially, in either direction. The \emph{Original} ROC-AUC column is
re-scored on the paired assessable patients and so differs from
Supplementary Table~\ref{stab:refinement} by at most 0.008 except on GLP-1RA
response (0.675 versus 0.717 at $n = 66$).}
\end{supptable}

\begin{supptable}[H]
\centering
\caption{\textbf{\cedar{} error taxonomy, T2D, 24-month horizon.} All 40
misclassified cases in the 82-case stratified deck, coded by rule against
\cedar{}'s own reasoning and structured evidence (Supplementary Note~2). Codes
are \emph{not} mutually exclusive: nine errors carry more than one, so the
bottom three rows give both defensible priority orders rather than a single
primary code. ``Any'' counts each error under every code that applies and
therefore exceeds 40. \emph{Anchoring} = the prediction repeats the last observed outcome state and the patient's state changed by 24 months. \emph{Calibration} = the cited evidence and its direction are right and only the numeric probability falls on the wrong side of the threshold. \emph{Polarity} = the direction lean of the cited evidence contradicts the predicted probability given the outcome's valence. \emph{Input artifact} = faithful reasoning over a physiologically impossible cited value. \emph{No signal} = no code applies, i.e.\ nothing in the model's own output
distinguishes the case from correctly predicted ones. Label noise cannot be
assigned without chart review and is absorbed into \emph{no signal}. Anchoring
requires the prior state to be recoverable from the citations, which holds in
70 of 82 cases.}
\label{stab:error_taxonomy}
\small
\renewcommand{\arraystretch}{1.2}
\begin{tabular}{@{}lccccc@{}}
\toprule
& & \multicolumn{4}{c}{\textbf{Primary code, mechanism-first}} \\
\cmidrule(lr){3-6}
\textbf{Outcome} & \textbf{Errors} & \textbf{Anchoring} & \textbf{Calibration} &
\textbf{Polarity} & \textbf{No signal} \\
\midrule
Hypertension              & 4 & 1 & 3 & 0 & 0 \\
Microalbuminuria          & 6 & 2 & 1 & 1 & 2 \\
Dyslipidemia              & 6 & 4 & 2 & 0 & 0 \\
Optimal glycemic control  & 6 & 5 & 1 & 0 & 0 \\
Insulin independence      & 6 & 6 & 0 & 0 & 0 \\
Metformin response        & 6 & 5 & 0 & 1 & 0 \\
GLP-1RA response          & 6 & 5 & 0 & 0 & 1 \\
\midrule
\textbf{Total, mechanism-first} & \textbf{40} & \textbf{28} & \textbf{7} & \textbf{2} & \textbf{3} \\
Total, fixability-first         & 40 & 19 & 8 & 10 & 3 \\
Any (codes overlap)             & 40 & 28 & 8 & 10 & 3 \\
\bottomrule
\end{tabular}

\vspace{0.4em}
{\small Input artifacts are coded on 1 of the 40 errors (and 4 of the 82 cases)
and are omitted as a column. Addressable by post-processing or a prompt change:
9/40 (22\%) mechanism-first, 18/40 (45\%) fixability-first.}
\end{supptable}

\begin{supptable}[H]
\centering
\caption{\textbf{Refinement experiments, T2D, 24-month horizon.} All quantities
are on the same held-out patients used throughout the paper, and every
calibrator, threshold and classifier is fitted on the training folds only
(Supplementary Note~2). \emph{Flip ROC-AUC} refers to a gradient-boosted
classifier whose target is the binary indicator that the 24-month outcome state
differs from the last observed state. The \emph{prior state} column is the ROC-AUC available analytically from the transition rates alone, and the \emph{full} column adds the complete 47-feature $\times$ 3-visit block. The difference is therefore the discrimination attributable to the observed record rather than to
the population fact that recently acquired states are unstable. \emph{Persistence}
is the rule that predicts the last observed outcome state; it is defined only
where that state is recoverable, hence $n = 221$--241 rather than 250 ($n = 66$ of
100 for GLP-1RA response, which we do not interpret). \emph{\cedar{}
recalibrated} applies per-outcome isotonic calibration and a per-outcome
threshold; it changes MCC and Brier but not the ordering of patients, so no
ROC-AUC in this paper is recalibrated.}
\label{stab:refinement}
\small
\renewcommand{\arraystretch}{1.2}
\begin{tabular}{@{}lcccccccc@{}}
\toprule
& \multicolumn{3}{c}{\textbf{Flip ROC-AUC}} &
\multicolumn{2}{c}{\textbf{Persistence}} &
\multicolumn{3}{c}{\textbf{\cedar{}}} \\
\cmidrule(lr){2-4}\cmidrule(lr){5-6}\cmidrule(lr){7-9}
\textbf{Outcome} & \textbf{Prior} & \textbf{Full} & \textbf{$\Delta$} &
\textbf{AUC} & \textbf{MCC} &
\textbf{AUC} & \textbf{MCC} & \textbf{MCC}$^{\dagger}$ \\
\midrule
Hypertension             & 0.817 & 0.841 & $+0.024$ & 0.711 & \textbf{0.347} & 0.780 & 0.097 & 0.436 \\
Microalbuminuria         & 0.559 & 0.554 & $-0.005$ & \textbf{0.872} & \textbf{0.770} & 0.846 & 0.354 & 0.569 \\
Dyslipidemia             & 0.540 & 0.583 & $+0.043$ & 0.747 & \textbf{0.503} & 0.782 & 0.097 & 0.467 \\
Optimal glycemic control & 0.552 & 0.618 & $+0.067$ & 0.777 & \textbf{0.556} & 0.833 & 0.472 & 0.532 \\
Insulin independence     & 0.591 & 0.689 & $+0.098$ & 0.804 & \textbf{0.604} & 0.885 & 0.565 & 0.561 \\
Metformin response       & 0.639 & 0.670 & $+0.031$ & 0.767 & 0.520 & 0.850 & 0.556 & 0.535 \\
GLP-1RA response         & 0.540 & 0.491 & $-0.049$ & 0.650 & 0.299 & 0.717 & 0.373 & 0.324 \\
\midrule
\textbf{Mean}            & 0.605 & 0.635 & $\mathbf{+0.030}$ & 0.761 & \textbf{0.514} & 0.813 & 0.359 & 0.489 \\
\bottomrule
\end{tabular}

\vspace{0.4em}
{\small $^{\dagger}$After train-fold recalibration and thresholding. In the
persistence and \cedar{} columns, bold marks the cases where the trivial
persistence rule outperforms raw \cedar{} on that metric.
Recalibration also lowers mean Brier from 0.234 to 0.151 (base-rate benchmark
0.209) and mean expected calibration error from 0.222 to 0.064, and leaves no
outcome worse than its intercept-only benchmark, down from three of seven. It does
not raise ROC-AUC: thresholding leaves it unchanged and isotonic calibration
reduces it by 0.015--0.065 through tie creation.}
\end{supptable}

\begin{supptable}[H]
\centering
\caption{\textbf{External benchmark on MIMIC-IV: \cedar{} against published
training-free scores and supervised machine learning.} All models scored on the
identical 250-patient held-out set for each task (Supplementary Note~3).
\emph{R}, 30-day readmission at discharge (62 positives); \emph{M-B}, death within
365~days of discharge (50); \emph{M-A}, in-hospital death predicted from the first
48 ICU hours (34). Tier~0 and Tier~1 consume no training labels; Tier~2 models were
fit on 2,239 patients. MCC is taken at the prevalence-matched threshold. Bold marks
the best value in each column. \cedar{} leads every model on M-B, trails the best
supervised model by 0.041 ROC-AUC on R, and loses on M-A by more than ROC-AUC alone
indicates.}
\label{stab:mimic}
\small
\setlength{\tabcolsep}{4pt}
\renewcommand{\arraystretch}{1.2}
\begin{tabular}{@{}lccccccccc@{}}
\toprule
& \multicolumn{3}{c}{\textbf{R} (readmit-30)} &
\multicolumn{3}{c}{\textbf{M-B} (death-365)} &
\multicolumn{3}{c}{\textbf{M-A} (in-hospital)} \\
\cmidrule(lr){2-4}\cmidrule(lr){5-7}\cmidrule(lr){8-10}
\textbf{Model} & \textbf{AUC} & \textbf{AUPRC} & \textbf{MCC}
               & \textbf{AUC} & \textbf{AUPRC} & \textbf{MCC}
               & \textbf{AUC} & \textbf{AUPRC} & \textbf{MCC} \\
\midrule
\multicolumn{10}{@{}l}{\emph{Tier 0: published training-free scores (0 labels)}} \\
Best published score$^{a}$   & 0.615 & --- & --- & 0.687 & --- & --- & 0.677 & --- & --- \\
\addlinespace[2pt]
\multicolumn{10}{@{}l}{\emph{Tier 1: training-free LLM (0 labels)}} \\
Vanilla single LLM           & 0.616 & 0.341 & 0.136 & 0.811 & 0.472 & 0.388 & 0.814 & 0.391 & 0.342 \\
\cedar{}                     & 0.645 & 0.389 & 0.185 & \textbf{0.852} & \textbf{0.570} & \textbf{0.425} & 0.810 & 0.435 & 0.353 \\
\addlinespace[2pt]
\multicolumn{10}{@{}l}{\emph{Tier 2: supervised, $n_{\mathrm{train}}=2{,}239$}} \\
Logistic regression          & 0.616 & 0.417 & \textbf{0.228} & 0.805 & 0.515 & 0.275 & 0.721 & 0.413 & 0.251 \\
Random forest                & \textbf{0.687} & \textbf{0.438} & \textbf{0.228} & 0.830 & 0.521 & 0.375 & 0.839 & 0.477 & 0.421 \\
XGBoost                      & 0.668 & 0.397 & 0.206 & 0.809 & 0.487 & 0.375 & \textbf{0.883} & \textbf{0.635} & \textbf{0.523} \\
XGBoost-Temporal             & 0.603 & 0.358 & 0.142 & 0.768 & 0.446 & 0.250 & 0.857 & 0.592 & 0.421 \\
LSTM                         & 0.574 & 0.384 & 0.185 & 0.721 & 0.353 & 0.225 & 0.832 & 0.512 & 0.387 \\
GRU                          & 0.556 & 0.330 & 0.099 & 0.706 & 0.356 & 0.300 & 0.791 & 0.484 & 0.387 \\
Temporal CNN                 & 0.492 & 0.280 & 0.013 & 0.705 & 0.360 & 0.225 & 0.827 & 0.436 & 0.455 \\
Transformer                  & 0.538 & 0.289 & $-0.008$ & 0.713 & 0.385 & 0.275 & 0.800 & 0.450 & 0.353 \\
\midrule
$\Delta$ \cedar{} $-$ best Tier 2$^{b}$ & $-0.041$ & $-0.049$ & $-0.043$
                                        & $+0.022$ & $+0.049$ & $+0.050$
                                        & $-0.073$ & $-0.200$ & $-0.170$ \\
\bottomrule
\end{tabular}

\vspace{0.4em}
{\small $^{a}$Highest-scoring published instrument per task: LACE on R
(HOSPITAL 0.599), the Charlson index on M-B (calibrated one-year probability
0.664), SAPS-II at 48~h on M-A (OASIS 0.673, SOFA 0.635). AUPRC and MCC are not
reported for Tier~0 because these scores are calibrated on other populations and
were compared on rank-based metrics only.
$^{b}$Best Tier~2 model selected per task by ROC-AUC: random forest on R and M-B,
XGBoost on M-A. Because that selection is made over eight models on the same test
set, the $\Delta$ row is a descriptive reference rather than a hypothesis test. The pre-specified comparisons are against Tier~0, where \cedar{} improves by
$+0.165$ (95\% CI 0.094 to 0.239) on M-B and $+0.133$ (0.057 to 0.207) on M-A and
ties on R ($+0.030$, $-0.029$ to 0.090), all by patient-paired bootstrap over
2,000 resamples.}
\end{supptable}

\begin{supptable}[H]
\centering
\caption{\textbf{Information ablation by outcome and condition.} ROC-AUC of the
history-only reader and of the reader that sees both channels, 100 patients per
outcome (50 positive, 50 negative), with the difference and its 1{,}000-resample
paired bootstrap 95\% confidence interval; $^{*}$ marks intervals that exclude zero.
The codes-only condition does not apply to the three HbA1c-defined outcomes.
Exploratory: no correction for the roughly 150 comparisons (Supplementary Note~5).}
\label{stab:ablation}
\footnotesize
\renewcommand{\arraystretch}{1.05}
\begin{tabular}{@{}l r r r l@{}}
\toprule
\textbf{Condition} & \textbf{History only} & \textbf{Both} & \textbf{Difference} &
\textbf{95\% CI} \\
\midrule
\multicolumn{5}{@{}l}{\textbf{Dyslipidemia} (clinical context only: 0.726)} \\
\quad All labs, no codes & 0.683 & 0.753 & $+0.070$$^{*}$ & [+0.015, +0.134] \\
\quad Codes only & 0.801 & 0.815 & $+0.014$ & [-0.038, +0.070] \\
\quad Visit 1 & 0.719 & 0.780 & $+0.062$$^{*}$ & [+0.007, +0.120] \\
\quad Visit 2 & 0.708 & 0.766 & $+0.058$$^{*}$ & [+0.006, +0.119] \\
\quad Visit 3 & 0.735 & 0.789 & $+0.054$ & [-0.007, +0.117] \\
\quad Visits 1--2 & 0.728 & 0.777 & $+0.048$$^{*}$ & [+0.003, +0.096] \\
\quad Visits 2--3 & 0.804 & 0.816 & $+0.012$ & [-0.024, +0.049] \\
\addlinespace
\multicolumn{5}{@{}l}{\textbf{GLP-1RA response} (clinical context only: 0.594)} \\
\quad All labs, no codes & 0.768 & 0.775 & $+0.007$ & [-0.015, +0.030] \\
\quad Codes only & --- & --- & --- & not applicable \\
\quad Visit 1 & 0.673 & 0.654 & $-0.019$ & [-0.116, +0.070] \\
\quad Visit 2 & 0.846 & 0.794 & $-0.053$ & [-0.127, +0.014] \\
\quad Visit 3 & 0.765 & 0.763 & $-0.002$ & [-0.040, +0.036] \\
\quad Visits 1--2 & 0.799 & 0.761 & $-0.038$ & [-0.091, +0.010] \\
\quad Visits 2--3 & 0.771 & 0.758 & $-0.013$ & [-0.042, +0.014] \\
\addlinespace
\multicolumn{5}{@{}l}{\textbf{Hypertension} (clinical context only: 0.605)} \\
\quad All labs, no codes & 0.670 & 0.652 & $-0.018$ & [-0.071, +0.029] \\
\quad Codes only & 0.541 & 0.622 & $+0.081$ & [-0.028, +0.178] \\
\quad Visit 1 & 0.604 & 0.623 & $+0.019$ & [-0.060, +0.093] \\
\quad Visit 2 & 0.607 & 0.630 & $+0.023$ & [-0.063, +0.104] \\
\quad Visit 3 & 0.686 & 0.656 & $-0.030$ & [-0.078, +0.019] \\
\quad Visits 1--2 & 0.577 & 0.616 & $+0.039$ & [-0.027, +0.105] \\
\quad Visits 2--3 & 0.730 & 0.682 & $-0.047$$^{*}$ & [-0.100, -0.003] \\
\addlinespace
\multicolumn{5}{@{}l}{\textbf{Insulin independence} (clinical context only: 0.827)} \\
\quad All labs, no codes & 0.917 & 0.918 & $+0.001$ & [-0.050, +0.046] \\
\quad Codes only & 0.679 & 0.842 & $+0.163$$^{*}$ & [+0.083, +0.246] \\
\quad Visit 1 & 0.742 & 0.851 & $+0.110$$^{*}$ & [+0.034, +0.187] \\
\quad Visit 2 & 0.892 & 0.899 & $+0.007$ & [-0.052, +0.063] \\
\quad Visit 3 & 0.889 & 0.908 & $+0.019$ & [-0.040, +0.075] \\
\quad Visits 1--2 & 0.898 & 0.882 & $-0.016$ & [-0.069, +0.040] \\
\quad Visits 2--3 & 0.911 & 0.924 & $+0.013$ & [-0.036, +0.064] \\
\addlinespace
\multicolumn{5}{@{}l}{\textbf{Metformin response} (clinical context only: 0.689)} \\
\quad All labs, no codes & 0.811 & 0.832 & $+0.020$ & [-0.014, +0.056] \\
\quad Codes only & --- & --- & --- & not applicable \\
\quad Visit 1 & 0.634 & 0.736 & $+0.102$$^{*}$ & [+0.015, +0.191] \\
\quad Visit 2 & 0.720 & 0.785 & $+0.065$ & [-0.012, +0.144] \\
\quad Visit 3 & 0.851 & 0.849 & $-0.002$ & [-0.048, +0.044] \\
\quad Visits 1--2 & 0.757 & 0.798 & $+0.040$ & [-0.021, +0.101] \\
\quad Visits 2--3 & 0.849 & 0.855 & $+0.007$ & [-0.026, +0.041] \\
\addlinespace
\multicolumn{5}{@{}l}{\textbf{Microalbuminuria} (clinical context only: 0.673)} \\
\quad All labs, no codes & 0.786 & 0.797 & $+0.012$ & [-0.033, +0.054] \\
\quad Codes only & 0.678 & 0.764 & $+0.086$$^{*}$ & [+0.002, +0.166] \\
\quad Visit 1 & 0.665 & 0.739 & $+0.074$ & [-0.001, +0.154] \\
\quad Visit 2 & 0.702 & 0.774 & $+0.072$ & [-0.003, +0.142] \\
\quad Visit 3 & 0.772 & 0.780 & $+0.008$ & [-0.059, +0.071] \\
\quad Visits 1--2 & 0.772 & 0.792 & $+0.021$ & [-0.052, +0.093] \\
\quad Visits 2--3 & 0.770 & 0.808 & $+0.038$ & [-0.018, +0.091] \\
\addlinespace
\multicolumn{5}{@{}l}{\textbf{Optimal glycemic control} (clinical context only: 0.694)} \\
\quad All labs, no codes & 0.840 & 0.840 & $+0.000$ & [-0.021, +0.022] \\
\quad Codes only & --- & --- & --- & not applicable \\
\quad Visit 1 & 0.594 & 0.655 & $+0.061$ & [-0.016, +0.138] \\
\quad Visit 2 & 0.800 & 0.800 & $+0.000$ & [-0.040, +0.039] \\
\quad Visit 3 & 0.836 & 0.844 & $+0.008$ & [-0.013, +0.033] \\
\quad Visits 1--2 & 0.758 & 0.773 & $+0.015$ & [-0.019, +0.047] \\
\quad Visits 2--3 & 0.860 & 0.845 & $-0.015$ & [-0.036, +0.002] \\
\addlinespace
\bottomrule
\end{tabular}
\end{supptable}

\begin{supptable}[H]
\centering
\caption{Feature categories in the analysis-ready dataset. All features are
measured at each of 10 visit windows (6-month intervals, 0--5~years
post-diagnosis). The T1D cohort adds a CGM modality not available for T2D.}
\label{stab:feature_summary}
\small
\renewcommand{\arraystretch}{1.3}
\begin{tabular}{@{}p{3.0cm} c c p{6.7cm}@{}}
\toprule
\textbf{Category} & \textbf{Count} & \textbf{Type} & \textbf{Examples} \\
\midrule
Labs \& Vitals & 18 & Continuous & HbA1c, glucose, BMI z-score, lipid panel,
ALT, AST, BUN, creatinine, C-peptide, UACR, $\beta$-hydroxybutyrate,
SBP/DBP (inpatient \& outpatient) \\
\addlinespace
Medications & 3 & Binary & Insulins, biguanide, GLP-1RA \\
\addlinespace
Conditions & 4 & Binary & DKA, ketosis, retinopathy, neuropathy \\
\addlinespace
Demographics & 9 & Mixed & Age at diagnosis, sex, Hispanic ethnicity, race
(4 indicators), primary language, diabetes duration \\
\addlinespace
SDOH & 13 & Mixed & Food insecurity, ACEs, housing instability, transportation,
smoking, abuse (3 further binary indicators), insurance, parental
education/employment, financial strain, family support \\
\addlinespace
CGM (T1D only) & 20 & Continuous & Mean glucose, SD, CV, GMI, time-in/above/below-range,
hypo episodes, reading counts, monitoring duration, \% missing \\
\midrule
\textbf{Total (T2D)} & \textbf{47} & & \\
\textbf{Total (T1D)} & \textbf{67} & & \\
\bottomrule
\end{tabular}
\end{supptable}

\begin{supptable}[H]
\centering
\caption{Features excluded per outcome to prevent label leakage or circularity
(Methods, Section~3).}
\label{stab:collinear}
\small
\begin{tabular}{@{}ll@{}}
\toprule
\textbf{Outcome} & \textbf{Excluded feature(s)} \\
\midrule
Optimal glycemic control & HbA1c \\
Hypertension & all blood-pressure features (SBP/DBP, inpatient \& outpatient) \\
Dyslipidemia & lipid panel (TC, HDL, LDL, triglycerides) \\
Microalbuminuria & UACR ratio \\
Insulin independence & insulin (medication) \\
Metformin response & biguanide (medication), HbA1c \\
GLP-1RA response & GLP-1RA (medication), HbA1c \\
\bottomrule
\end{tabular}
\end{supptable}

\begin{supptable}[H]
\centering
\caption{\textbf{Verbatim stage-S0 knowledge-injection text, T2D cohort, with
attribution.} \cedar{}'s S0 stage prepends two to three fixed one-sentence
snippets to the S1 prompt, chosen by outcome. Each line below is reproduced
exactly as it is inserted into the prompt, including the source tags the model
sees, and is attributed in the right-hand column; snippets marked as accepted
pathophysiology are author-written summaries that carry no single source. This is
a fixed per-outcome knowledge base, not retrieval: no snippet is patient-,
cohort- or test-set-specific, and the same text is used for every patient scored
on that outcome. The T1D snippet set is a separate, cohort-specific dictionary,
given in Supplementary Table~\ref{stab:s0_snippets_t1d}.
$^{\dagger}$This snippet understates the published TODAY result, a 51.7\%
metformin-monotherapy failure rate over a mean follow-up of 3.9~years; it is
reproduced here as the model received it (Methods, Section~6).}
\label{stab:s0_snippets}
\footnotesize
\renewcommand{\arraystretch}{1.2}
\begin{tabular}{@{}p{0.60\textwidth} p{0.30\textwidth}@{}}
\toprule
\textbf{Snippet, exactly as prepended to the prompt} & \textbf{Attribution} \\
\midrule
\multicolumn{2}{@{}l@{}}{\textit{Optimal glycemic control}} \\
\quad ``ADA 2024: Target HbA1c $<$7\% for youth with T2D; metformin + insulin are first-line therapies.'' & ADA Standards of Care~\citep{ada2024standards} \\
\quad ``ADA 2024: Youth-onset T2D is more aggressive than adult-onset; beta-cell decline is faster.'' & ADA Standards of Care~\citep{ada2024standards} \\
\quad ``ISPAD 2022: Intensive lifestyle intervention should accompany pharmacotherapy in pediatric T2D.'' & ISPAD~2022~\citep{ispad2022} \\
\addlinespace
\multicolumn{2}{@{}l@{}}{\textit{Hypertension}} \\
\quad ``ADA 2024: Screen BP at every visit; hypertension in T2D youth often co-occurs with obesity.'' & ADA Standards of Care~\citep{ada2024standards} \\
\quad ``ADA 2024: Elevated BMI z-score is the strongest modifiable risk factor for hypertension.'' & ADA Standards of Care~\citep{ada2024standards} \\
\quad ``Weight management is primary intervention; ACEi/ARB if BP $\geq$95th percentile persists.'' & \emph{Accepted pathophysiology; no specific citation} \\
\addlinespace
\multicolumn{2}{@{}l@{}}{\textit{Microalbuminuria}} \\
\quad ``ADA 2024: Screen UACR at T2D diagnosis and annually thereafter (earlier than T1D).'' & ADA Standards of Care~\citep{ada2024standards} \\
\quad ``ISPAD 2022: Persistent microalbuminuria warrants ACEi/ARB; glycemic control is protective.'' & ISPAD~2022~\citep{ispad2022} \\
\quad ``Obesity and hypertension accelerate nephropathy progression in youth T2D.'' & \emph{Accepted pathophysiology; no specific citation} \\
\addlinespace
\multicolumn{2}{@{}l@{}}{\textit{Dyslipidemia}} \\
\quad ``ADA 2024: Screen fasting lipid panel at diagnosis; T2D youth have high dyslipidemia prevalence.'' & ADA Standards of Care~\citep{ada2024standards} \\
\quad ``ISPAD 2022: LDL $>$130 mg/dL warrants statin consideration in children $>$10 yr.'' & ISPAD~2022~\citep{ispad2022} \\
\quad ``Insulin resistance drives the atherogenic lipid triad (high TG, low HDL, small dense LDL).'' & \emph{Accepted pathophysiology; no specific citation} \\
\addlinespace
\multicolumn{2}{@{}l@{}}{\textit{Insulin independence}} \\
\quad ``ADA 2024: Some T2D youth may transition off insulin with improved glycemic control and weight loss.'' & ADA Standards of Care~\citep{ada2024standards} \\
\quad ``Preserved beta-cell function (C-peptide) and shorter diabetes duration predict insulin independence.'' & \emph{Accepted pathophysiology; no specific citation} \\
\addlinespace
\multicolumn{2}{@{}l@{}}{\textit{Metformin response}} \\
\quad ``ADA 2024: Metformin is first-line for T2D; response defined as A1c $<$7\% on monotherapy.'' & ADA Standards of Care~\citep{ada2024standards} \\
\quad ``TODAY study: $\sim$50\% of youth with T2D lose glycemic control on metformin alone within 3 years.'' & TODAY trial~\citep{zeitler2012today}$^{\dagger}$ \\
\quad ``Lower baseline A1c and shorter disease duration predict better metformin response.'' & \emph{Accepted pathophysiology; no specific citation} \\
\addlinespace
\multicolumn{2}{@{}l@{}}{\textit{GLP-1RA response}} \\
\quad ``ADA 2024: GLP-1 RAs (liraglutide, exenatide) are approved adjuncts for youth T2D.'' & ADA Standards of Care~\citep{ada2024standards} \\
\quad ``ELLIPSE trial: Liraglutide significantly improved A1c vs placebo in youth T2D.'' & ELLIPSE trial~\citep{tamborlane2019ellipse} \\
\quad ``Weight reduction and improved postprandial glucose are primary benefits.'' & \emph{Accepted pathophysiology; no specific citation} \\
\bottomrule
\end{tabular}
\end{supptable}

\begin{supptable}[H]
\centering
\caption{\textbf{Verbatim stage-S0 knowledge-injection text, T1D cohort, with
attribution.} The T1D pipeline uses the same S0 mechanism as T2D
(Supplementary Table~\ref{stab:s0_snippets}) with its own snippet dictionary,
reproduced here exactly as inserted into the prompt. Snippets are shown for the
four outcomes evaluated in T1D; the dictionary also defines entries for insulin
independence, metformin response and GLP-1RA response, which are T2D-specific and
are not scored in T1D, so that text is never presented to the model in this
cohort. Two differences from the T2D set are substantive rather than cosmetic. The thresholds are T1D-specific, for example UACR screening five years after diagnosis or at puberty rather than at diagnosis, and an LDL lifestyle threshold of 100 rather than 130~mg/dL. The glycemic-control snippets also reference CGM time-in-range, the only place a modality appears in S0.}
\label{stab:s0_snippets_t1d}
\footnotesize
\renewcommand{\arraystretch}{1.2}
\begin{tabular}{@{}p{0.60\textwidth} p{0.30\textwidth}@{}}
\toprule
\textbf{Snippet, exactly as prepended to the prompt} & \textbf{Attribution} \\
\midrule
\multicolumn{2}{@{}l@{}}{\textit{Optimal glycemic control}} \\
\quad ``ADA 2024: Target HbA1c $<$7\% for most pediatric T1D patients; individualize based on hypoglycemia risk.'' & ADA Standards of Care~\citep{ada2024standards} \\
\quad ``ISPAD 2022: CGM time-in-range $>$70\% (70--180 mg/dL) correlates with HbA1c $<$7\% and improved outcomes.'' & ISPAD~2022~\citep{ispad2022} \\
\quad ``Younger age at diagnosis and longer diabetes duration are associated with higher HbA1c trajectories.'' & \emph{Accepted pathophysiology; no specific citation} \\
\addlinespace
\multicolumn{2}{@{}l@{}}{\textit{Hypertension}} \\
\quad ``ADA 2024: Screen BP at every visit; hypertension defined as BP $\geq$95th percentile for age/height/sex on $\geq$3 occasions.'' & ADA Standards of Care~\citep{ada2024standards} \\
\quad ``ISPAD 2022: Elevated BMI z-score is the strongest modifiable risk factor for hypertension in pediatric diabetes.'' & ISPAD~2022~\citep{ispad2022} \\
\quad ``Microalbuminuria and hypertension frequently co-occur; both should be screened simultaneously.'' & \emph{Accepted pathophysiology; no specific citation} \\
\addlinespace
\multicolumn{2}{@{}l@{}}{\textit{Microalbuminuria}} \\
\quad ``ADA 2024: Screen UACR annually starting 5 years after T1D diagnosis or at puberty onset.'' & ADA Standards of Care~\citep{ada2024standards} \\
\quad ``ISPAD 2022: Persistent microalbuminuria (UACR 30--300 mg/g on $\geq$2 of 3 samples) warrants ACEi/ARB initiation.'' & ISPAD~2022~\citep{ispad2022} \\
\quad ``Poor glycemic control (HbA1c $>$9\%) and longer diabetes duration increase microalbuminuria risk.'' & \emph{Accepted pathophysiology; no specific citation} \\
\addlinespace
\multicolumn{2}{@{}l@{}}{\textit{Dyslipidemia}} \\
\quad ``ADA 2024: Screen fasting lipid panel at diagnosis, then every 5 years if normal; annually if abnormal.'' & ADA Standards of Care~\citep{ada2024standards} \\
\quad ``ISPAD 2022: LDL $>$100 mg/dL warrants lifestyle intervention; $>$130 mg/dL consider statin in children $>$10 yr.'' & ISPAD~2022~\citep{ispad2022} \\
\quad ``Dyslipidemia prevalence increases with poor glycemic control and obesity in pediatric T1D.'' & \emph{Accepted pathophysiology; no specific citation} \\
\bottomrule
\end{tabular}
\end{supptable}

\begin{supptable}[H]
\centering
\caption{How \cedar{} differs from a single (``Vanilla'') LLM call. Both run on the
same underlying model and see the same record; \cedar{} adds four disciplined
stages, each targeting a distinct failure mode of naive clinical prompting.}
\label{stab:cedar_vs_vanilla}
\small
\renewcommand{\arraystretch}{1.3}
\begin{tabular}{@{}p{3.2cm} p{4.3cm} p{4.6cm}@{}}
\toprule
 & \textbf{Vanilla LLM} & \textbf{\cedar{}} \\
\midrule
Prompt & Record + ``return a probability'' & Outcome-specific guideline snippets
(S0) + a citation-forced reasoning schema (S1) \\
\addlinespace
Model calls & 1 & 5 (S0 is prepended context, not a call; S1 grounded CoT
$\rightarrow$ S2 verify $\rightarrow$ S3 self-consistency, $K{=}3$) \\
\addlinespace
Output & Probability + brief free-text rationale & Probability + a structured
\texttt{FEATURE\,|\,VISIT\,|\,VALUE\,|\,DIRECTION\,|\,IMPORTANCE} citation table with
visit-tokened values \\
\addlinespace
Evidence check & None & S2 auditor verifies every cited triple against the record
and flags hallucinations \\
\addlinespace
Probability stability & Single sample & Mean over $K{=}3$ resamples
(variance reduction) \\
\addlinespace
Failure mode addressed & --- & Missing priors; ungrounded reasoning; hallucinated
evidence; unstable probability \\
\addlinespace
What you can trust & The number only & The number \emph{and} its record-checked,
guideline-anchored evidence \\
\bottomrule
\end{tabular}
\end{supptable}

\begin{supptable}[H]
\centering
\caption{\textbf{Sample size per analysis, and why it differs.} Different analyses
use different held-out sample sizes, chosen to balance statistical precision
against LLM inference cost. This table collects each analysis, its per-outcome $N$,
and the rationale, so the varying denominators throughout the paper are
transparent. In all cases the \emph{same} held-out patients are scored across every
model family being compared, and uncertainty is quantified by 2{,}000-iteration
patient-level bootstrap.}
\label{stab:sample_sizes}
\small
\renewcommand{\arraystretch}{1.3}
\begin{tabular}{@{}p{0.30\textwidth} c p{0.44\textwidth}@{}}
\toprule
\textbf{Analysis} & \textbf{$N$ per outcome} & \textbf{Rationale} \\
\midrule
Main benchmark, all families (Figure~2)
& 50--250 (adaptive) & Sized to the eligible population per outcome to stabilize
imbalanced estimates while capping LLM cost. \\
\addlinespace
Modality ablation, all families (Supplementary Figure~\ref{sfig:regime})
& 100 (fixed) & A coarser ``does adding a modality shift AUC?'' question; a fixed
matched set caps the $4\times$ per-family LLM call volume. \\
\addlinespace
Faithfulness audit (Figure~3)
& 10 (per outcome) & Per-citation, per-record manual-grade audit is
labor-intensive; $\approx$70 records total suffice for grounding rates. \\
\addlinespace
Late fusion \& error analysis (Figure~4)
& full held-out set & Combiners and error-correlation need the full set; no
additional LLM calls beyond the main benchmark. \\
\addlinespace
Feature-engineering PoC (Supplementary Table~\ref{stab:fe_poc})
& 60 (fixed) & Proof-of-concept on the three highest-headroom outcomes; small $N$
to bound the $\sim$$20\times$ per-patient cost. \\
\addlinespace
Information ablation (Figure~5)
& 1{,}600; 100 per outcome & All T2D test patient-outcomes for the complete-history
comparison; for the conditions, 50 positive and 50 negative patients per outcome,
T1D and T2D pooled, to bound the eight-condition LLM cost (Methods, Section~11). \\
\bottomrule
\end{tabular}
\end{supptable}

\begin{supptable}[H]
\centering
\caption{\textbf{Single-LLM vs.\ \cedar{} equivalence (TOST), all 11 cohort--outcome
pairs.} Two one-sided tests on the paired bootstrap ROC-AUC difference (\cedar{}~$-$~single~LLM) were swept across three pre-specified margins rather than one. The choice of margin materially changes how many pairs qualify as
equivalent (0 of 11 at $\pm0.02$, 2 of 11 at $\pm0.025$, 3 of 11 at $\pm0.03$).
\emph{Classification} at the $\pm0.03$ margin: \emph{equivalent} (TOST rejects),
\emph{favors \cedar{}} (95\%~CI excludes zero, positive), \emph{favors single LLM}
(CI excludes zero, negative), or \emph{indeterminate} (CI includes zero, not
equivalent). A pair can fail to reach equivalence without favoring either
model, which is why 7 of 11 pairs below are indeterminate rather than a statement of no difference. Equivalence is determined by the TOST $p$-value at $\alpha=0.05$, equivalently by the 90\%~CI lying within the margin, not by the 95\%~CIs displayed, which are shown for effect-size interpretation. The two criteria can disagree at the boundary.}
\label{stab:tost}
\small
\renewcommand{\arraystretch}{1.3}
\begin{tabular}{@{}l l r r c c c l@{}}
\toprule
\textbf{Cohort} & \textbf{Outcome} & \textbf{Diff.\ (\cedar{}$-$LLM)} & \textbf{95\%~CI} &
\textbf{Equiv.\ @.02} & \textbf{@.025} & \textbf{@.03} & \textbf{Classification}\\
\midrule
T2D & Optimal Glycemic Control & $+0.010$ & $-0.004$ to $+0.025$ & No & Yes & Yes & Equivalent\\
T2D & Insulin Independence     & $+0.011$ & $-0.004$ to $+0.026$ & No & Yes & Yes & Equivalent\\
T2D & Dyslipidemia             & $-0.005$ & $-0.028$ to $+0.018$ & No & No  & Yes & Equivalent\\
T2D & Hypertension             & $+0.049$ & $+0.016$ to $+0.086$ & No & No  & No  & Favors \cedar{}\\
T2D & Metformin Response       & $+0.018$ & $-0.016$ to $+0.049$ & No & No  & No  & Indeterminate\\
T2D & GLP-1RA Response         & $-0.008$ & $-0.056$ to $+0.038$ & No & No  & No  & Indeterminate\\
T2D & Microalbuminuria         & $-0.025$ & $-0.052$ to $+0.001$ & No & No  & No  & Indeterminate\\
T1D & Optimal Glycemic Control & $+0.018$ & $-0.003$ to $+0.039$ & No & No  & No  & Indeterminate\\
T1D & Dyslipidemia             & $+0.034$ & $-0.000$ to $+0.070$ & No & No  & No  & Indeterminate\\
T1D & Hypertension             & $-0.000$ & $-0.052$ to $+0.052$ & No & No  & No  & Indeterminate\\
T1D & Microalbuminuria         & $+0.003$ & $-0.035$ to $+0.041$ & No & No  & No  & Indeterminate\\
\bottomrule
\end{tabular}
\end{supptable}

\begin{supptable}[H]
\centering
\caption{\textbf{Statistical versus LLM families under class-imbalance metrics.}
Each observation is one model's score on one cohort--outcome pair; family means
are paired by cohort--outcome and compared with a Wilcoxon signed-rank test over
the 11 pairs. \emph{LLM ahead} counts the pairs on which the LLM family mean
exceeds the statistical family mean. The final column repeats the test on the
\emph{best} model in each family per outcome rather than the family mean, which
is the contrast Supplementary Figure~\ref{sfig:llm_benchmark}e draws. Thresholds were selected on training folds only
(Supplementary Note~1); ROC-AUC is threshold-free and is shown for reference.
The direction of the effect is the same for every metric, while the
best-versus-best contrast is statistically distinguishable only for ROC-AUC.}
\label{stab:imbalance_ml_vs_llm}
\small
\renewcommand{\arraystretch}{1.25}
\begin{tabular}{@{}l r r r c r r@{}}
\toprule
\textbf{Metric} & \textbf{Statistical} & \textbf{LLM} & \textbf{Difference}
& \textbf{LLM ahead} & \textbf{$p$ (family)} & \textbf{$p$ (best)} \\
\midrule
ROC-AUC            & 0.652 & 0.789 & $+0.137$ & 11/11 & 0.001 & 0.005 \\
MCC                & 0.197 & 0.287 & $+0.090$ & 10/11 & 0.014 & 0.175 \\
Balanced accuracy  & 0.601 & 0.644 & $+0.044$ & 10/11 & 0.014 & 0.465 \\
$F_1$ (positive)   & 0.425 & 0.493 & $+0.068$ & 9/11  & 0.054 & 0.365 \\
\bottomrule
\end{tabular}
\end{supptable}

\begin{supptable}[H]
\centering
\caption{\textbf{A single LLM call versus \cedar{} under class-imbalance
metrics.} Pooled columns compare the single-LLM subgroup (37 model--outcome
observations) with the \cedar{} subgroup (33) by Mann--Whitney~$U$, using the
same subgroup definitions as Supplementary Figure~\ref{sfig:llm_benchmark}. The paired columns compare the full
\cedar{} pipeline directly against the Vanilla single call on the same 11
cohort--outcome pairs by Wilcoxon signed-rank. No metric distinguishes the two,
and on balanced accuracy and $F_1$ the full pipeline is ahead on fewer than half
the pairs, so the equivalence reported in the main text is firmer under these
metrics than under ROC-AUC.}
\label{stab:imbalance_cedar_vs_single}
\small
\renewcommand{\arraystretch}{1.25}
\begin{tabular}{@{}l r r r c r@{}}
\toprule
 & \multicolumn{3}{c}{\textbf{Pooled subgroups}}
 & \multicolumn{2}{c}{\textbf{Paired, \cedar{} vs Vanilla}} \\
\cmidrule(lr){2-4}\cmidrule(lr){5-6}
\textbf{Metric} & \textbf{Single LLM} & \textbf{\cedar{}} & \textbf{$p$}
& \textbf{\cedar{} ahead} & \textbf{$p$} \\
\midrule
ROC-AUC            & 0.782 & 0.793 & 0.264 & 7/11 & 0.175 \\
MCC                & 0.275 & 0.285 & 0.759 & 7/11 & 0.695 \\
Balanced accuracy  & 0.642 & 0.640 & 0.944 & 4/11 & 0.922 \\
$F_1$ (positive)   & 0.485 & 0.486 & 0.953 & 4/11 & 0.922 \\
\bottomrule
\end{tabular}
\end{supptable}

\begin{supptable}[H]
\centering
\caption{\textbf{TRIPOD+AI reporting checklist.} Each item of the TRIPOD+AI
statement~\citep{collins2024tripodai} with the place in the main text (Methods
sections are numbered) or Supplementary Information where it is reported. Items the
study does not address are stated as such.}
\label{stab:tripod_ai}
\scriptsize
\renewcommand{\arraystretch}{1.05}
\begin{tabular}{@{}p{0.06\linewidth}p{0.33\linewidth}p{0.55\linewidth}@{}}
\toprule
\textbf{Item} & \textbf{Topic} & \textbf{Where reported} \\
\midrule
1 & Title & Title page \\
2 & Abstract & Abstract \\
3a & Background and rationale & Introduction, paragraphs 1--5 \\
3b & Target population, intended use and users & Introduction; Discussion, paragraph 1 \\
3c & Known health inequalities & Introduction, paragraph 1 \\
4 & Objectives & Introduction, the four questions and final paragraph \\
5a & Data sources & Methods, Section~1; external data, Supplementary Note~3 \\
5b & Dates of data collection & Diagnosis years 2011--2025; LLM access dates, Methods, Section~5 \\
6a & Setting and centres & Methods, Section~1 (single centre) \\
6b & Eligibility criteria & Methods, Section~1 \\
6c & Treatments & Methods, Sections~2 and~3 (medications as features and in outcome definitions) \\
7 & Data preparation & Methods, Sections~1--2; Supplementary Tables~\ref{stab:feature_summary} and~\ref{stab:feature_availability} \\
8a & Outcome definition and horizon & Methods, Section~3; Figure~1 \\
8b & Outcome assessors & Not applicable: outcomes are rule-based on laboratory values, codes and medications \\
8c & Blinding of outcome assessment & Not applicable (rule-based outcomes); clinician review blinding, Methods, Section~10 \\
9a & Choice of predictors & Methods, Section~2; Supplementary Table~\ref{stab:feature_summary} \\
9b & Predictor definitions & Methods, Section~2; Supplementary Tables~\ref{stab:feature_summary} and~\ref{stab:collinear} \\
9c & Subjectively assessed predictors & Methods, Section~2 (LLM extraction of social determinants with expert review) \\
10 & Sample size & Methods, Section~10 (all eligible patients; adaptive held-out test set; no formal calculation) \\
11 & Missing data & Methods, Sections~3 and~4 \\
12a & Data partitioning & Methods, opening paragraph and Section~10 \\
12b & Predictor handling & Methods, Section~4 \\
12c & Model type, building, hyperparameters & Methods, Sections~4--6 (fixed hyperparameters); Supplementary Note~1 \\
12d & Heterogeneity across clusters & Not applicable (single centre) \\
12e & Performance measures & Methods, Section~10; Supplementary Note~1; Supplementary Figure~\ref{sfig:calibration} \\
12f & Model updating & Supplementary Note~2 (recalibration) \\
12g & How predictions are calculated & Methods, Sections~5--6; Code availability \\
13 & Class imbalance & Methods, Section~4; Supplementary Note~1 \\
14 & Fairness & Not assessed \\
15 & Model output and thresholds & Methods, Section~6; Supplementary Note~1 \\
16 & Development versus evaluation data & Methods, opening paragraph and Section~9 \\
17 & Ethical approval and consent & Methods, Section~1 \\
18a & Funding & Acknowledgments \\
18b & Conflicts of interest & Competing interests \\
18c & Protocol & No protocol was prepared \\
18d & Registration & Not registered \\
18e & Data sharing & Data availability \\
18f & Code sharing & Code availability \\
19 & Patient and public involvement & None \\
20a & Participant flow & Figure~1; Supplementary Table~\ref{stab:outcome_prevalence} \\
20b & Participant characteristics & Supplementary Tables~\ref{stab:cohort_composition} and~\ref{stab:feature_availability} \\
20c & Development versus evaluation distributions & Supplementary Note~1 (temporal split) \\
21 & Participants and events per analysis & Figure~1; Supplementary Table~\ref{stab:sample_sizes} \\
22 & Full model specification & Methods, Sections~4--6; Supplementary Table~\ref{stab:s0_snippets}; Code availability \\
23a & Performance with confidence intervals & Figure~2; Supplementary Tables~\ref{stab:pooled_stats} and~\ref{stab:cedar_vs_best_ml}; subgroups not reported \\
23b & Heterogeneity across clusters & Not applicable (single centre) \\
24 & Model updating results & Supplementary Note~2; Supplementary Tables~\ref{stab:refinement} and~\ref{stab:refinement2} \\
25 & Interpretation & Discussion \\
26 & Limitations & Discussion; Supplementary Note~1 \\
27a & Poor-quality or missing inputs & Results (information ablation); Discussion, paragraph 2 \\
27b & User interaction and expertise & Results (clinician review); Methods, Section~10 \\
27c & Next steps & Discussion; Supplementary Note~1 \\
\bottomrule
\end{tabular}
\end{supptable}

\section*{Supplementary Figures}

The supplementary figures extend the main analyses along the family/model
axis: the T1D-cohort counterpart of the faithfulness figure, the standardized
modality ablation across model families (Supplementary Figure~\ref{sfig:regime})
and its per-model breakdown, and a
temporal (train-on-past, validate-on-recent) robustness check compared against
the original analysis across all methods. Descriptive clinical and
epidemiological analyses of this cohort (longitudinal profiles, trajectory
phenotypes, cumulative incidence, and univariate risk factors) are reported in a
companion study of the same dataset (in preparation) and are not reproduced here.

\begin{suppfigure}[H]
\centering
\includegraphics[width=0.75\linewidth]{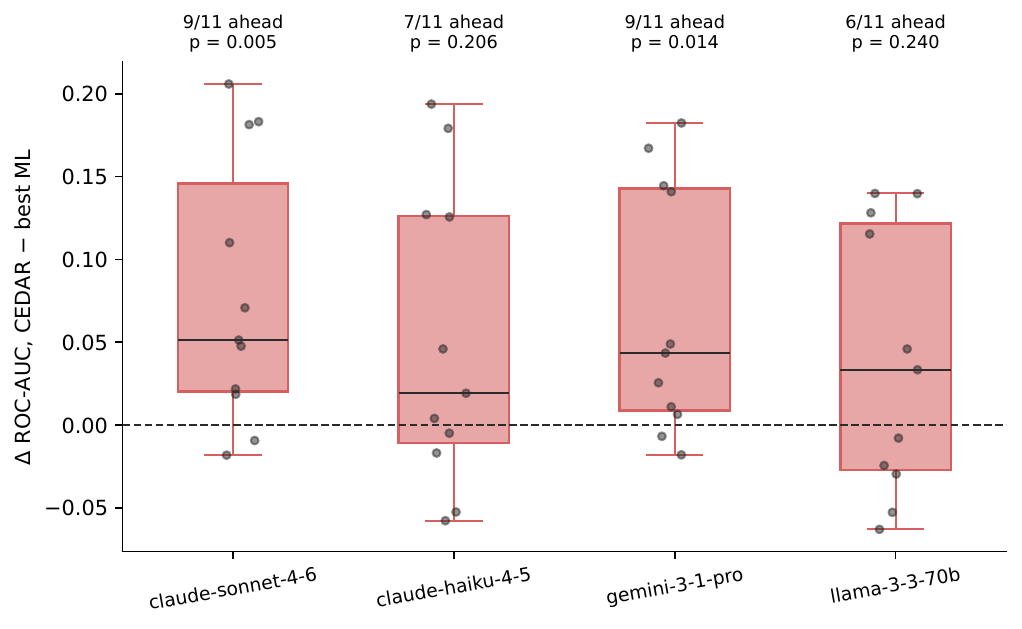}
\caption{\textbf{\cedar{} against the best statistical model on four base models}
(Supplementary Note~4). For each base model, the ROC-AUC of the full \cedar{}
pipeline minus that of the best classical or temporal model on the same
cohort--outcome pair, one point per pair (11 per base model). The statistical
models do not depend on the base model, so one best-ML value per pair serves every
base model. Annotations give the number of pairs on which \cedar{} is ahead and
the uncorrected Wilcoxon signed-rank $p$; after Holm--Bonferroni correction across
the four base models the lead remains significant on \texttt{claude-sonnet-4-6} and
\texttt{gemini-3-1-pro}.}
\label{sfig:backbone_vs_ml}
\end{suppfigure}

\begin{suppfigure}[H]
\centering
\includegraphics[width=\linewidth]{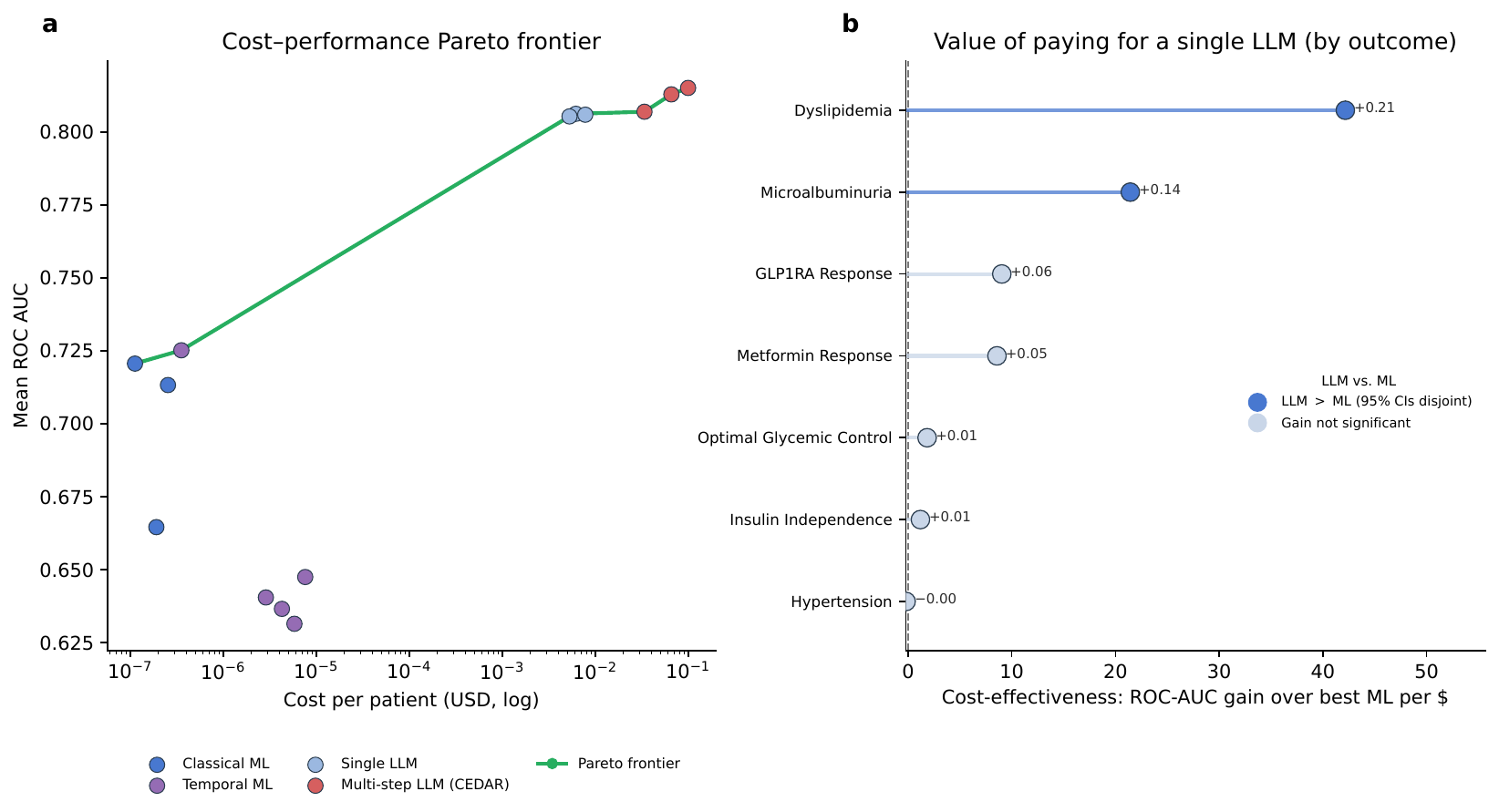}
\caption{\textbf{Clinical economics: what paying for an LLM buys (T2D).} These are
cost \emph{estimates} rather than measured values (Methods, Section~8), which is
why the analysis appears here rather than in the main text. They are list prices
per token and per CPU-second at one point in time and exclude engineering, review
and infrastructure effort. Actual spend depends on the provider, the contract, the
model tier and any caching or batching applied, none of which are properties of
the method, so these panels separate the tiers by order of magnitude and should
not be read as a budget. The quantity that transfers between sites is the number
of model calls per patient: one for a single LLM, five for \cedar{}.
(a)~Cost--performance Pareto frontier (log cost per patient against mean
ROC-AUC): classical ML anchors the cheap end, and both LLM tiers reach the top
accuracy band several orders of magnitude above it, with \cedar{} above a single LLM at the same accuracy, its extra calls buying auditability rather than discrimination. (b)~Per-outcome cost-effectiveness of paying for a single LLM. The comparison is deliberately family-level, a single LLM against the best classical or temporal ML model, rather than a search for the single best model, because the
benchmark of Supplementary Note~4 finds that the inexpensive single LLM already
captures most of the LLM accuracy advantage (Supplementary
Figure~\ref{sfig:llm_benchmark}a). For each outcome the ratio is the ROC-AUC gain of
the single LLM ($\sim$\$0.006/patient) over the best ML model,
divided by the single-LLM cost per patient, i.e.\ ROC-AUC points gained per
dollar. Outcomes are sorted best to worst, each stem is annotated with the
underlying ROC-AUC gain, and color indicates whether that gain is statistically
distinguishable from ML (non-overlapping 95\% confidence intervals, solid) or
numerical only (pale). We report the raw gain-per-dollar ratio rather than
applying a fixed ``worth-it'' AUC cut, which would be arbitrary. This panel
differs from Supplementary Figure~\ref{sfig:llm_benchmark}e, which compares best-ML and best-LLM AUC per outcome
without a cost axis.}
\label{sfig:economics}
\end{suppfigure}

\begin{suppfigure}[H]
\centering
\includegraphics[width=\linewidth]{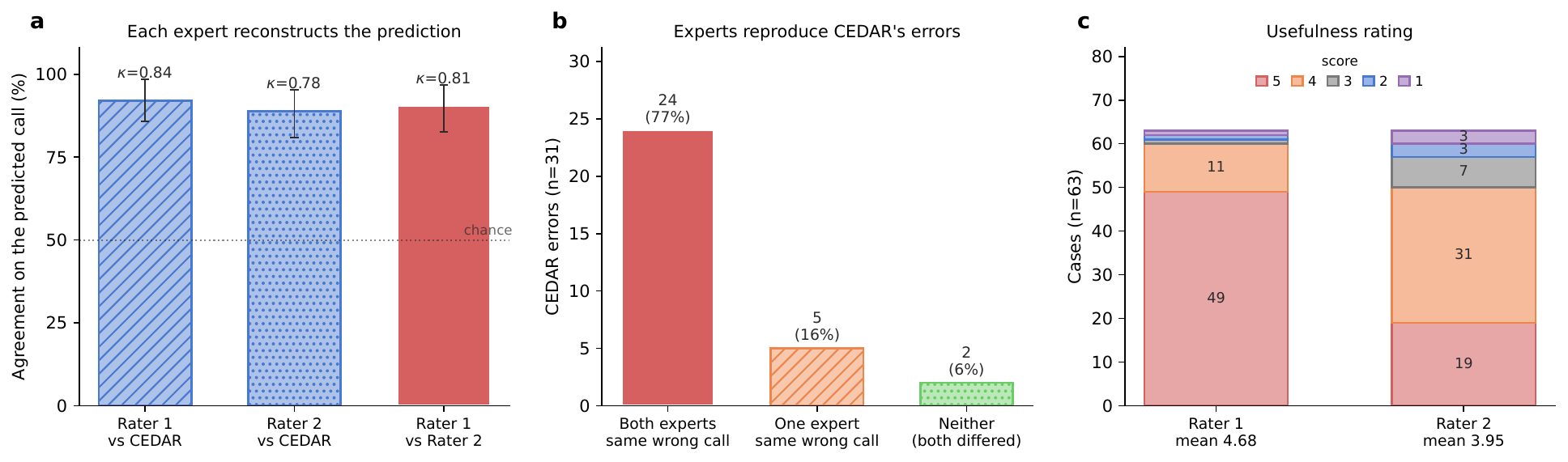}
\caption{\textbf{Two-clinician review of \cedar{}'s reasoning chains}
(supporting the clinician review reported in the Results). Two board-certified
pediatric endocrinologists independently reviewed the same 63 T2D patient-outcome
cases, sampled across the four confusion cells, each shown as \cedar{}'s
citation table and reasoning chain with the predicted probability withheld and
with no indication of the recorded outcome or of whether \cedar{} was correct.
(a)~Agreement on the predicted call: each rater against \cedar{}, and the two
raters against each other, with Cohen's $\kappa$ above each bar and
2{,}000-iteration bootstrap 95\% confidence intervals; the dotted line marks
chance. An expert reading only the cited evidence arrives at \cedar{}'s
conclusion about nine times in ten, and the two raters agree with each other
about as often as either agrees with the model. (b)~The 31 cases \cedar{} scored
incorrectly, split by how many raters independently reached the same incorrect
call. Only 2 of 31 errors were avoided by both raters, so \cedar{}'s residual
errors sit where two clinicians reading the same record also fail.
(c)~Distribution of the 1--5 clinical-usefulness rating for each rater, with
means in the axis labels. Ratings are high for both raters and, consistent
with~(b), do not mark out the cases \cedar{} got wrong ($p=0.58$ and $p=0.62$,
Mann--Whitney). Because the 63 cases are a stratified draw across the confusion
cells, this subset supports agreement and usefulness comparisons but not an
accuracy estimate for either the raters or \cedar{} (Methods, Section~10).}
\label{sfig:annotation}
\end{suppfigure}

\begin{suppfigure}[H]
\centering
\includegraphics[width=\linewidth]{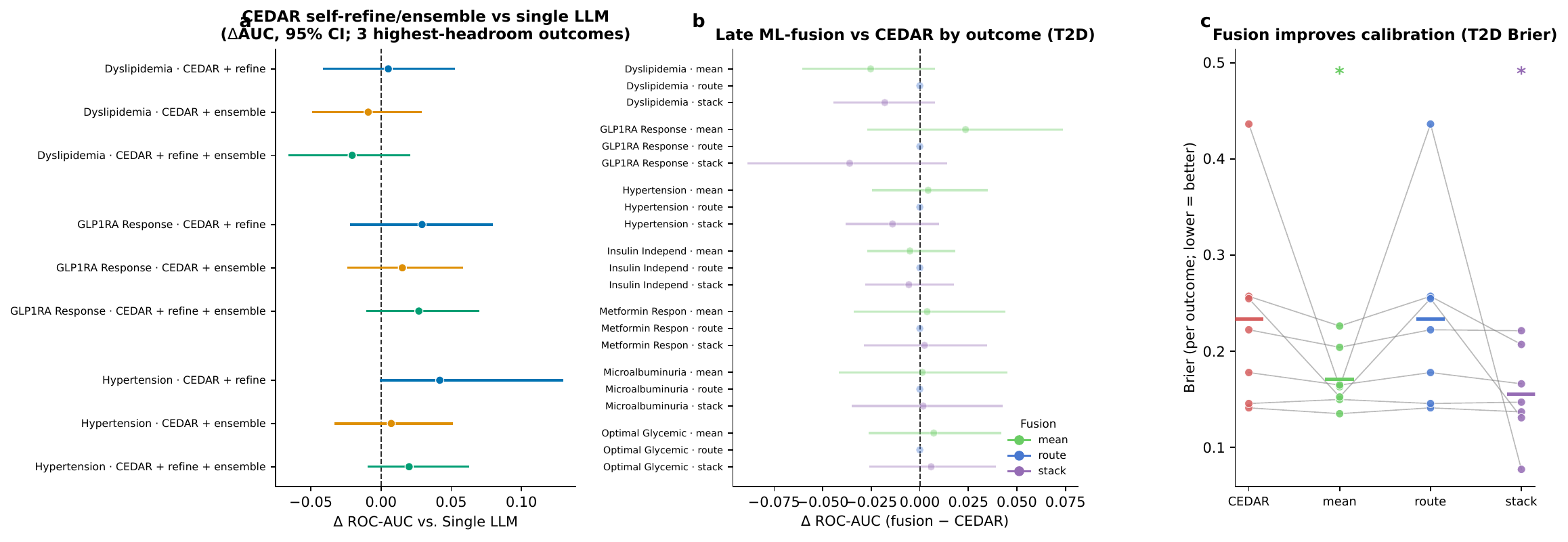}
\caption{\textbf{Additional \cedar{}-enhancement configurations tested, none of
which improves discrimination over a single LLM call (T2D).} The main text already
compares all model families, so the former pooled-bars panel is omitted here.
(a)~Pre-specified ``\cedar{}~v2'' screen (verifier-gated self-refinement, a
diverse-reasoner ensemble, and their combination) on the three highest-headroom T2D
outcomes (N=60): paired $\Delta$AUC vs.\ a single LLM. Every interval crosses zero.
The refinement step almost never triggered ($n_{\text{refine}}=0$) because
first-pass grounding is already $\geq$0.9999 (Figure~3), consistent with reports
that self-correction does not reliably help without external
feedback~\citep{huang2024selfcorrect, kamoi2024selfcorrect} and that multi-agent
debate/ensembling does not reliably beat simpler prompting~\citep{smit2024mad}.
(b)~Late ML-fusion: per-outcome paired $\Delta$AUC vs.\ \cedar{} for all three
pre-specified combiners (fixed mean, logistic stacking, and prevalence-aware
routing) across the seven T2D outcomes (bold = 95\%~CI excludes zero); wins and
losses cancel and no combiner separates from \cedar{} on discrimination.
(c)~Late fusion nonetheless improves \emph{calibration}: per-outcome Brier for
\cedar{} vs.\ each combiner (points = outcomes, bar = mean; gray lines pair the same
outcome across systems). Pooled over records the Brier score falls from 0.234 to
0.155 in T2D and from 0.217 to 0.115 in T1D. The mean and stacking combiners lower Brier on 6 of 7
outcomes (Wilcoxon signed-rank vs.\ \cedar{} across the 7 outcomes: mean $p=0.031$,
stack $p=0.047$; marked~*), by recalibrating the LLM's overconfident probabilities
while preserving its ranking (hence no ROC-AUC change). Prevalence-aware routing sent
all seven T2D outcomes to the LLM and therefore simply recovers \cedar{} (no change,
no mark). Given only seven paired outcomes these calibration $p$-values are
suggestive rather than definitive.}
\label{sfig:omitted_configs}
\end{suppfigure}

\begin{suppfigure}[H]
\centering
\includegraphics[width=\linewidth]{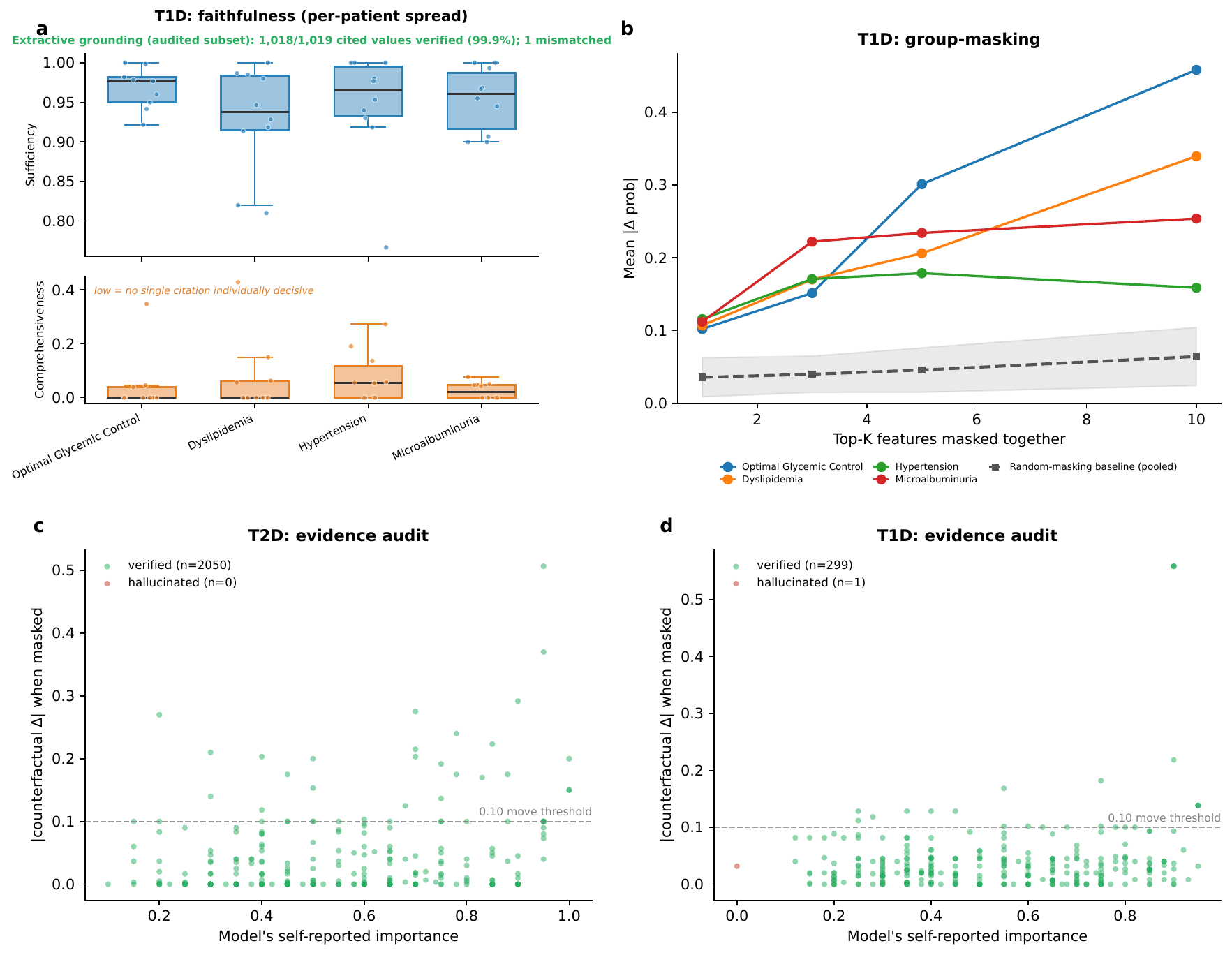}
\caption{\textbf{Faithfulness of \cedar{} reasoning: T1D cohort and per-citation
evidence audit for both cohorts} (companion to Figure~3).
(a)~T1D faithfulness per outcome across audited patients (box plots: sufficiency,
top; comprehensiveness, bottom; points are individual patients), with extractive
grounding reported as a count in the annotation, as in T2D (a near-ceiling fabrication check). (b)~T1D group-masking curve (mean $|\Delta p|$ vs.\ top-$K$ cited features masked), which rises as in T2D far above the size-matched random-masking baseline (dashed gray, pooled; $\approx$0.06 at
$K=10$). (c,~d)~Per-citation evidence audit for
T2D and T1D: each cited feature's self-reported importance versus its actual
counterfactual impact when masked, colored by verification status. Essentially all
citations match real record values: within this deeper audit every audited T2D
citation was verified and all but one in T1D, and the cohort-wide
grounding check found 4 mismatches in 47{,}803 T2D citations and none in 22{,}443
T1D citations (Results; dashed line, 0.10 move threshold). Consistent with the low
comprehensiveness, the median single-citation $|\Delta p|$ is small
($\approx$0.01). No one citation is individually decisive, but the cited set is
collectively relevant. \emph{Each panel plots a random subsample of that cohort's
audited citations for legibility, while the legend reports the totals.}}
\label{sfig:faithfulness}
\end{suppfigure}

\begin{suppfigure}[H]
\centering
\includegraphics[width=\linewidth]{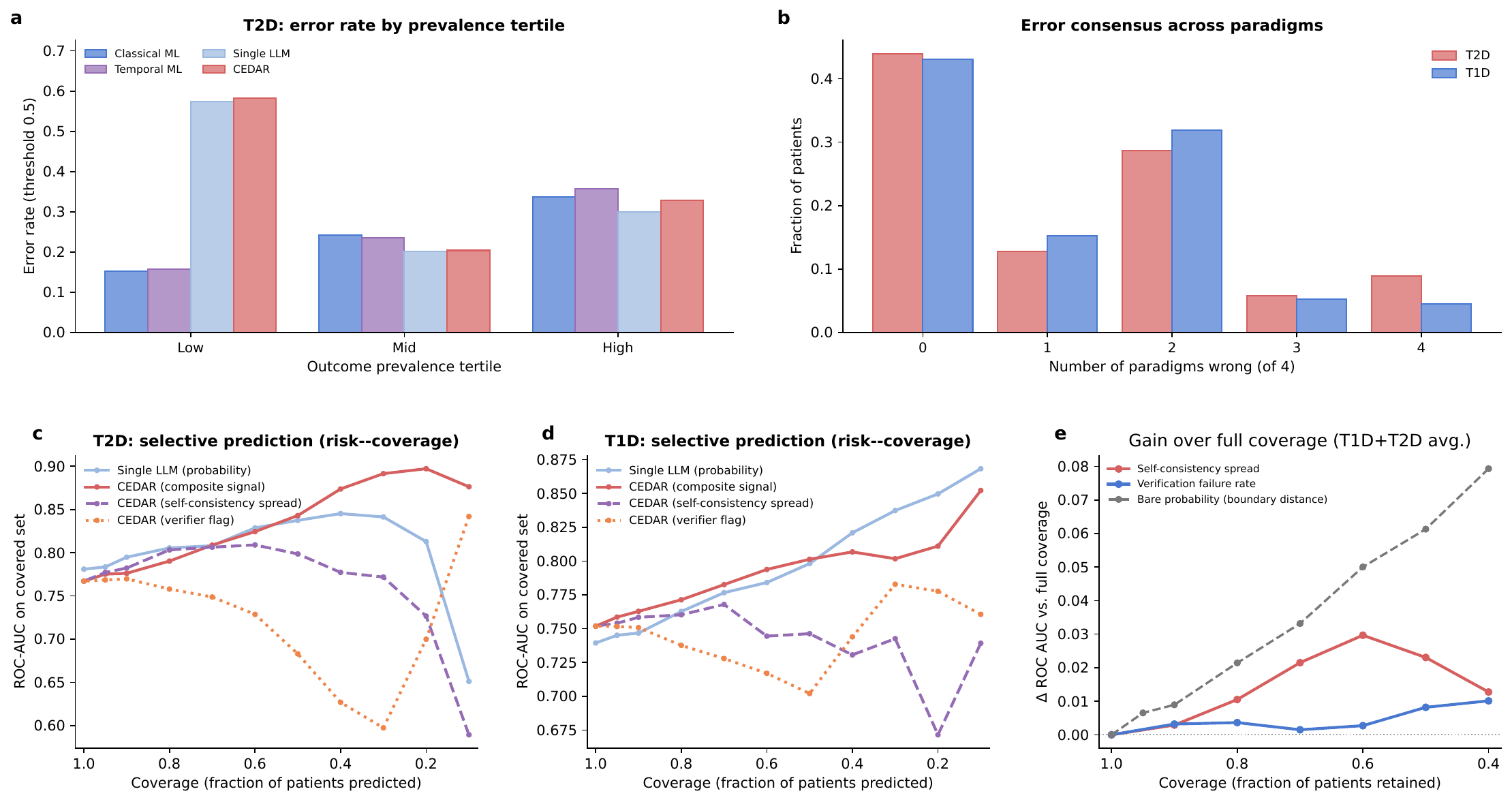}
\caption{\textbf{Error stratification and selective prediction} (supporting the
per-patient error analysis of Figure~4; the error-correlation
heatmap is in Figure~4a and not duplicated here).
(a)~Error rate of each family within outcome-prevalence tertiles (T2D). At a
0.5 threshold, classical and temporal ML have a lower error rate on
low-prevalence outcomes only because they safely predict the majority class,
a near-chance ranking, whereas the LLMs attempt the rare positive calls. This
majority-class effect, not an ML discrimination advantage, is the mechanism
behind the cross-family error decorrelation. (b)~Distribution of how many
of the four subgroups misclassify each patient: a large mass at 0 (all
correct) and a substantial ``exactly 2 wrong'' cluster, the two correlated
pairs failing as units.
(c,~d)~Selective prediction: risk--coverage curves (covered-set ROC-AUC
against the fraction of patients predicted, full coverage at left,
increasing abstention to the right), T2D and T1D. Four curves per cohort: a
single LLM's bare probability (its distance from 0.5), \cedar{}'s own bare
probability (``composite,'' which folds in the two signals below but is
dominated by the bare-probability term since grounding is near-perfect),
\cedar{}'s self-consistency spread, and \cedar{}'s verifier flag. Deferring
the least-confident patients raises covered-set AUC for the single LLM and for
\cedar{}'s bare probability, and the two track each other closely at every
coverage level. Self-consistency spread adds nothing beyond that shared
baseline in T2D and degrades it in T1D. The verifier flag is actively harmful as a confidence signal: in T2D, ranking on it alone lowers covered-set ROC-AUC by up to 0.14 at 40\% coverage, consistent with a signal that almost never
fires and is therefore mostly noise. Useful abstention is a generic benefit of
any usable probability rather than something \cedar{}'s verification or
self-consistency machinery adds. (e)~The same comparison as a gain over full
coverage, T1D and T2D averaged. Neither signal \cedar{} produces internally, the
spread of its $K=3$ self-consistency samples or the fraction of cited evidence its
verifier flags, improves on its own predicted probability read as a distance from
0.5 (gray dashed). Curves stop at 40\% coverage, below which the retained subset
is too small to be reliable.}
\label{sfig:error_analysis}
\end{suppfigure}

\begin{suppfigure}[H]
\centering
\includegraphics[width=\linewidth]{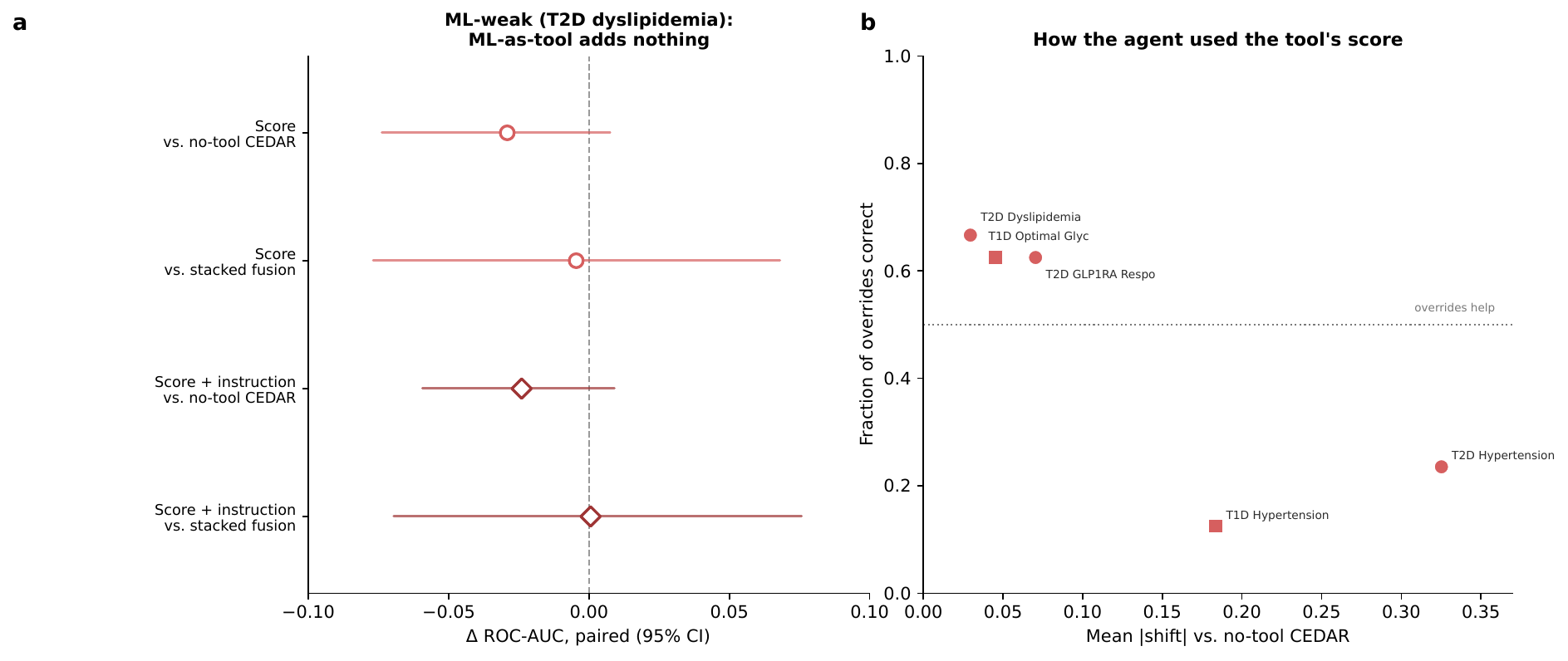}
\caption{\textbf{How the agent used the injected ML score (mechanism for the
targeted tool-use result in Figure~4b).} (a)~The counterpart of Figure~4b on an outcome on which ML performs
poorly. On T2D dyslipidemia, supplying the best-per-outcome ML model's frozen
score to \cedar{} as an in-context tool changes ROC-AUC significantly against
neither no-tool \cedar{} nor the stacked late-fusion combiner (paired bootstrap; a
filled marker would mean the 95\% confidence interval excludes zero). (b)~The
behavioral signal: the mean shift the injected score induced in \cedar{}'s output
relative to the no-tool run (x) versus the fraction of resulting overrides that were
correct (y), per outcome and cohort (circles T2D, squares T1D). The agent shifted
little and overrode correctly ($>$50\%) when the tool was informative (upper-left),
but shifted strongly and incorrectly on the noisiest, lowest-prevalence outcomes
(hypertension; lower-right). It therefore uses a good tool and largely resists a
bad one, though it can over-trust a noisy score. This is a hypothesis-generating result on a
single ML-competitive outcome, not a general accuracy claim; it motivates the future
direction of designing outcome- and modality-specific tools for the agent
(Discussion).}
\label{sfig:hybrid_b}
\end{suppfigure}

\begin{suppfigure}[H]
\centering
\includegraphics[width=0.62\linewidth]{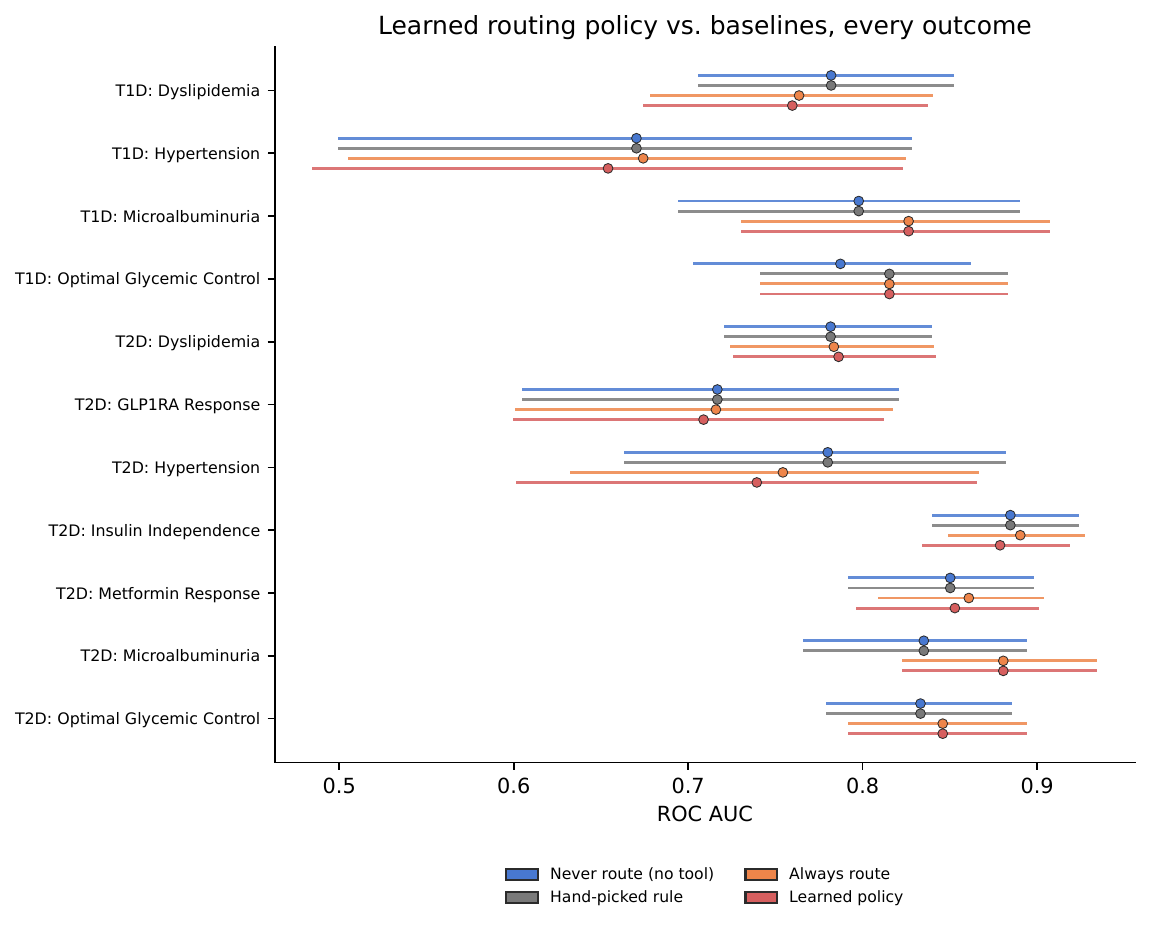}
\caption{\textbf{A learned routing policy, evaluated on every outcome}
(supporting Figure~4c). The hand-picked routing result of Figure~4b generalized
into a learned per-patient policy, gated on \cedar{}'s self-consistency spread or
the ML tool's own confidence exceeding a threshold fit on the training folds, and
evaluated on every outcome in both cohorts against always-route, never-route and
the original hand-picked rule. It tracks those baselines everywhere, including the
ten outcomes where the tool carries nothing, so the earlier gain is not an artifact
of outcome selection. Per-outcome intervals are wide, so the panel shows that the
policy does not hurt rather than carrying the tool-use effect itself.}
\label{sfig:routing_policy}
\end{suppfigure}

\begin{suppfigure}[H]
\centering
\includegraphics[width=\linewidth]{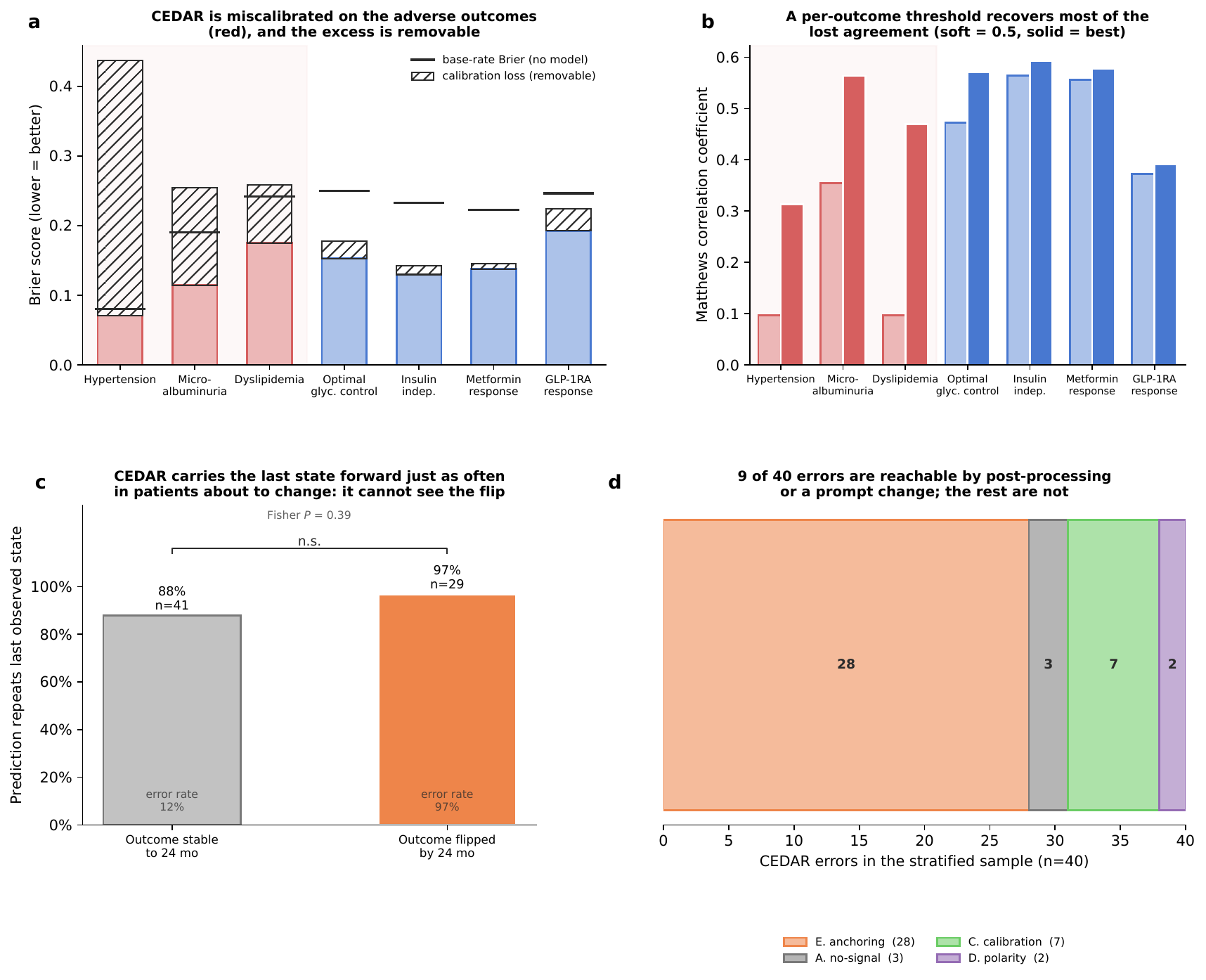}
\caption{\textbf{\cedar{} error analysis and refinement ceiling, T2D}
(Supplementary Note~2). \cedar{} (\texttt{model\_d\_full}) only, 24-month
horizon. Adverse outcomes are shaded in (a) and (b).
(a)~Brier score per outcome, decomposed on the reliability bins into a
refinement term (solid, the Brier score that would remain after a perfect
per-outcome recalibration) and a removable calibration loss (hatched). The horizontal rule on each bar marks the Brier score of an intercept-only model
that predicts the base rate. \cedar{} is worse than that no-model benchmark on
all three adverse outcomes and better on all four beneficial ones.
(b)~Matthews correlation coefficient at the 0.5 threshold used throughout the
paper (pale) and at the best threshold for that outcome (solid). Both the
recalibration in (a) and the threshold in (b) are fitted on the same held-out
bins they are evaluated on, so each is an optimistic ceiling rather than an
achievable gain, and neither operation can change ROC-AUC.
(c)~Rate at which \cedar{}'s predicted label simply repeats the last observed
outcome state, split by whether that state actually changed between the last
observed window (18~months) and the target (24~months). The two rates are
statistically indistinguishable, i.e.\ \cedar{} carries the state forward just
as readily in the patients who are about to change: it cannot identify them in
advance. The error rates printed inside the bars follow from this and are
partly true by construction (an anchored prediction on a flipping patient is an
error by definition), which is why the test shown is the anchoring rate rather
than the error rate.
(d)~The 40 misclassified cases in the stratified deck by primary error code,
mechanism-first (Supplementary Table~\ref{stab:error_taxonomy}); anchoring is
the largest class under either priority order.}
\label{sfig:cedar_refinement}
\end{suppfigure}

\begin{suppfigure}[H]
\centering
\includegraphics[width=\linewidth]{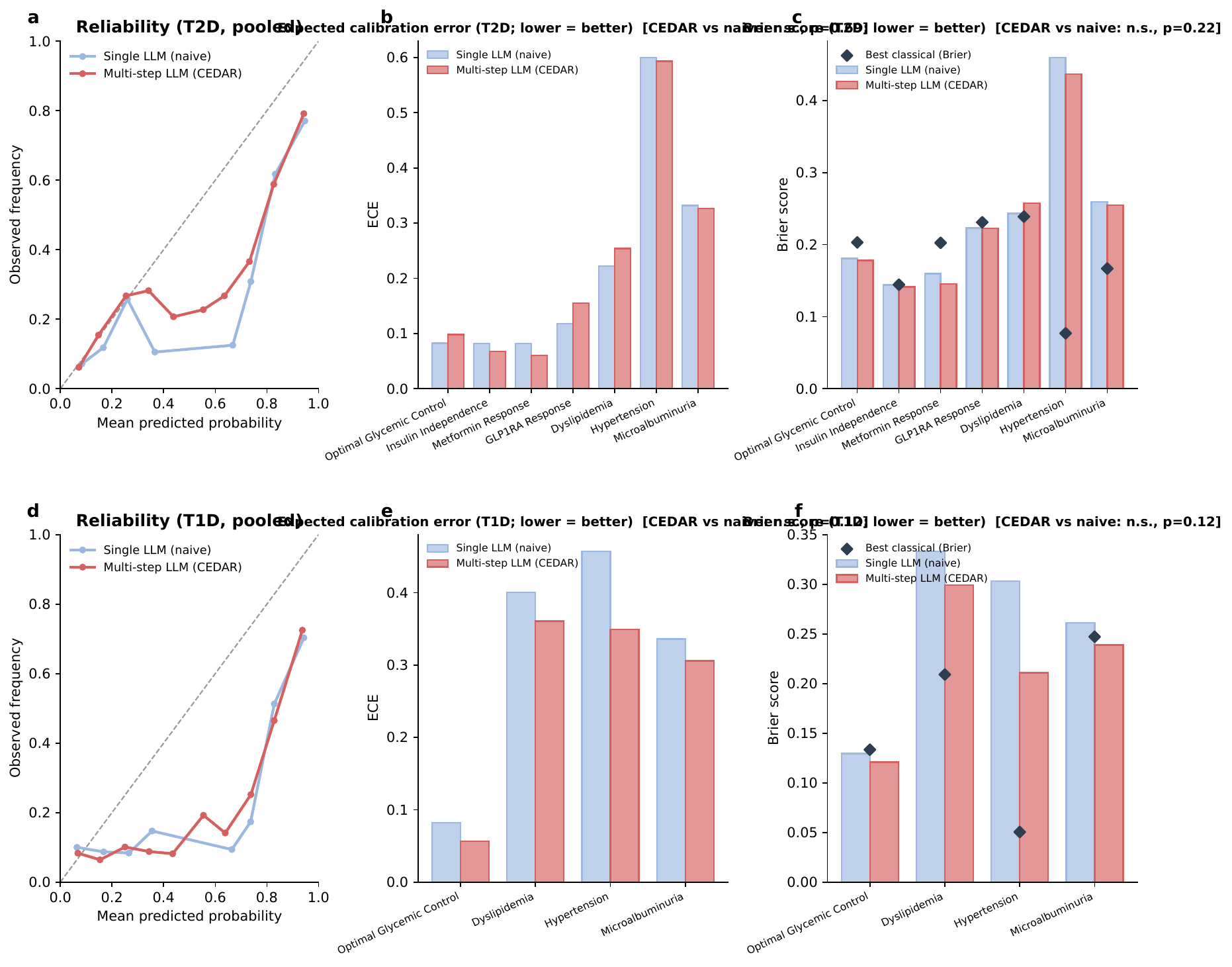}
\caption{\textbf{Calibration of \cedar{} vs.\ the naive single LLM.}
Reliability curves (pooled over outcomes, count-weighted), expected calibration
error (ECE), and Brier score for the naive single LLM and \cedar{}, with the best classical model's Brier shown as a reference diamond, for (a--c)~T2D and (d--f)~T1D. All are computed on the shared held-out test set.
Self-consistency does not produce a statistically significant calibration
improvement in either cohort. In T2D, \cedar{} and the naive LLM are comparably
(mis)calibrated (mean ECE 0.222 vs.\ 0.217, Wilcoxon signed-rank $p=0.69$; mean
Brier 0.234 vs.\ 0.238, $p=0.22$). In T1D, \cedar{} is numerically better on both
metrics (mean ECE 0.268 vs.\ 0.319; mean Brier 0.217 vs.\ 0.257), but with only
four evaluable outcomes the paired test cannot reach significance ($p=0.125$ for
both, the smallest attainable value at $n=4$), so we do \emph{not} claim a
calibration advantage. We therefore treat self-consistency primarily as variance
reduction (main text) and report calibration transparently rather than as a
headline claim.}
\label{sfig:calibration}
\end{suppfigure}

\begin{suppfigure}[H]
\centering
\includegraphics[width=\linewidth]{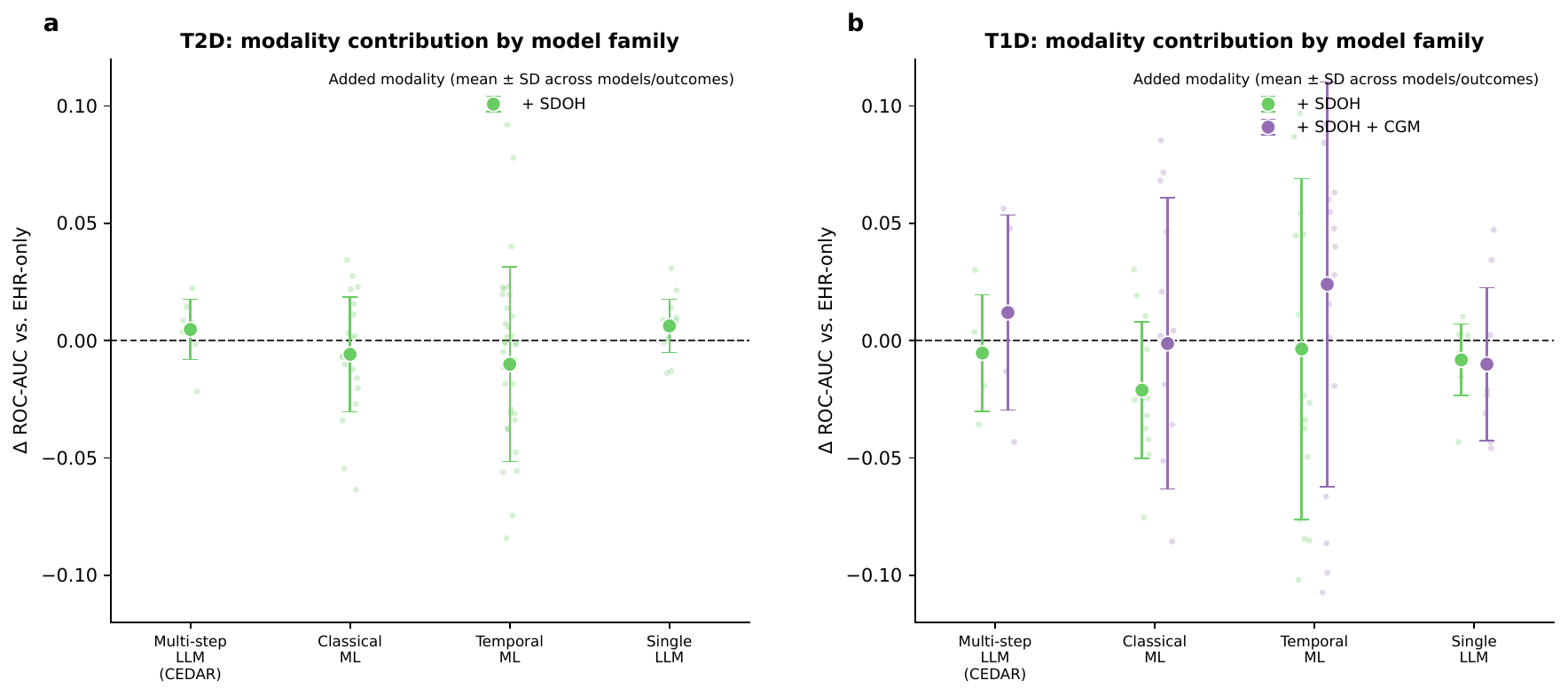}
\caption{\textbf{Modality contribution across all model families.}
Standardized ablation on a fixed held-out set (N=100 patients per outcome),
run identically for every family, classical ML, temporal ML, single LLM, and
\cedar{}. Bars show the change in mean ROC-AUC (averaged over models and outcomes
within each family) from adding a modality on top of EHR-only. (a)~T2D: adding
SDOH changes discrimination by essentially zero in every family, including
\cedar{}. (b)~T1D: adding SDOH is again near-zero, while adding CGM gives a small
positive contribution for the classical, temporal, and \cedar{} families (driven
by optimal glycemic control) but not the single LLM. No modality changes a
family's mean ROC-AUC by more than 0.03; the largest single effect is a 0.021
decrease from adding SDOH to classical ML in T1D. A per-model breakdown is given in
Supplementary Figure~\ref{sfig:modality}.}
\label{sfig:regime}
\end{suppfigure}

\begin{suppfigure}[H]
\centering
\includegraphics[width=\linewidth]{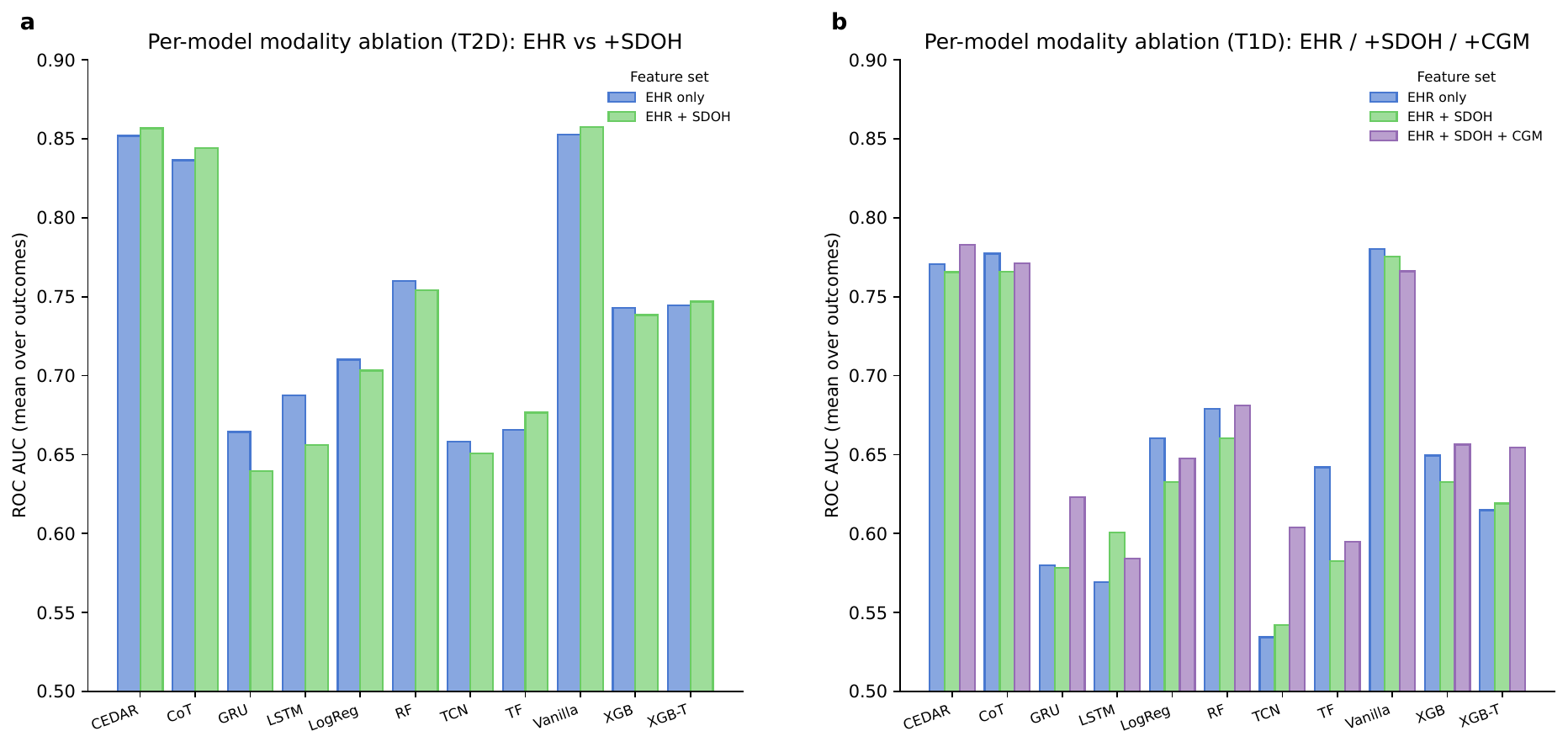}
\caption{\textbf{Per-model modality ablation (all families).} The
by-family ablation in Supplementary Figure~\ref{sfig:regime}, broken out for each individual
model (mean ROC-AUC over outcomes per model $\times$ feature set), confirming the
pattern is consistent across every estimator, including the LLMs (Vanilla, CoT,
\cedar{}), and not an averaging artefact. Models span classical ML (LogReg, RF,
XGB), temporal ML (XGB-T, LSTM, GRU, TCN, TF), the single LLM (Vanilla, CoT), and
\cedar{}. (a)~T2D: EHR-only vs.\ EHR+SDOH; (b)~T1D: EHR-only vs.\ EHR+SDOH vs.\
EHR+SDOH+CGM, on the fixed N=100 held-out set. Adding SDOH leaves discrimination
essentially unchanged for every model, while adding CGM gives most T1D models
(including \cedar{}) a small lift driven by glycemic control. Dashed line
= chance.}
\label{sfig:modality}
\end{suppfigure}

\begin{suppfigure}[H]
\centering
\includegraphics[width=\linewidth]{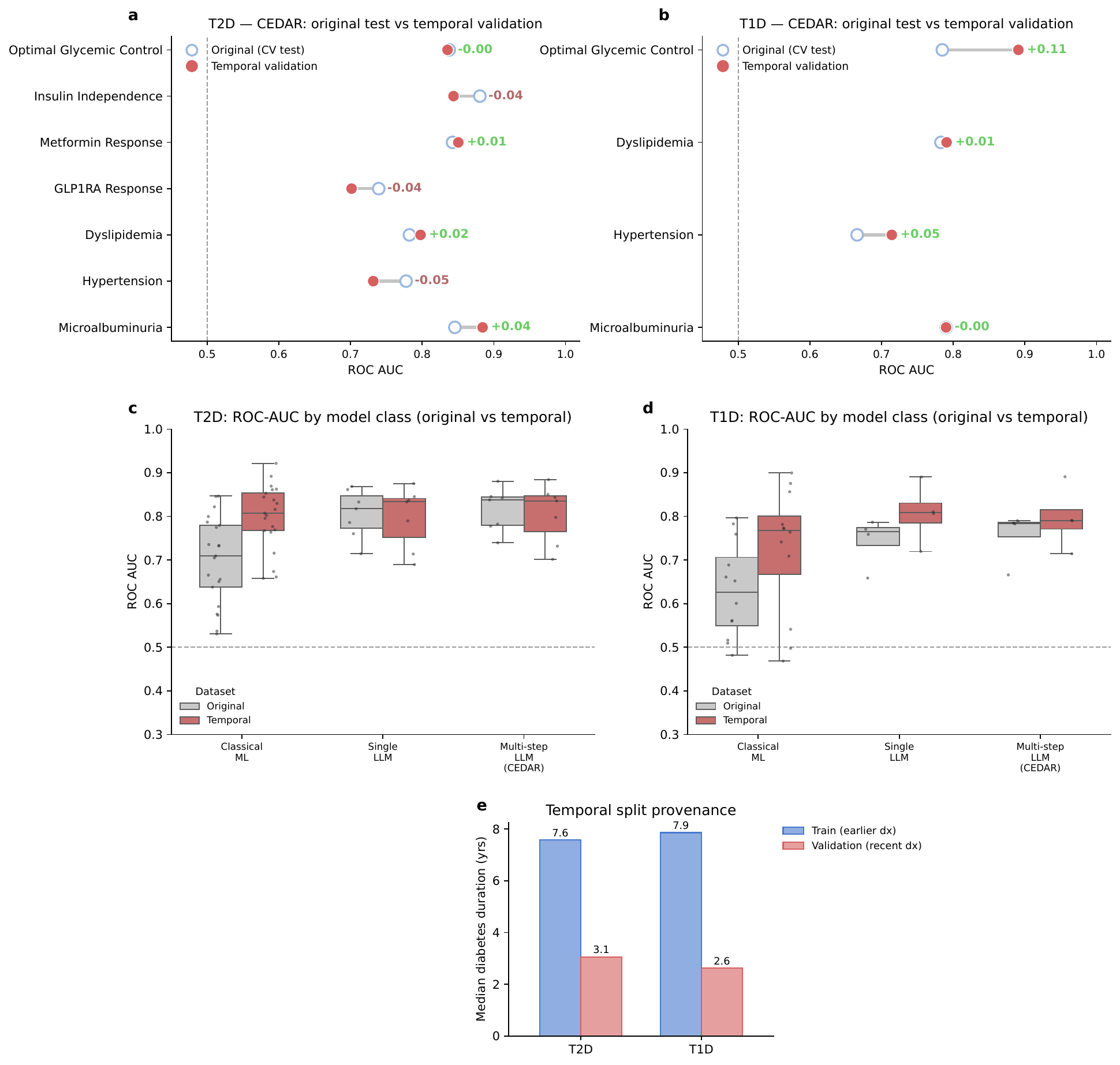}
\caption{\textbf{Temporal (prospective-style) validation: original analysis vs.\
temporal split.} Because no separate validation cohort was available, patients
were split by diagnosis recency: models were trained on the earlier-diagnosed
cohort and evaluated on the more-recently-diagnosed cohort (``train on the past,
validate on the future''). (a,~b)~\cedar{}'s original held-out test ROC-AUC (open blue) vs.\ its temporal-validation ROC-AUC (filled red) per outcome, with the change annotated (green = holds/improves, red = drops), for (a)~T2D and (b)~T1D. Discrimination is preserved under the temporal split: most outcomes move by less than 0.05 ROC-AUC and several improve, e.g.\ T1D optimal glycemic control by $+0.11$. (c,~d)~ROC-AUC distributions by model family in the original vs.\ temporal analysis, for (c)~T2D and (d)~T1D, showing that the agreement holds \emph{per family} and not just for \cedar{}. Each box pools one ROC-AUC per model and outcome within a family, and the original analysis is restricted to the models present in the temporal run so that the comparison is like for like.
Every family, classical ML, single LLM, and the multi-step LLM
(\cedar{}), occupies a similar ROC-AUC range under both datasets, with \cedar{}
the tightest and most stable, indicating the family ordering reproduces across
the temporal shift. (Temporal-ML and multi-agent models were not included in the
temporal-validation export and are therefore absent from (c,~d).) (e)~Split
provenance: median diabetes duration of the train (earlier-diagnosed) vs.\
validation (recent-diagnosed) subsets for each cohort, confirming a genuine
temporal separation.}
\label{sfig:temporal_val}
\end{suppfigure}

\begin{suppfigure}[H]
\centering
\includegraphics[width=\linewidth]{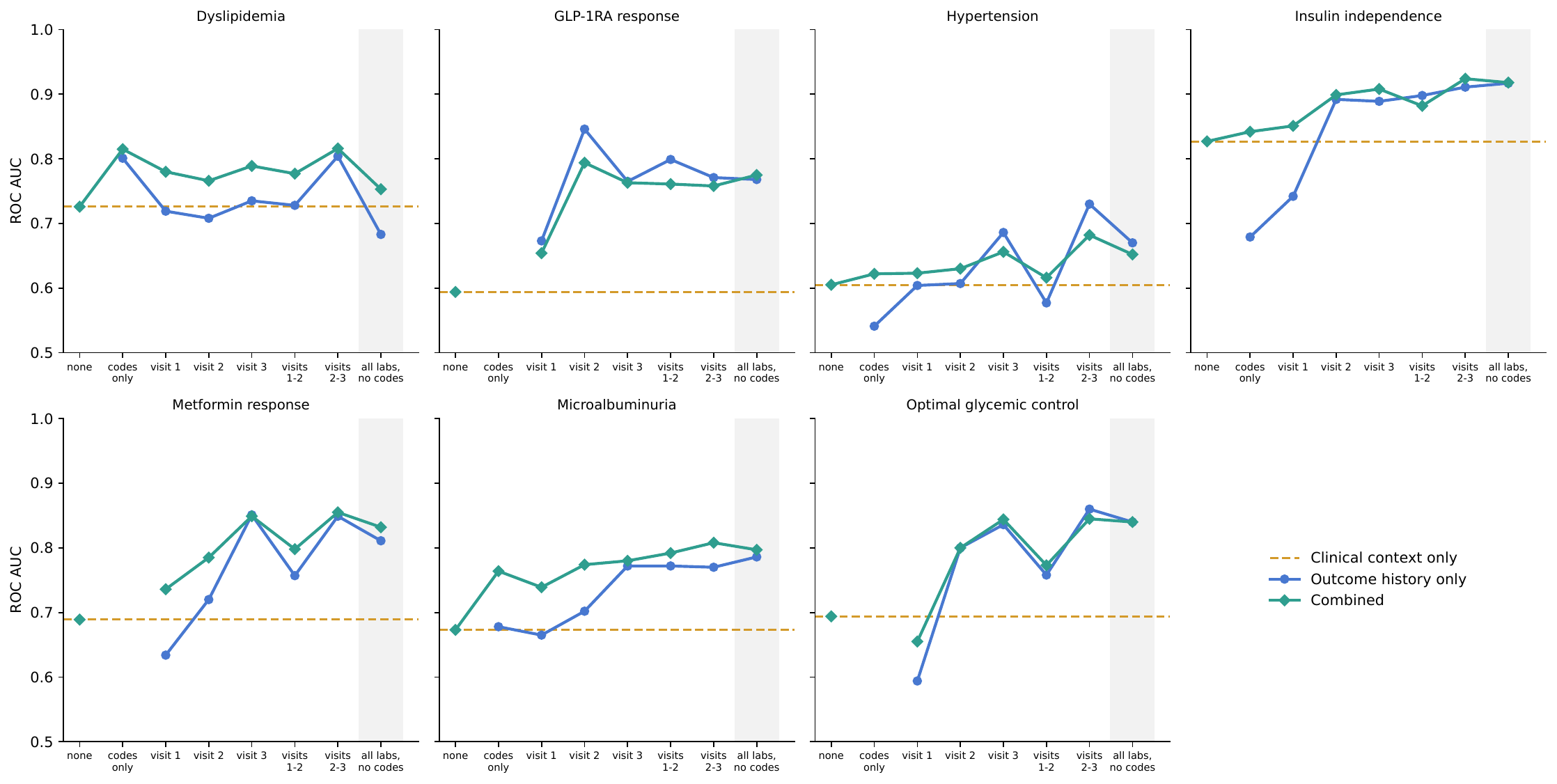}
\caption{\textbf{Information ablation for all seven outcomes} (supporting
Figure~5e). ROC-AUC by condition for the history-only reader and the reader that
sees both channels, with the context-only level dashed, 100 patients per outcome.
Conditions run from no history to the full laboratory history (shaded), which is not
nested with the visit-restricted conditions. Estimates and confidence intervals are
in Supplementary Table~\ref{stab:ablation}, and the exceptions and limits are in
Supplementary Note~5.}
\label{sfig:ablation_all}
\end{suppfigure}

\begin{suppfigure}[H]
\centering
\includegraphics[width=\linewidth]{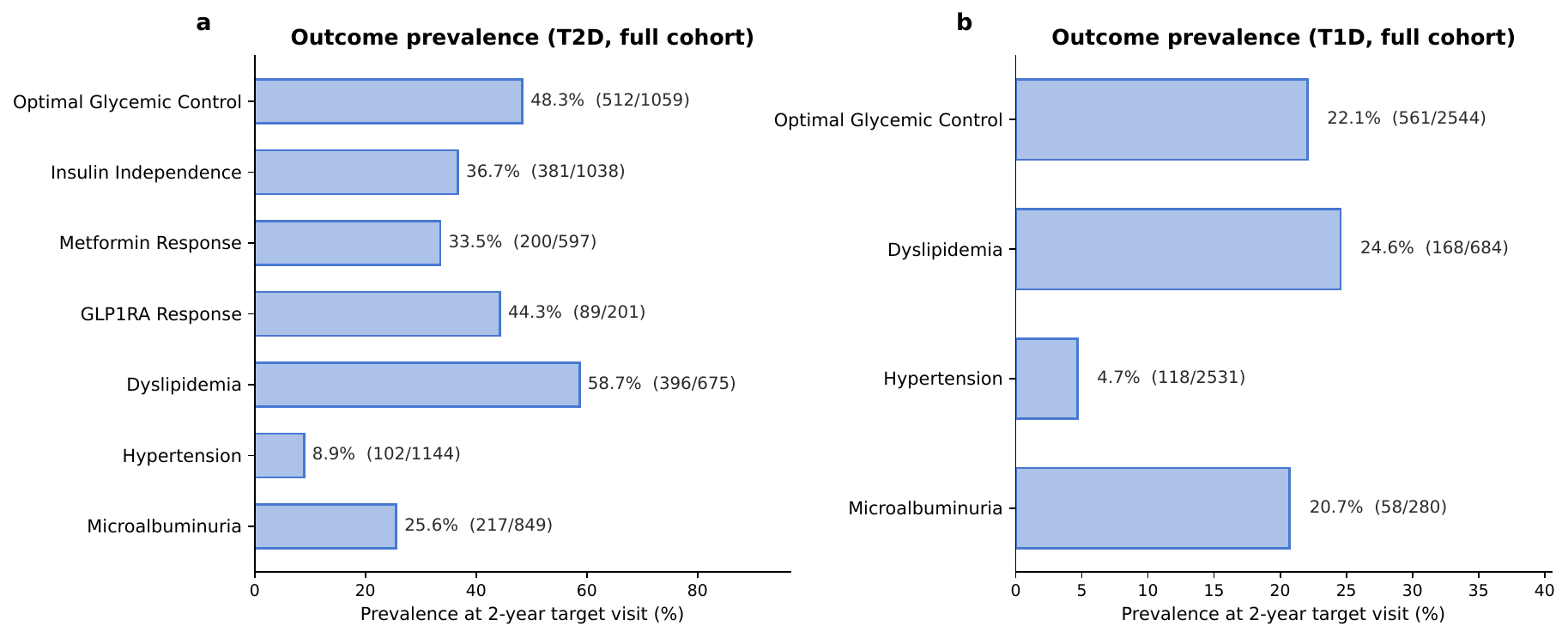}
\caption{\textbf{Outcome prevalence and assessable sample size, by cohort.}
Prevalence at the 2-year target visit for each outcome, annotated with
event/assessable counts, for (a)~T2D and (b)~T1D. The two cohorts occupy different regimes. T1D is evaluated on four outcomes (insulin independence, metformin response, and GLP-1RA response are not clinically applicable in T1D) and generally at lower prevalence (e.g.\ optimal glycemic control 22\% vs.\ 48\%, hypertension 5\% vs.\ 9\%), while assessable $N$ varies substantially across outcomes (e.g.\ T1D microalbuminuria $N=280$). These differences, not a change
in method, underlie the different per-outcome numbers between cohorts.}
\label{sfig:prevalence}
\end{suppfigure}

\begin{suppfigure}[H]
\centering
\includegraphics[width=\linewidth]{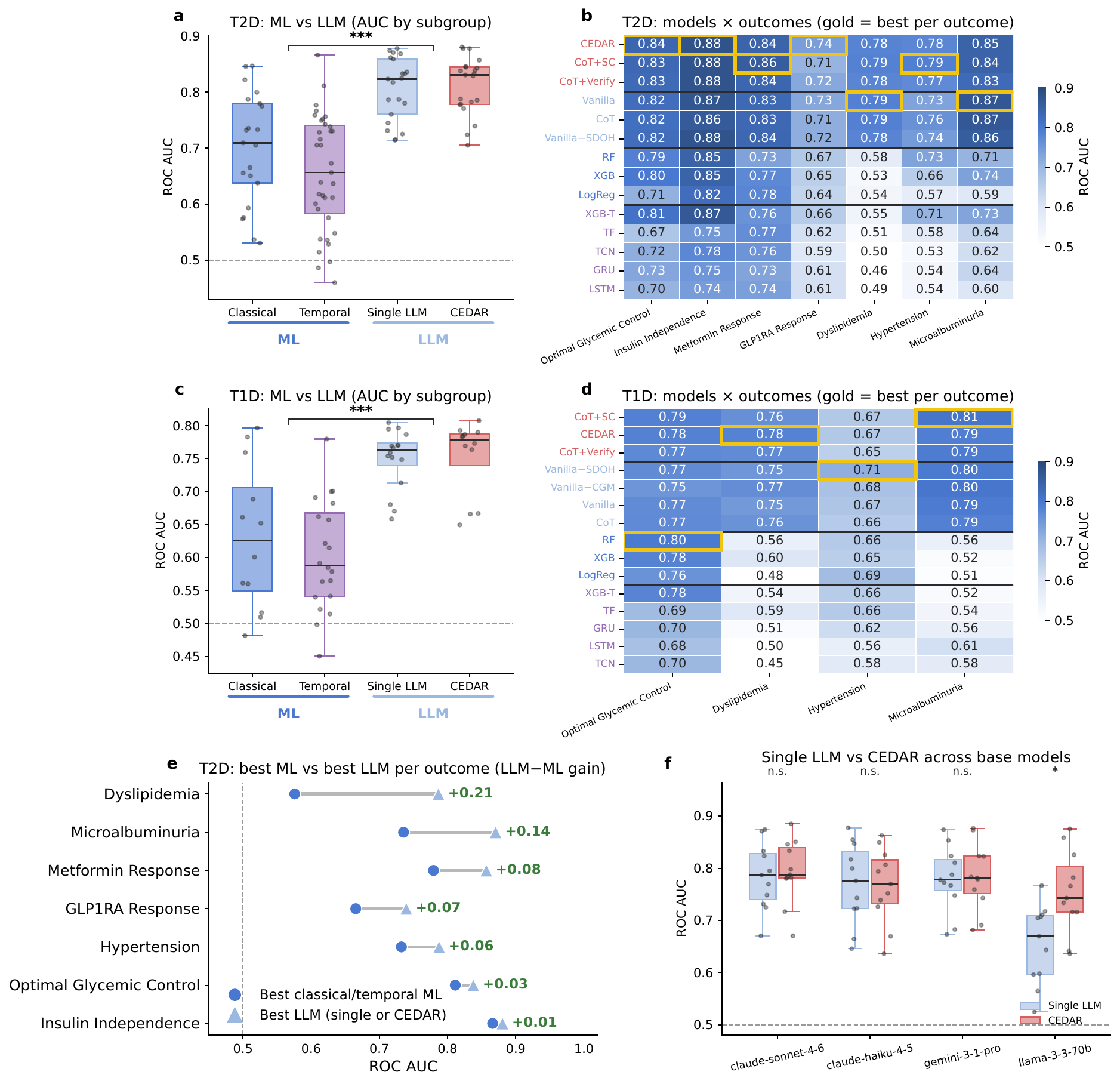}
\caption{\textbf{Full benchmark including naive single-call LLM implementations}
(Supplementary Note~4). Models are grouped into two families, ML and LLM, each split
into two subgroups: classical and temporal ML, and the single LLM (Vanilla, CoT and
modality-ablated Vanilla) and \cedar{} (the full pipeline with its CoT+Verify and
CoT+SC ablations). \textbf{(a,~b)~T2D; (c,~d)~T1D.} (a,~c)~ROC-AUC by subgroup with
family brackets (Mann--Whitney~$U$, $^{***}p<0.001$). Within the LLM family the
single LLM and \cedar{} are statistically indistinguishable, $p=0.71$ in T2D and
$p=0.56$ in T1D. (b,~d)~Model~$\times$~outcome ROC-AUC matrix with the best model
per outcome outlined in gold. (e)~Best ML model versus best LLM configuration per
T2D outcome. (f)~Single LLM versus \cedar{} on four base models, pooling the 11
cohort--outcome pairs, paired Wilcoxon signed-rank test per base model,
Holm--Bonferroni corrected. The equivalence holds on the three commercial base
models and not on \texttt{llama-3-3-70b}.}
\label{sfig:llm_benchmark}
\end{suppfigure}

\begin{suppfigure}[H]
\centering
\includegraphics[width=\linewidth]{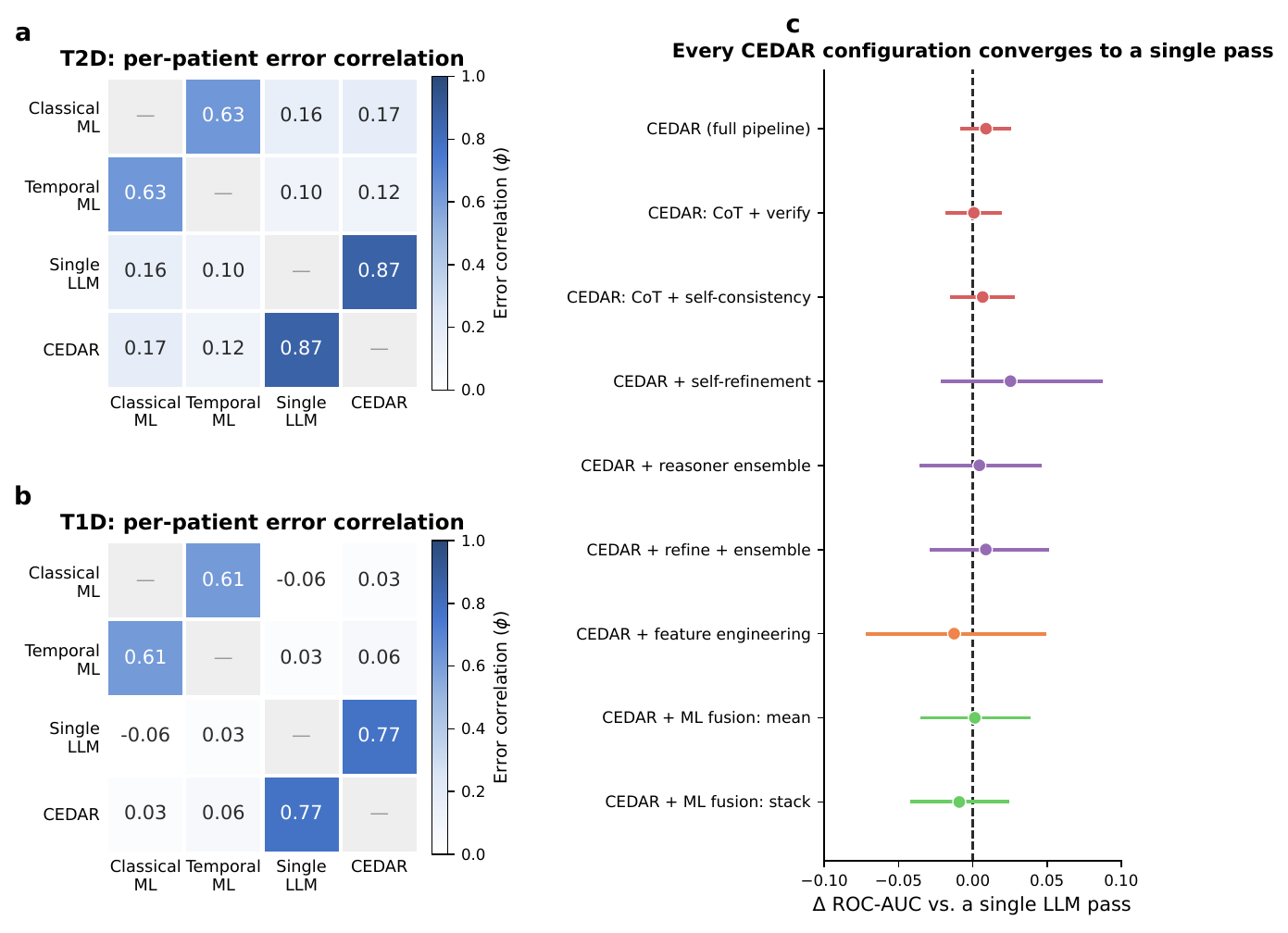}
\caption{\textbf{Error overlap with a single call, and the mechanisms that did not
improve on it} (Supplementary Note~4). (a,~b)~Per-patient error correlation among
all four subgroups, T2D and T1D, as $\phi$ of the misclassification indicators at a
0.5 threshold. The single LLM and \cedar{} err on almost the same patients
($\phi=0.87$ and 0.77), so the added stages do not reshape the error set. (c)~Every
\cedar{} configuration evaluated, as a paired mean $\Delta$ROC-AUC against a single
LLM pass (point $\pm$95\%~CI), colored by the kind of change: internal mechanism,
more reasoning, LLM-guided feature engineering and late ML fusion. All cluster on
zero.}
\label{sfig:llm_errors}
\end{suppfigure}

\clearpage
\bibliography{references}